\documentclass[useAMS,usenatbib]{mn2e}

\usepackage{graphicx}
\usepackage{graphics}        
\usepackage[english]{babel}
\usepackage{url}
\usepackage{amssymb}
\usepackage{amsmath}
\usepackage{wasysym}
\usepackage{pdflscape}
\usepackage{color}
\usepackage{longtable}
\usepackage{rotating}
\usepackage{array}
\usepackage{float}
\usepackage{caption}
\usepackage{natbib}
\usepackage{multicol}

\title[Far-IR Environment of 4C+41.17]{The \textit{Herschel} view of the environment of the radio galaxy 4C+41.17 at $\mathbf{z = 3.8}$}
\author[D. Wylezalek et al.]{D. Wylezalek,$^{1}$\thanks{E-mail:
dwylezal@eso.org}\thanks{{\it Herschel} is an ESA space observatory with science instruments provided by European-led Principal Investigator consortia and with important participation from NASA.} 
J. Vernet,$^{1}$
C. De Breuck,$^{1}$
D. Stern,$^{2}$
A. Galametz,$^{3}$
N. Seymour,$^{4}$ \newauthor
M. Jarvis,$^{5,6}$  
P. Barthel,$^{7}$ 
G. Drouart,$^{1,8}$
T.R. Greve,$^{9}$
M. Haas,$^{10}$
N. Hatch,$^{11}$ \newauthor
R. Ivison,$^{12,13}$ 
M. Lehnert,$^{14}$ 
K. Meisenheimer,${^{15}}$ 
G. Miley,$^{16}$ 
N. Nesvadba,$^{17}$ \newauthor
H.J.A. R\"ottgering,$^{16}$ and
J.A. Stevens$^{5}$ \\
$^{1}$ European Southern Observatory, Karl-Schwarzschildstr.2, D-85748 Garching bei M\"{u}nchen, Germany\\
$^{2}$ Jet Propulsion Laboratory, California Institute of Technology, 4800 Oak Grove Dr., Pasadena, CA 91109, USA \\
$^{3}$ INAF - Osservatorio di Roma, Via Frascati 33, I-00040, Monteporzio, Italy \\
$^{4}$ CASS, PO Box 76, Epping, NSW, 1710, Australia\\
$^{5}$ Centre for Astrophysics Research, STRI, University of Hertfordshire, Hatfield, AL10 9AB, UK \\
$^{6}$ Physics Department, University of the Western Cape, Bellville 7535, South Africa \\
$^{7}$ Kapteyn Astronomical Institute, University of Groningen, PO Box 800, 9700 AV Groningen, The Netherlands \\
$^{8}$ Institut d Astrophysique de Paris, 98bis Bd Arago, 75014 Paris, France \\
$^{9}$ Department of Physics and Astronomy, University College London, Gower Street, London WC1E 6BT, UK \\
$^{10}$ Astronomisches Institut, Ruhr-Universit\"{a}t Bochum, Universit\"{a}tsstr. 150, Geb\"{a}ude NA 7/173, D-44780 Bochum, Germany \\
$^{11}$ School of Physics and Astronomy, University of Nottingham, University Park, Nottingham, NG7 2RD, UK \\
$^{12}$ UK Astronomy Technology Centre, Royal Observatory, Blackford Hill, Edinburgh, EH9 3HJ, UK \\
$^{13}$ Institute for Astronomy, University of Edinburgh, Royal Observatory, Edinburgh, EH9 3HJ, UK \\
$^{14}$ GEPI, Observatoire de Paris, UMR 8111, CNRS, UniversitŽ Paris Diderot, 5 place Jules Janssen, 92190, Meudon, France \\
$^{15}$ Max-Planck-Institut f\"{u}r Astronomie, K\"{o}nigstuhl 17, 69117 Heidelberg, Germany \\
$^{16}$ Leiden Observatory, University of Leiden, P.O. Box 9513, 2300 RA Leiden, Netherlands \\
$^{17}$ Institut d'Astrophysique Spatiale, CNRS, UniversitŽ Paris-Sud, 91405, Orsay, France \\
}

\date{Accepted 2012 October 22. Received 2012 Oct 8 ; in original form 2012 July 19}

\pagerange{\pageref{firstpage}--\pageref{lastpage}} \pubyear{2002}

\def\LaTeX{L\kern-.36em\raise.3ex\hbox{a}\kern-.15em
    T\kern-.1667em\lower.7ex\hbox{E}\kern-.125emX}

\begin{document}
\maketitle

\label{firstpage}

\begin{abstract}
We present \textit{Herschel} observations at 70, 160, 250, 350 and 500 $\mu$m of the environment of the radio galaxy 4C+41.17 at $z = 3.792$. About 65\% of the extracted sources are securely identified with mid-IR sources observed with the \textit{Spitzer Space Telescope} at 3.6, 4.5, 5.8, 8 and 24 $\mu$m. We derive simple photometric redshifts, also including existing 850 $\mu$m and 1200 $\mu$m data, using templates of AGN, starburst-dominated systems and evolved stellar populations. We find that most of the \textit{Herschel} sources are foreground to the radio galaxy and therefore do not belong to a structure associated with 4C+41.17. We do, however, find that the SED of the closest ($\sim 25^{\prime\prime}$ offset) source to the radio galaxy is fully consistent with being at the same redshift as 4C+41.17. We show that finding such a bright source that close to the radio galaxy at the same redshift is a very unlikely event, making the environment of 4C+41.17 a special case. We demonstrate that multi-wavelength data, in particular on the Rayleigh-Jeans side of the spectral energy distribution, allow us to confirm or rule out the presence of protocluster candidates that were previously selected by single wavelength data sets. 
\end{abstract}

\section{Introduction}

\subsection{High-Redshift Radio Galaxies as Tracers of Protoclusters}

High-redshift radio galaxies (HzRGs) are galaxies in the distant universe ($z > 1$) showing enormous radio luminosities \citep*[L$_{500\rm{MHz}} > 10^{27}$WHz$^{-1}$,][]{Miley_2008}. They are extremely rare objects, with number densities $\sim 10^{-8}$ Mpc$^{-3}$ in the redshift range $2 < z < 5$ \citep{Dunlop_1990, Willot_2001, Venemans_2007}. Investigating their spectral energy distribution (SED) reveals features of their stellar, dust and AGN components. In particular, studies of the stellar and dust component have shown that HzRGs are amongst the most massive galaxies in the early universe \citep[e.g.,][]{Seymour_2007, Bryant_2009, Breuck_2010}. According to the hierarchical model of galaxy assembly \citep*{White_1978}, this implies that they reside in  peaks of dark matter overdensities. As galaxy clusters represent the most massive structures in the universe, HzRGs are expected to preferentially reside in sites of galaxy cluster formation. At $z = 2$ the universe is only $\sim$ 3.2 Gyr old and galaxy clusters are likely still forming but have not had time to virialize. For this reason we refer to these matter overdensities as protoclusters.  Observations have indeed shown that HzRGs preferentially reside in overdense environments \citep[e.g.][]{Stevens_2003, Falder_2010, Stevens_2010, Galametz_2010, Galametz_2012, Mayo_2012} and protoclusters are very likely to be found in the vicinity of these objects. As HzRGs are found up to very high redshift, they serve as efficient beacons for identifying very high redshift galaxy clusters. The fields of HzRGs are therefore unique laboratories to study the formation and evolution of the first galaxies and galaxy structures.

\subsection{The HeRG\'E Project}

With the launch of the \textit{Herschel} satellite \citep{Pilbratt_2010}, it is possible for the first time to obtain full coverage of the far-IR SED for a large sample of HzRGs. The \textit{Herschel} Radio Galaxy Evolution project (HeRG\'E) makes use of the two imaging instruments on board \textit{Herschel}: PACS,  the Photodetecting Array Camera \citep{Poglitsch_2010} and SPIRE, the Spectral and Photometric Imaging Receiver \citep{Griffin_2010}. These instruments cover a wavelength range of 70 $- 500 \mu$m and thus constrain the far-IR dust peak very well for a range of redshifts. 
The project was granted $\sim$ 27h of OT1 observing time (PI: N. Seymour) allowing 71 HzRGs to be observed in five bands in PACS and SPIRE (PACS: 70/100 $\mu$m, 160 $\mu$m; SPIRE: 250 $\mu$m, 350 $\mu$m, 500 $\mu$m). In addition to studying the radio galaxies themselves in more detail \citep{Ivison_2012, Rocca_2012, Seymour_2012} project HeRG\'E allows us, for the first time, to systematically study the environments of the radio galaxies at these wavelengths, reaching out 1-3$^{\prime}$ from the HzRGs. This complements our statistical studies of the HzRG environments in the mid-IR \citep{Galametz_2012, Mayo_2012}. Reaching out to longer wavelengths allows us to constrain the dust peak of the SEDs and derive photometric redshift estimates to confirm or rule out overdensities associated with the HzRG. \newline This work reports our pilot study of the well known HzRG 4C+41.17. This analysis will be expanded systematically to the whole data set in future work.

\subsection{4C+41.17}

4C+41.17 at $z = 3.792$ is one of the best studied HzRGs. It was discovered by \citet*{Chambers_1990}. The steep radio spectrum ($\alpha \sim-1.3$) together with extended optical continuum emission and the large rest frame Ly$\alpha$ equivalent width ($\sim 270$ \AA) identified 4C+41.17 as a HzRG. Its high far-infrared luminosity, $L_{\rm{FIR}} \sim10^{13} L_{\astrosun}$ \citep{Benford_1999, Humphrey_2011}, large dust mass \citep{Dunlop_1994} and molecular gas reservoir \citep{Breuck_2005} make this radio galaxy a very likely site of an enormous starburst at high redshift. Similar far-infrared luminosities have also been found for other high redshift radio galaxies \citep{Barthel_2012, Seymour_2012} accumulating the evidence for massive starburst in these galaxies. Deep observations at 450 $\mu$m and 850 $\mu$m carried out with SCUBA  \citep[Submillimeter Common-User Bolometer Array, ][]{Holland_1999} by \citet{Ivison_2000} in the field centered on 4C+41.17 show an order-of-magnitude overdensity of luminous sub-mm galaxies within a 2.5 arcmin diameter region centered on the radio galaxy. From tentative redshift constraints based on the 450 to 850 $\mu$m and the 850 $\mu$m to 1.4 GHz flux density ratios of sources then available, \citet{Ivison_2000} conclude that the overdensity is consistent with lying at the same redshift as the radio source, 4C+41.17, and therefore suggests a likely protocluster. However, photometric redshifts estimated from the 1.6 $\mu$m stellar bump by \citet{Greve_2007} place at least two out of the five sub-mm sources reported by \citet{Ivison_2000} at redshifts lower than 1.3. \citet{Greve_2007} also present deep SHARC-II \citep{Dowell_2003} 350$\mu$m and MAMBO \citep[Max-Planck Millimeter Bolometer Array, ][]{Kreysa_1998} 1200 $\mu$m imaging of the field around 4C+41.17 and combine them with multi-wavelength data at 3.6, 4.5, 5.8, 8 $\mu$m from \textit{Spitzer} IRAC \citep[Infrared Array Camera, ][]{Fazio_2004}, 24 and 70 $\mu$m data from \textit{Spitzer} MIPS \citep[Multiband Imaging Photometer, ][]{Rieke_2004} and 850 $\mu$m observations from SCUBA. They find a surface density of $\sim$ 0.24 1200 $\mu$m sources per arcmin$^{-2}$ to a depth of $\sim$ 2 mJy, consistent with the blank field source density at this wavelength. From cross-correlation analysis and estimation of photometric redshifts, \citet{Greve_2007} conclude that at least half of the sub-mm galaxies are foreground sources and are not, in fact, associated with 4C+41.17. \newline In this paper, we present a multi-wavelength study of the environment of the HzRG 4C+41.17, recently observed within the HeRG\'E project in five PACS and SPIRE bands. \citet{Rocca_2012} present a full, in-depth study of the SED of the radio galaxy itself. We make use of the data at hand to derive photometric redshifts and to confirm or rule out the companionship of the galaxies in the field with the HzRG. Section 2 describes the observations and reduction of the multi wavelength data. Section 3 gives details of the source extraction and cross-correlation. In section 4 we present our analysis and draw conclusions in \S 5. Throughout the paper we assume H$_{0}$ = 70 km s$^{-1}$ Mpc$^{-1}$, $\Omega_{\rm{matter}}$ = 0.3, $\Omega_{\Lambda}$ = 0.7.

\section{Observations and Data Reduction}

\subsection{Far-Infrared Observations}

Observations at 70 and 160 $\mu$m were obtained with the Herschel/PACS instrument on UT 2010 October 12. The image covers $\sim$ 20 arcmin$^{2}.$ We retrieved the Level 0 data from the \textit{Herschel} Science Archive and processed it using version 7.3.0 of the \textit{Herschel} Interactive Processing Environment, HIPE \citep{Ott_2010}. The data were taken to Level 1 following the standard pipelines provided in HIPE. To create Level 2 products, we slightly adapted the standard pipeline to correct for the slew to target data and to improve the point source sensitivity by decreasing the high pass filter radius\footnote{http://herschel.esac.esa.int/twiki/pub/Public/\newline PacsCalibrationWeb/bolopsfv1.01.pdf}. \newline The \textit{Herschel}/SPIRE instrument observed a region covering $\sim$ 80 arcmin$^{2}$ around 4C+41.17 on UT 2010 September 21 with all three bands, at 250 $\mu$m, 350 $\mu$m and 500 $\mu$m. The exposure times for the PACS/SPIRE observations were 2$\times$1404s and 721s, respectively, reaching an average 1$\sigma $ depth of  6.0, 6.4, 10.2, 9.6, 11.2 mJy at 70, 160, 250, 350 and 500 $\mu$m, respectively. Both the SPIRE and PACS observations are part of the guaranteed time key program \textit{The Dusty Young Universe: Photometry and Spectroscopy of Quasars at z $>$ 2} (Observation ID: 1342206336/7 and 1342204958, PI: Meisenheimer). 

\subsection{Mid-Infrared Data}

In addition, we include \textit{Spitzer} IRAC and MIPS observations in the analysis from \citet{Seymour_2007}. The field was deeply mapped using all four IRAC bands (3.6, 4.5, 5.8, 8 $\mu$m - referred to as channels 1, 2, 3 and 4), covering an area of 5.3 $\times$ 5.3 arcmin$^{2}$, and all three MIPS bands  (24, 70 and 160 $\mu$m), covering an area of $\sim 8.0 \times 7.4$ arcmin$^{2}$. The exposure times were 5000s for the IRAC observations and 267 s, 67 s and 2643 s for the three MIPS bands, in order of increasing wavelength. The data were reduced using the \textit{Spitzer} reduction package, MOPEX.  In this work, we only use the 24 $\mu$m images given the deeper PACS observations at longer wavelengths. The 3$\sigma$ depths reached were 0.8, 1.1, 3.2 and 4.3 $\mu$Jy for the IRAC channels 1, 2, 3 and 4, respectively and 30 $\mu$Jy for the MIPS 24 $\mu$m image \citep{Greve_2007}.

\subsection{(Sub)millimetre Data}

A field covering $\sim$ 58 arcmin$^{2}$ around 4C+41.17 was imaged at 1200 $\mu$m with MAMBO. Details of the observations, data reduction and analysis are reported by \citet{Greve_2007}. Positions and flux densities of the extracted sources are taken from there.
\newline
4C+41.17 was also observed at 850 $\mu$m with SCUBA \citep[covering an area $\sim$ 2.5 arcmin in diameter, ][]{Ivison_2000, Stevens_2003}. The data were initially published by \citet{Ivison_2000} and details can be found there. 
\begin{landscape}
\begin{table}
\vspace{5cm}
\caption{Flux densities and $1\sigma$ uncertainties of sources with at least two \textit{Herschel} detections}
\begin{center}
\begin{tabular}{c c c c c c c c c c c c c }
\hline\hline
Source &  $S_{3.6\mu m} $&$S_{4.5\mu m}$ & $S_{5.8\mu m} $& $S_{8\mu m}$ &$ S_{24\mu m}$ &$ S_{70\mu m} $& $S_{160\mu m} $& $S_{250\mu m}$ & $S_{350\mu m} $& $S_{500\mu m}$ & $S_{850\mu m} $&$ S_{1200\mu m} $ \\
 & [$ \mu$Jy] & $ [\mu$Jy] &$  [\mu$Jy] &$ [\mu$Jy] & [mJy] & [mJy] & [mJy] & [mJy] & [mJy] & [mJy] & [mJy] & [mJy]  \\
\hline
1 & $-$ & $-$ & $-$ & $-$ & $-$ & $-$ & $-$ &  $54 \pm   15$ & $  61 \pm 13$ & $ 42 \pm 11$ & $-$ & $-$ \\
2 & $41 \pm 4 $ & $ 46 \pm  5 $ &  $44 \pm  5 $ & $ 51 \pm 5 $ & $ 0.37 \pm 0.03$ & $-$ & $-$ & $47 \pm 11$ & $ 38 \pm 11$ & $36 \pm 16$ & $-$ & $3.0 \pm 0.6$ \\
4 &  $-$ & $-$ & $-$ & $-$ & $0.48 \pm  0.03$ & $-$ & $-$ & $42 \pm  13$ & $22 \pm 7 $ & $-$ & $-$ & $-$ \\
5 & $16 \pm  2 $ & $ 20 \pm 2 $ & $ 26 \pm 3 $ & $ 13 \pm 2 $ & $ 0.31 \pm 0.02$ & $-$ & $-$ & $ 19 \pm  3$  & $  26 \pm  8 $ & $ 10 \pm  3 $ & $-$ & $7.5 \pm 0.6$ \\
7 & $130 \pm 13$ & $ 104 \pm 10$ & $ 72 \pm 8$ & $  193 \pm  20 $ & $ 0.39 \pm 0.03$ & $-$ & $-$ & $31 \pm 13 $ & $ 24 \pm 11 $ & $  19 \pm  9 $ & $-$ & $- $ \\
9 & $ 99 \pm 10 $ & $73 \pm 7 $  & $79 \pm  9 $& $97 \pm 10 $& $ 0.98 \pm 0.05 $ & $  19  \pm  5 $ & $ 67 \pm  6 $ & $ 64 \pm  11 $ & $ 31 \pm 6 $ & $-$ & $-$ & $- $ \\
11 & $29 \pm  3 $ & $ 29 \pm 3 $ & $ 24 \pm 3 $ & $ 26 \pm 3 $ & $ 0.35 \pm 0.03$ & $-$ & $ 24 \pm  7 $ & $  44 \pm  10 $ & $ 39 \pm  8 $ & $ 33 \pm 9 $ & $ 9 \pm 1 $ & $ 4.6 \pm 0.4$ \\
12 & $-$ & $-$ & $-$ & $-$ & $-$ & $-$ &  $19 \pm 7$ & $ 27 \pm  9 $ & $ 17 \pm 4 $ & $-$ & $-$ & $3.0 \pm 0.6$ \\ 
13 & $-$ & $27 \pm  3 $ & $-$ & $ 45 \pm  5$ & $ 0.32 \pm  0.02 $ & $-$ & $-$ & $15 \pm 6 $ & $30 \pm 14 $ & $-$ & $-$ & $-$ \\
16 & $11  \pm 1 $ & $  14 \pm  2 $ & $  24 \pm 3$ & $ 45 \pm 5 $ & $ 0.47 \pm  0.03$ & $-$ & $ 15 \pm  6 $ & $ 42 \pm 5 $ & $48 \pm 4 $ & $ 39 \pm 4$ & $12 \pm 1 $ & $3.8Ê\pm 0.4$ \\
4C+41.17 & $ 17 \pm 2 $ & $ 20 \pm 2 $ & $ 27 \pm 3 $ & $ 31 \pm 4 $ & $ 0.36 \pm 0.03 $ & $-$ & $16 \pm 7 $ & $ 36 \pm 4 $ & $ 43 \pm  4 $ & $ 38 \pm 5$ & $12 \pm 1 $ & $ 4.4 \pm 0.4$ \\
18 & $-$ &$-$ &$-$ &$-$ &$-$ & $-$ & $27 \pm 6 $ & $ 33 \pm 8 $ & $  21Ê\pm 6 $ & $-$ & $-$ & $-$ \\
19 & $ 128 \pm  13$ & $ 106 \pm 11$ & $ 98 \pm 11 $ & $ 131 \pm 13 $ & $ 0.62 \pm 0.03$ & $ 8 \pm  3 $ & $23 \pm 7 $ & $29 \pm 11 $ & $-$ & $-$ & $-$ & $-$ \\
21 &  $21 \pm	2 $ & $ 26 \pm 3 $ & $28 \pm 3 $  & $ 22 \pm 2 $ & $-$ & $-$ & $-$ & $ 15 \pm 3 $ & $36 \pm 15 $ & $21 \pm 9 $ & $-$ & $2.6 \pm 0.6$ \\   
24 &  $38 \pm 4 $ & $-$ & $51 \pm 5 $ & $-$ & $-$ & $-$ & $-$ & $ 10 \pm  4 $ & $11 \pm 5 $ & $-$ & $-$ & $3.6 \pm 0.6$ \\
28 & $-$ & $-$ & $-$ & $-$ & $0.35 \pm  0.03$ & $-$ & $-$ & $37 \pm  4 $ & $ 22 \pm  8 $ & $-$ & $-$ & $-$ \\
29 & $130 \pm 13 $ & $137 \pm 14 $ & $ 117 \pm  12 $ & $368 \pm 37 $ & $ 1.41 \pm  0.07$ & $-$ & $-$ &  $ 13  \pm   4 $ & $17 \pm  5 $ & $-$ & $-$ & $- $ \\
\hline
\end{tabular}
\end{center}
\label{fluxes}
\end{table}
\end{landscape}

\section{Source Extraction and Cross-correlation Analysis} 
\subsection{Source Extraction}
\subsubsection{PACS/SPIRE Source Extraction}

\begin{figure}
\begin{center}
 \parbox{25mm}{
    \centering 
    \includegraphics[width=25mm, height =25mm]{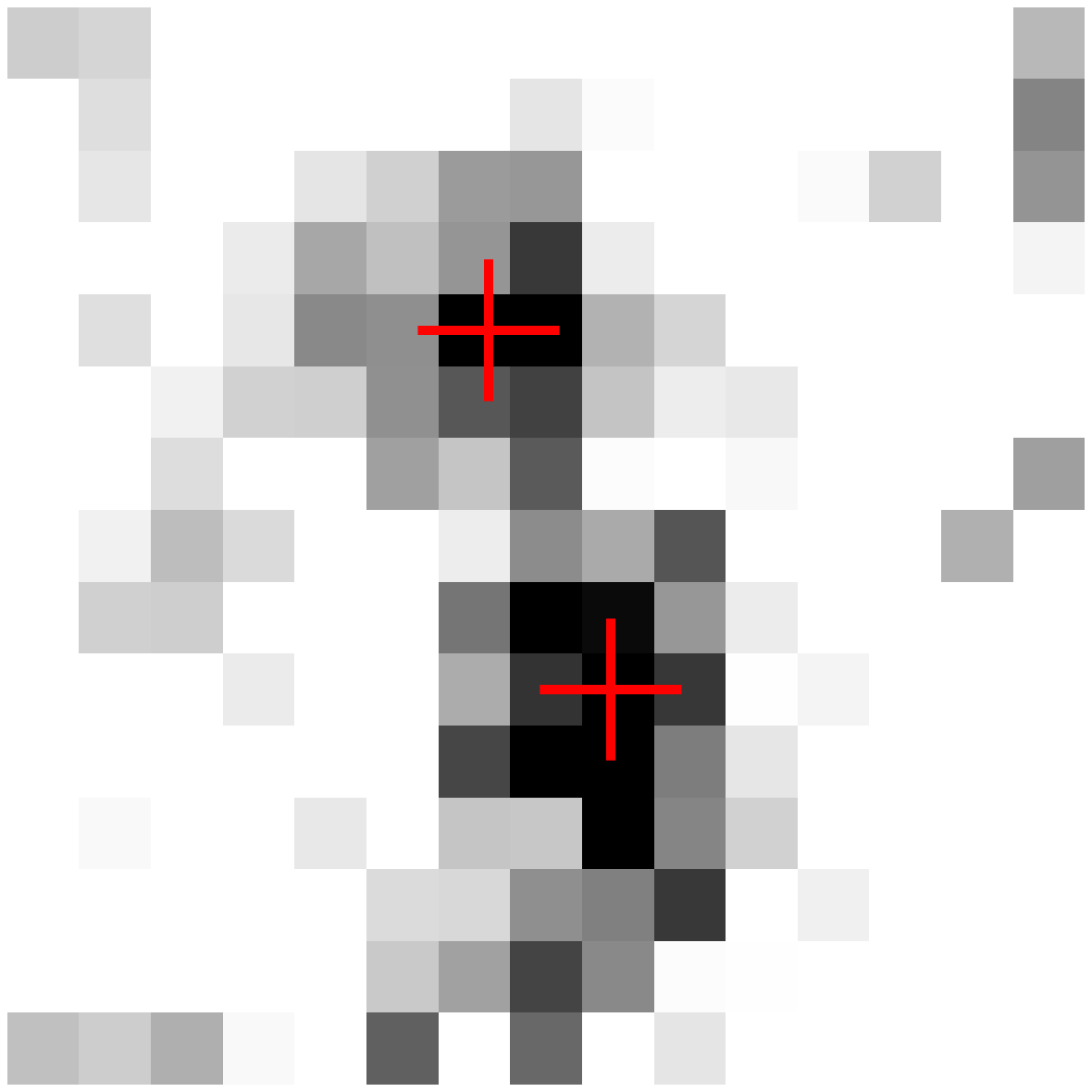}
  }
  \parbox{25mm}{
    \centering
    \includegraphics[width=25mm,height =25mm]{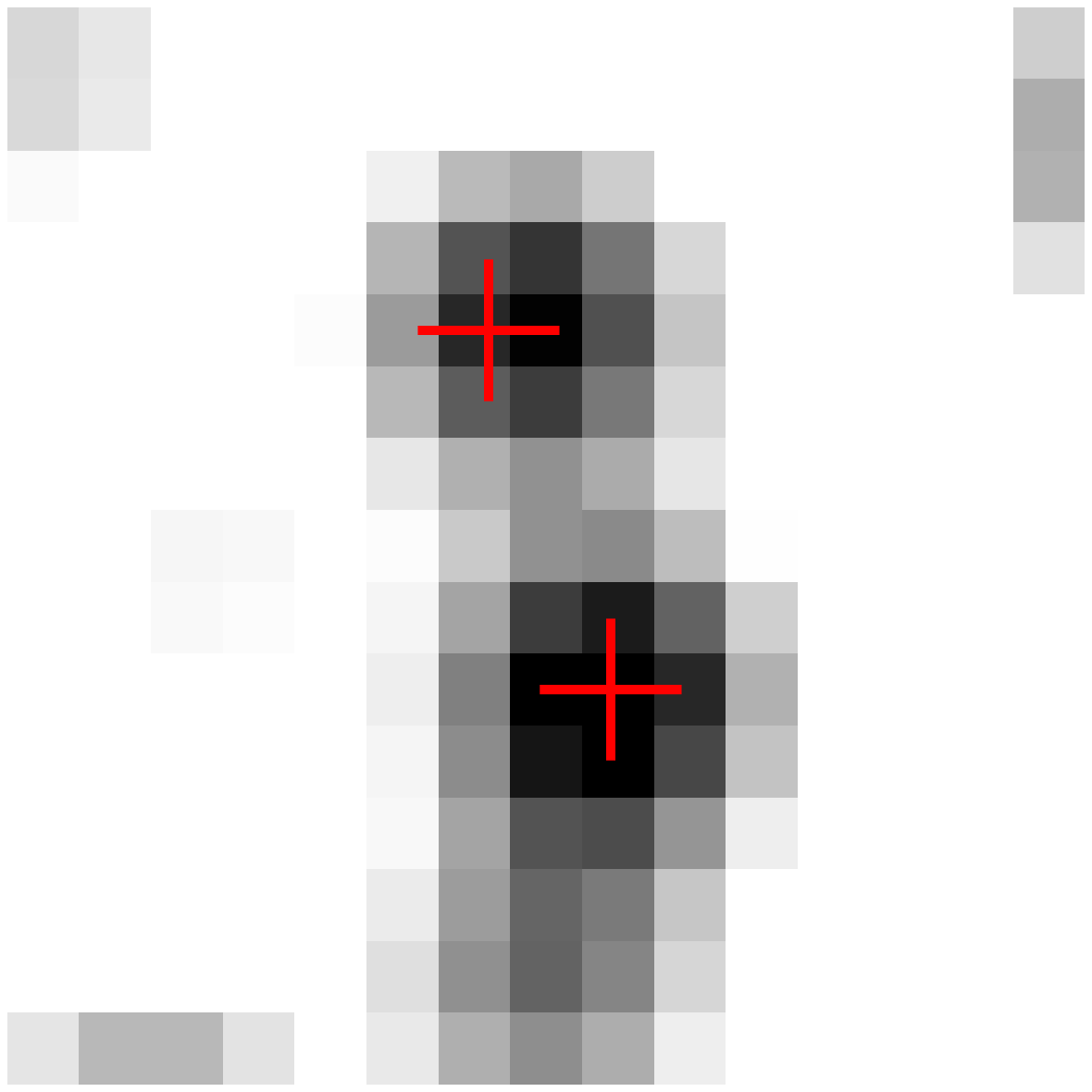}
  }
   \parbox{25mm}{
    \centering
    \includegraphics[width=25mm, height =25mm]{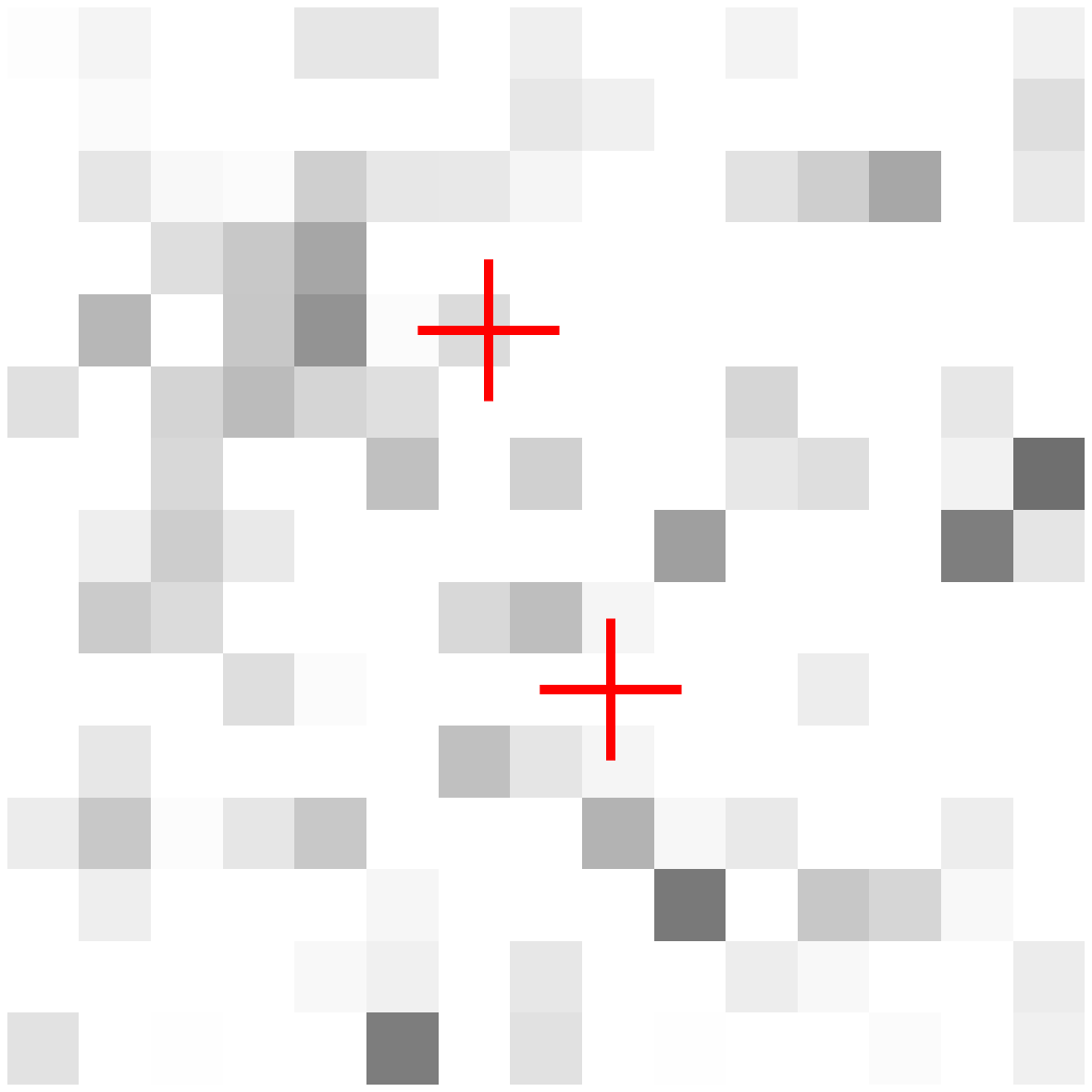}
   }
   
\parbox{25mm}{
   \centering
    \includegraphics[width=25mm, height =25mm]{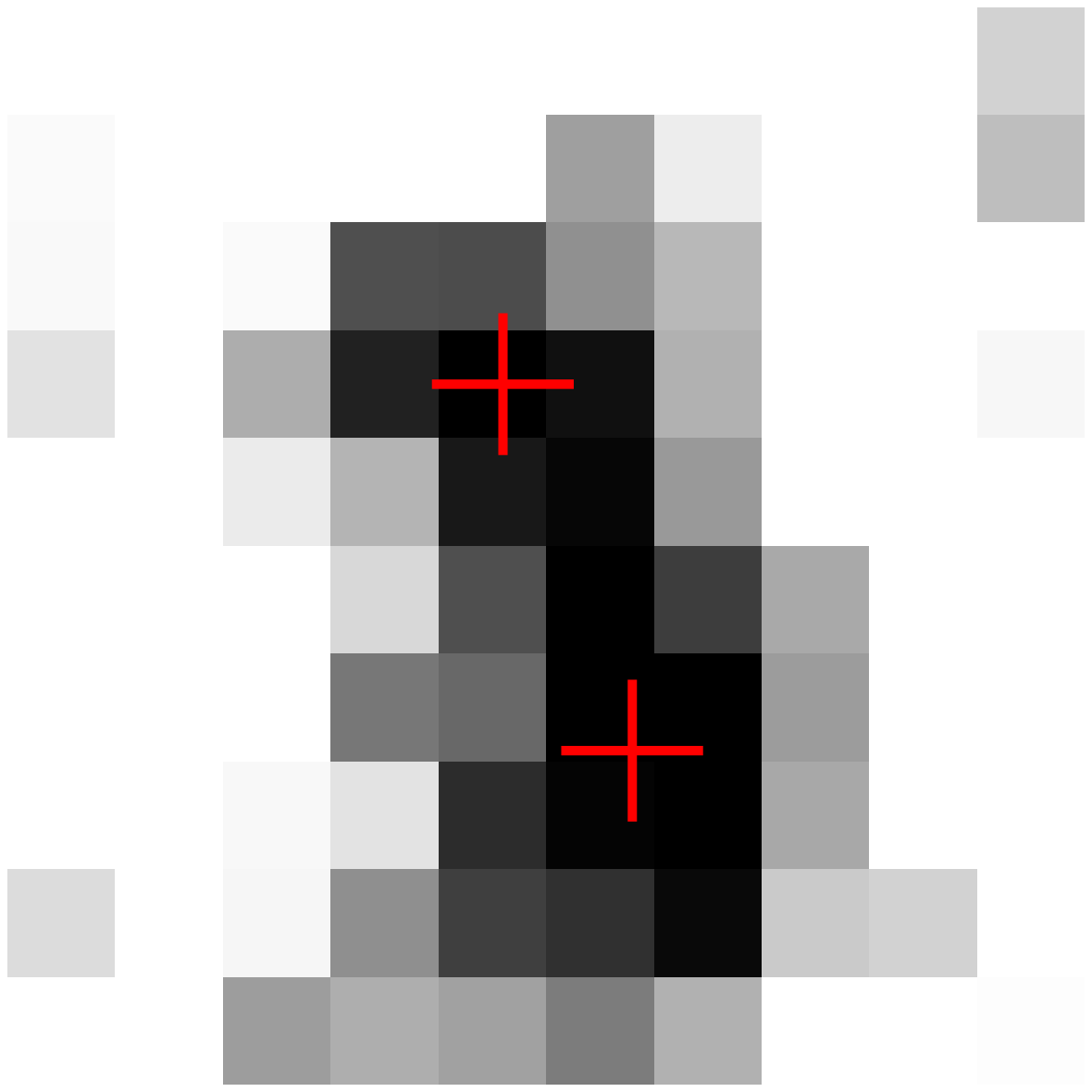}
   }
\parbox{25mm}{
    \centering
    \includegraphics[width=25mm, height =25mm]{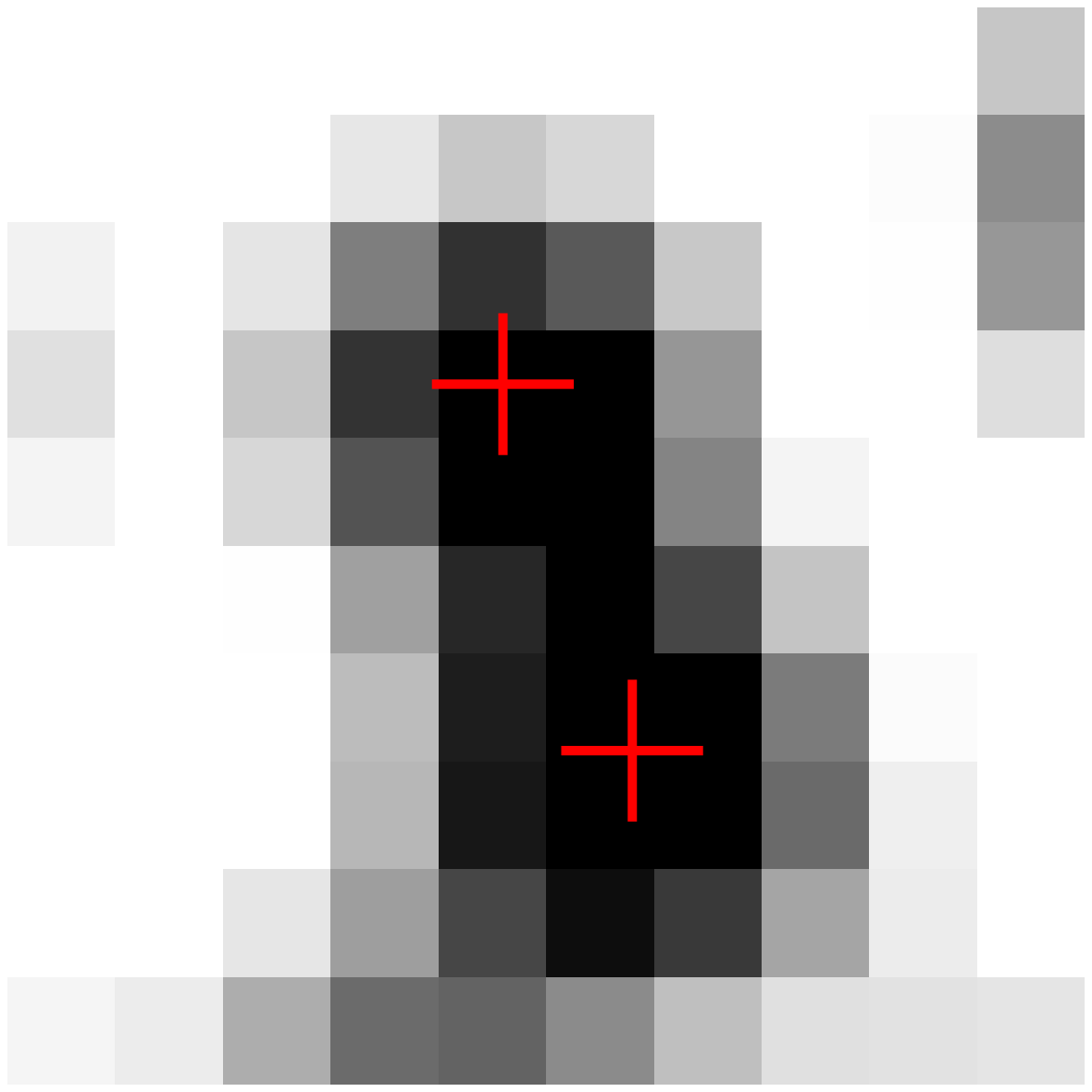}
   } 
\parbox{25mm}{
    \centering
    \includegraphics[width=25mm, height =25mm]{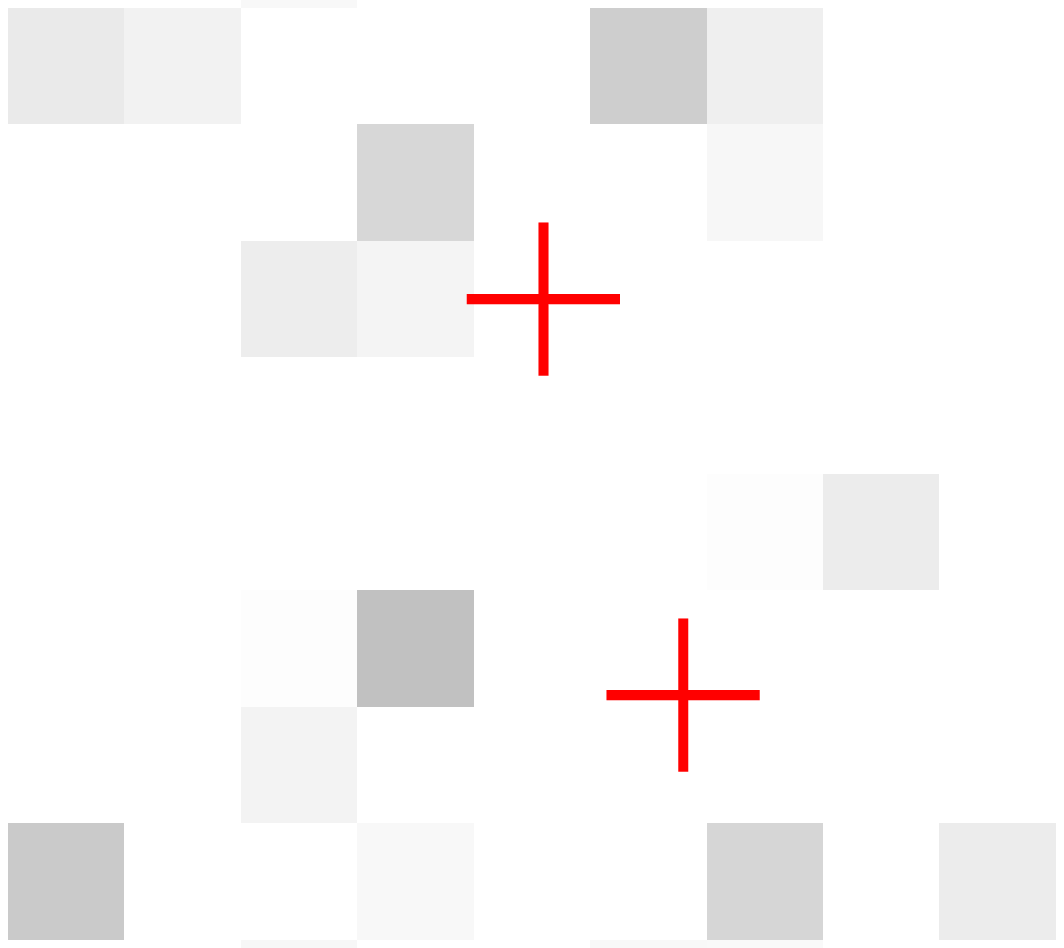}
   } 

 \parbox{25mm}{
    \centering 
    \includegraphics[width=25mm, height =25mm]{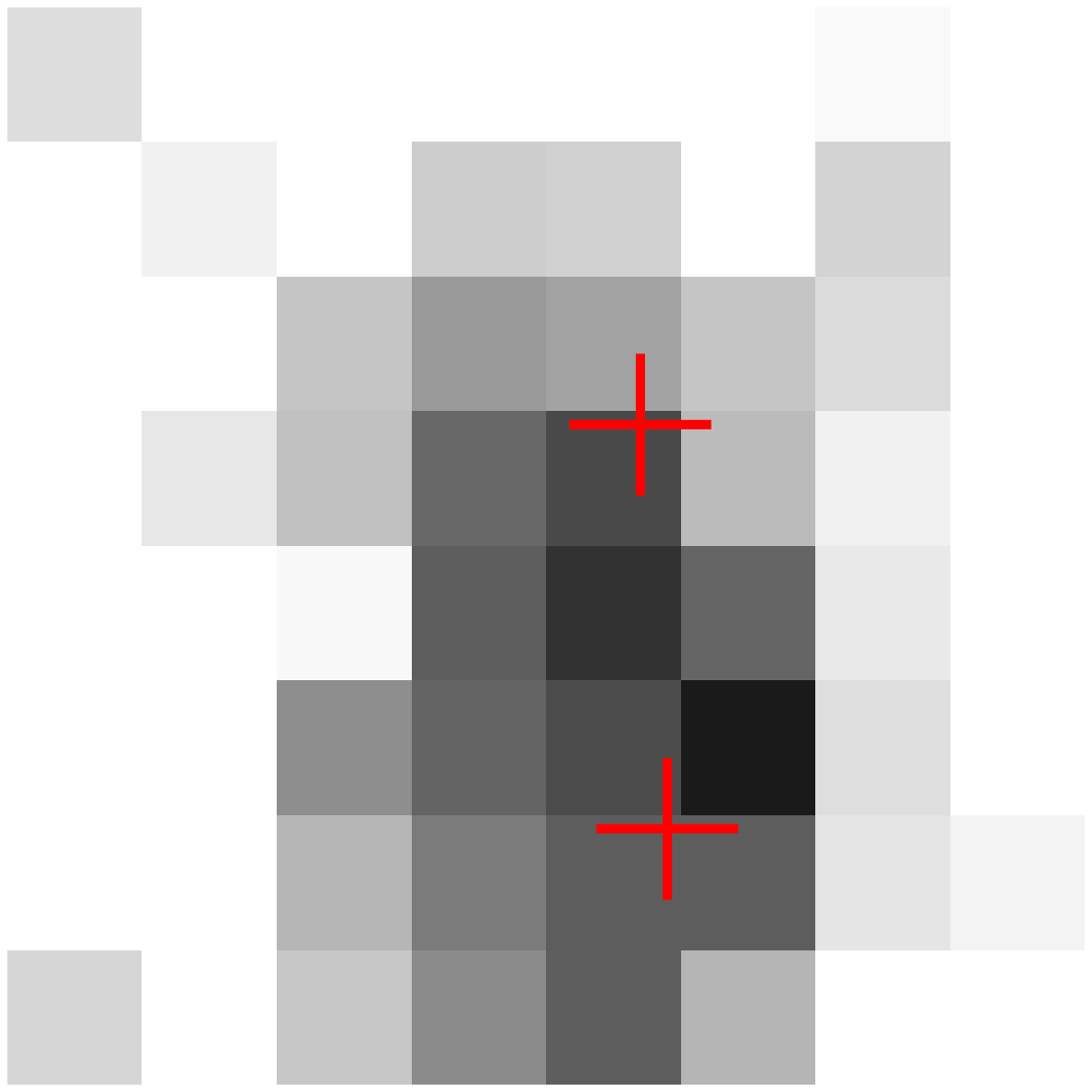}
  }
  \parbox{25mm}{
    \centering
    \includegraphics[width=25mm, height =25mm]{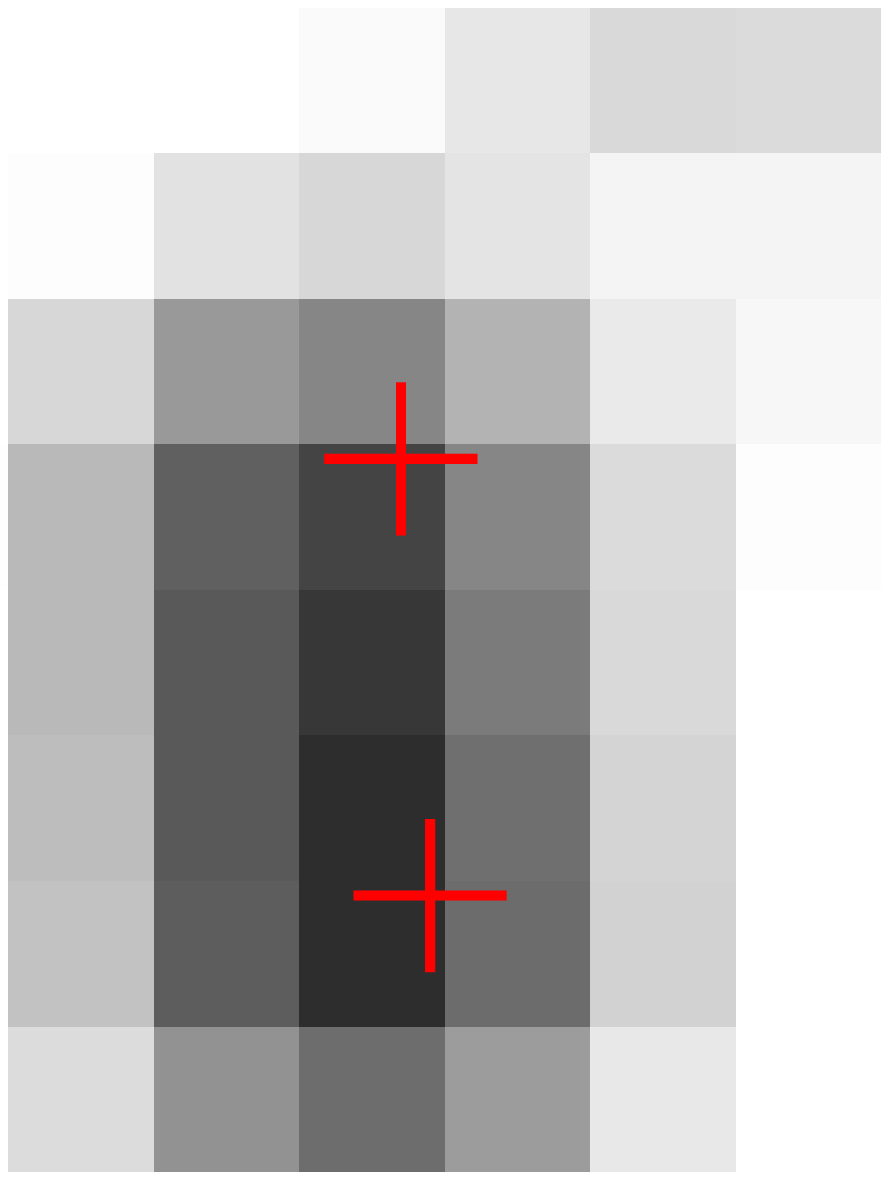}
  }
   \parbox{25mm}{
    \centering
    \includegraphics[width=25mm, height =25mm]{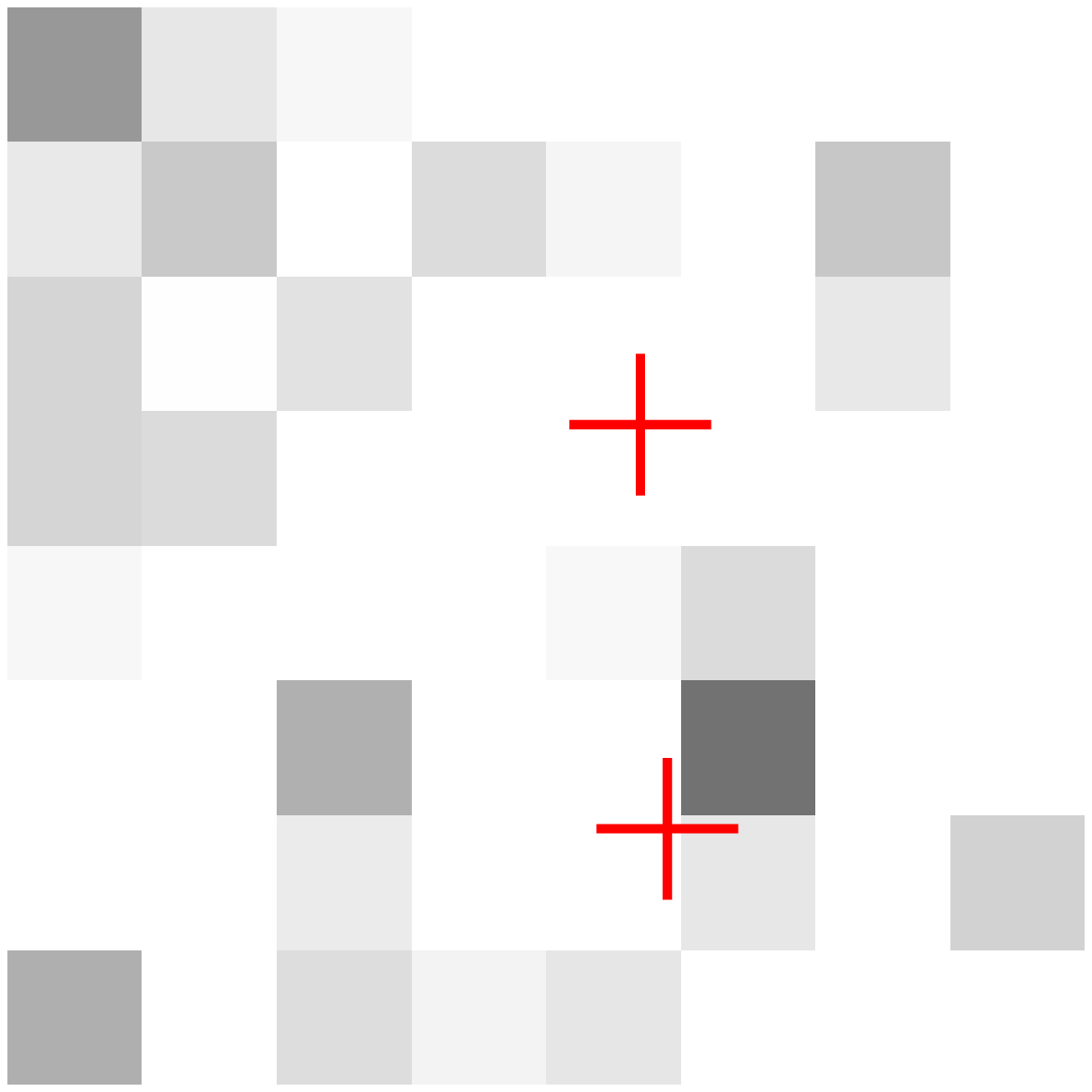}
   }
   \end{center}
\caption{$0.6^{\prime} \times 0.6^{\prime}$ postage stamps of the data (left), synthetic image derived by StarFinder (center) and residuals (right) for source 16 and 4C+41.17. From top to bottom the images at 250, 350 and 500 $\mu$m are shown. The sources are blended in all three images but the homogeneous residual image shows the good deblending with StarFinder. Red crosses indicate the positions of 4C+41.17 (upper source) and source 16 (lower source).}
\label{blend}
\end{figure}

Source extraction in the PACS and SPIRE images is performed using the tool sourceExtractorDaophot that is included in HIPE. The FWHM of the different bands are taken from the \textit{PACS Observer's Manual}\footnote{http://www.iac.es/proyecto/herschel/pacs/pacs\_om.pdf, p. 13} and are 5.2, 12, 18.1, 25.2 and 36.3$^{\prime\prime}$ for 70, 160, 250, 350 and 500 $\mu$m, respectively. The parameters for the source extraction, such as shape parameters roundness and sharpness, are tuned such that false detection rates and source blending is minimized (Table \ref{extraction}). We extract sources at a significance $\geqslant 2.5 \sigma$ within a circle of 3.3$^{\prime}$ (corresponding to 34.2 arcmin$^{2}$) radius around the radio galaxy. Due to the scanning mode the coverage is inhomogenous further away from the image center. We extract two sources from the PACS 70 $\mu$m image, 8 sources from the PACS 160 $\mu$m image, 27 sources from the SPIRE 250 $\mu$m image, 16 sources from the SPIRE 350 $\mu$m image and 8 sources from the SPIRE 500 $\mu$m image. The extracted source positions and given names are listed in Table \ref{sources} in order of increasing RA. We derive aperture photometry for the extracted sources applying an aperture correction of 1.45 and 1.44 to the blue and red PACS flux densities, respectively\footnote{http://herschel.esac.esa.int/Docs/PACS/html/pacs\_om.html}. Due to the inhomogeneous coverage in the \textit{Herschel} images the uncertainty on the flux densities is derived from sky annuli (see Table \ref{extraction}) around each source.  \newline Aperture photometry is, however, not applicable in the case of source 16 and 4C+41.17, which are blended. We therefore apply PSF photometry to those sources using StarFinder \citep{Diolaiti_2000}, a code designed to analyze images in very crowded fields. The deblending strategy in StarFinder consists of an iterative search for residuals around the object and subsequent fitting. We assumed the PSF to be Gaussian with a FWHM corresponding to the beam size. The positions of the two sources were determined independently in each \textit{Herschel} image as different material is probed at different wavelengths. The flux density measurements with StarFinder are consistent with the ones obtained with HIPE for unblended sources. Postage stamps of the image, synthetic image and residual image after deblending is shown in Figure \ref{blend}. No other sources in the field are blended and the fluxes densities and uncertainties \citep[including the $15\%$ and $7\%$ flux calibration uncertainties added in quadrature to the statistical uncertainties for PACS and SPIRE flux densities, respectively, ][]{Seymour_2012} are given in Table \ref{fluxes}.

\begin{table*}
\caption{HIPE parameters used to extract sources with sourceExtractorDaophot and details of the observations in different \textit{Herschel} bands. Roundness and sharpness parameters between -2 to 2 and -1.5 to 2 were used for the extraction in all bands.}
\begin{center}
\begin{tabular}{c c c c c c c c c}
\hline\hline
Band	& FWHM &  beam area & aperture radius& sky annulus& exposure time & average $1\sigma$ depth \\
 & [arcsec] &  [arcsec$^{2}$] &  [arcsec]  &  [arcsec] & & [mJy]\\
\hline
PACS 70 $\mu$m & 5.2 & 30.6 & 6 & 6-10 & $2\times1404$s & 6.0 \\
PACS 160 $\mu$m & 12 & 163.2 & 10 & 10-15  & $2\times1404$s & 6.4\\
SPIRE 250 $\mu$m & 18.1 & 373.3 & 22 & 22-32 & 721s & 10.2 \\
SPIRE 350 $\mu$m & 25.2 & 716.7 & 30 & 30-40 & 721s & 9.6 \\
SPIRE 500 $\mu$m & 36.3 & 1493.1 & 42 & 42-52 & 721s & 11.2\\
\hline
\end{tabular}
\end{center}
\label{extraction}
\end{table*}

\subsubsection{IRAC/MIPS Source Extraction}

Source extraction is performed using SExtractor \citep{Bertin_1996} in dual image mode using the 4.5 $\mu$m image for detection. We only report sources detected with a significance $\geqslant 3\sigma$. Unlike \citet{Greve_2007}, we use a smaller, 4$^{\prime\prime}$ diameter aperture for the IRAC images because of the close proximity of other sources in that crowded field. Tests with an aperture of 9.26$^{\prime\prime}$ diameter show that flux from neighbouring sources results in overestimated flux densities (e.g. for 4C+41.17 itself). We apply aperture corrections of 1.205, 1.221, 1.363, 1.571 to IRAC channels 1, 2, 3 and 4, respectively. MIPS 24 $\mu$m flux densities are measured in 5.25$^{\prime\prime}$ aperture radii. The aperture correction applied, 1.78, is calculated as described by the MIPS instrument handbook\footnote{http://irsa.ipac.caltech.edu/data/SPITZER/docs/mips/\newline mipsinstrumenthandbook/50/}. The uncertainties reported in Table \ref{fluxes} include the $10\%$ and $4.5\%$ systematic uncertainties for the IRAC and MIPS flux densities, respectively, that were added in quadrature to the statistical uncertainties to account for the absolute flux calibration and color correction uncertainties \citep{Seymour_2007}.

\begin{table*}
\tiny
\caption{Astrometry of the \textit{Herschel} sources in the 4C+41.17 field, listed in order of increasing RA. Positions are in the J2000 system.}
\begin{center}
\begin{tabular}{c c c c c c c c c c c}
\hline\hline
Source & RA$_{70\mu\rm{m}}$ & Dec.$_{70\mu\rm{m}}$ & RA$_{160\mu\rm{m}}$ & Dec.$_{160\mu\rm{m}}$ & RA$_{250\mu\rm{m}}$& Dec.$_{250\mu\rm{m}}$ & RA$_{350\mu\rm{m}}$ & Dec.$_{350\mu\rm{m}}$  &RA$_{500\mu\rm{m}}$  & Dec.$_{500\mu\rm{m}}$  \\
 &  &  &  &  &   &  \\
\hline
1&	&	&	&	&	6:50:37.0&	41:28:54&	6:50:36.9&	41:28:53&	6:50:36.6&	41:28:44\\
2&	&	&	&	&	6:50:40.5&	41:30:03&	6:50:40.7&	41:30:05&	6:50:40.7&	41:30:05\\
3&	&	&	&	&	6:50:40.9&	41:32:35&	&	&	&	\\
4&	&	&	&	&	6:50:42.6&	41:28:27&	6:50:42.8&	41:28:25&	&	\\
5&	&	&	&	&	6:50:43.1&	41:29:21&	6:50:43.7&	41:29:16&	6:50:43.2&	41:29:17\\
6&	&	&	&	&	6:50:44.0&	41:27:43&	&	&	&	\\
7&	&	&	&	&	6:50:45.6&	41:32:40&	6:50:44.9&	41:32:37&	6:50:46.4&	41:32:52\\
8&	&	&	&	&	6:50:47.1&	41:27:37&	&	&	&	\\
9&	6:50:47.4&	41:30:46&	6:50:51.1&	41:30:06&	6:50:47.2&	41:30:44&	6:50:47.3&	41:30:46&	&	\\
10&	&	&	&	&	6:50:47.4&	41:28:11&	&	&	&	\\
11&	&	&	6:50:48.9&	41:31:27&	6:50:48.8&	41:31:26&	6:50:49.1&	41:31:28&	6:50:49.0&	41:31:22\\
12&	&	&	6:50:50.1&	41:28:21&	6:50:50.1&	41:28:19&	6:50:50.0&	41:28:21&	&	\\
13&	&	&	&	&	6:50:50.3&	41:33:01&	6:50:50.5&	41:32:58&	&	\\
14&	&	&	&	&	6:50:50.3&	41:27:37&	&	&	&	\\
15&	&	&	&	&	6:50:51.2&	41:29:48&	&	&	&	\\
16&	&	&	6:50:51.2&	41:30:06&	6:50:51.4&	41:30:07&	6:50:50.9&	41:30:04&	6:50:51.1&	41:30:06\\
4C+41.17&	&	&	6:50:51.9&	41:30:32&	6:50:51.7&	41:30:32&	6:50:52.1&	41:30:31&	&	\\
18&	&	&	6:50:52.4&	41:28:53&	6:50:52.2&	41:28:51&	6:50:52.4&	41:28:55&	&	\\
19&	6:50:54.4&	41:29:33&	6:50:54.4&	41:29:33&	6:50:54.2&	41:29:34&	&	&	&	\\
20&	&	&	6:50:54.6&	41:30:47&	&	&	&	&	&	\\
21&	&	&	&	&	6:50:54.8&	41:32:36&	6:50:54.8&	41:32:29&	6:50:53.8&	41:32:42\\
22&	&	&	&	&	&	&	&	&	6:50:56.1&	41:33:35\\
23&	&	&	&	&	6:50:56.4&	41:34:57&	&	&	&	\\
24&	&	&	&	&	6:50:60.0&	41:27:58&	6:50:59.6&	41:27:58&	&	\\
25&	&	&	&	&	6:51:00.6&	41:29:07&	&	&	&	\\
26&	&	&	&	&	6:51:00.8&	41:31:16&	&	&	&	\\
27&	&	&	&	&	6:51:01.8&	41:33:03&	&	&	&	\\
28&	&	&	&	&	6:51:02.1&	41:31:59&	6:51:01.7&	41:31:57&	&	\\
29&	&	&	&	&	6:51:04.5&	41:29:44&	6:51:03.9&	41:29:49&	&	\\
\hline
\end{tabular}
\end{center}
\label{sources}
\end{table*}

\subsection{Cross-Correlation Between Bands}

After extracting sources in the different images with very different spatial resolutions we cross-correlate the sources in order to derive a clean, multi-wavelength source catalog. We only consider the 17 sources that have at least two detections in the \textit{Herschel} bands in order to minimize false detections. For the cross-correlation, we choose the SPIRE 250 $\mu$m whose 1$\sigma$ positional accuracy ($\sim 0.6 \times \frac{\rm{FWHM}}{\rm{SN}}$)\footnote{http://herschel.esac.esa.int/hcss-doc-8.0/load/hcss\_urm/html/herschel.ia.toolbox.srcext.\newline SourceExtractorDaophotTask.html} outperforms the other bands. Although the PACS images have an even better spatial resolution, they cannot be used systematically as reference images due to their shallowness and small field of view (see Figure \ref{zvert}). We then cross-correlate the cleaned source list with the sources detected at shorter and longer wavelengths. \newline We look for MIPS counterparts within 10$^{\prime\prime}$ of the 250 $\mu$m sources which corresponds to about the half-width at half maximum for the $250 \ \mu$m observations and which also corresponds to their $3\sigma$ positional error assuming the bulk of the 250 $\mu$m detections has a signal to noise ratio (SN) of $\sim$ 3 \citep{Magnelli_2012}. Following \citet*{Sutherland_1992}, we calculate the reliability $R=exp(-\pi r^{2}\sigma_{1}\sigma_{2}N)$ of finding no random source closer than the nearest candidate where $N$ is the number density of background objects, $r=\sqrt{(d_{1}/\sigma_{1})^{2}+(d_{2}/\sigma_{2})^{2}}$ is the normalized distance, $d_{1}$ and $d_{2}$ are the positional differences in each axis between the sources, and $\sigma_{1}$ and $\sigma_{2}$ are the standard deviations of the error ellipse. Since $\sigma_{1} = \sigma_{2}$ for our case, $R$ simplifies to $R=exp(-{\pi (d_{1}^{2}+d_{2}^{2})N}$).  \citet{Mayo_2012} find a surface density 0.549 arcmin$^{-2}$ at the depth of the MIPS image. This is in agreement with the density found by \citet{Papovich_2004} for the 23.1 $\mu$Jy depth of the MIPS image of 4C+41.17. We adopt this value for calculating the reliability of MIPS-250 $\mu$m sources counterparts.  Candidates with reliabilities $R \geqslant$ 90$\%$ are counted as correct identifications. We then cross-correlate the MIPS identifications with the IRAC catalogs and compute the reliability in the same way using a surface density of 2.80 arcmin$^{-2}$ from \citet{Galametz_2012}. For sources with no MIPS identification, we cross-correlated the 250 $\mu$m sources directly with the IRAC catalogs. Again, candidates with reliabilities $\geqslant$ 90$\%$ are counted as correct identifications. Although we require R $\geqslant$ 90$\%$, the reliabilities for MIPS identifications are all above 96\% and for IRAC identifications above 98\%. Table \ref{separation} lists the separations between the MIPS and 250 $\mu$m counterparts and the IRAC and MIPS counterparts together with the corresponding reliabilities of them to be the correct identifications. Following this methodology, out of the 17 \textit{Herschel} 250 $\mu$m sources, we identified MIPS counterparts for 11 sources (65\%), and IRAC counterparts for an overlapping, but not identical set of 11 far-IR sources (65\%). We checked the identifications where better resolved PACS, intermediate wavelength data were available and confirmed our identified sources. \newline In the Appendix we show postage stamps of all sources with SPIRE 250 $\mu$m sources that have high probability (R $>90$\%) counterparts in at least two IRAC bands and are detected in at least two \textit{Herschel} bands. We also overplot the sub-mm position from \citet{Greve_2007} where available. Except for sources 19 and 24, the \citet{Greve_2007} astrometry is in very good agreement with our mid-IR identifications. 

\begin{table}	
\caption{Separation between MIPS and 250 $\mu$m sources, between IRAC and MIPS sources, and calculated reliabilities for the nearest source to be the correct identification. Where no value is given, the IRAC and/or MIPS source is either outside the field of view or there are no detections within a 10$^{\prime\prime}$ radius. Italized numbers indicate that there are only IRAC2 and 4 or IRAC1 and 3 detections for the 250 $\mu$m source as the other bands are outside the field of view. In case of source 21 the IRAC-250 $\mu$m position and corresponding probability is listed.}
\begin{center}
\begin{tabular}{c c c c c }
\hline\hline
Source &  MIPS$-250\mu$m & $R_{24\mu m}$ & IRAC$-$MIPS & $R_{\rm{IRAC}}$ \\
 & separation & & separation & \\
  &  [$^{\prime\prime}$] &  & [$^{\prime\prime}$] & \\
\hline
1 & $-$ &  $-$ & $-$ & $-$ \\
2 & 9.18 &  0.96 &  2.81 & 0.98 \\
5 & 4.37 &      0.99 &   0.58 &      0.99 \\
7 &  3.70 &      0.99 &  0.89 &      0.99 \\
9 & 4.99    &  0.99 &  0.62     &  0.99\\
11 &  5.01 &      0.99 &  2.46 & 0.99 \\
12 & $-$ &  $-$ & $-$ & $-$ \\
13 & 2.54 &  0.99 &  \textit{0.51} &     \textit{ 0.99} \\
16 & 0.60 &      0.99 &       0.59 &      0.99 \\
4C+41.17 &  6.46 &   0.98 &     0.96 &     0.99 \\
19 & 4.86 &      0.99 &  0.71 &  0.99 \\
21 & $-$ & $-$ &  2.47 &      0.99 \\
28 &  4.87 &    0.99 & $-$ & $-$ \\
29 & 5.26 &     0.99 &  0.40 &  0.99 \\
\hline
\end{tabular}
\end{center}
\label{separation}
\end{table}

\section{Analysis}
\subsection{Photometric Redshifts}

We now derive photometric redshifts in order to investigate a physical connection between the radio galaxy and the objects in its vicinity. The following issues must be kept in mind when deriving photometric redshifts from combined sub-mm, far-IR and near-IR observations: 
\begin{enumerate} \renewcommand{\theenumi}{(\arabic{enumi})}
\item As the far-IR emission is of thermal origin, changing the dust temperature has the same effect on the sub-mm/mm colors as shifting the spectrum in redshift \citep{Blain_2002}. It is thus impossible to estimate the redshift from far-IR data alone; supporting observations are necessary to constrain a redshift.
\item The dust (at far-IR, sub-mm wavelengths) to stellar flux density (at near-IR wavelengths) ratio has a range of about 3 decades and varies with morphology, total IR luminosity and gas-phase metallicity  \citep{Skibba_2011}. This is not always well represented in the available template libraries. When deriving our own empirical templates, we therefore create templates with a wide range of dust to stellar ratios, ranging from 100 to 5000.
\item Differing error bars for near-IR and far-IR observations will introduce a bias in the photometric redshift fitting procedure, giving the high signal-to-noise ratio of the IRAC data more weight.
\end{enumerate}
For all 11 sources with detections in more than five wavelength bands we calculate photometric redshifts using the code {\tt hyperz} \citep*{Bolzonella_2000} which minimizes the reduced $\chi^{2}$ to find the best photometric redshift solution. We use both synthetic and empirical AGN and starburst templates from the SWIRE template library \citep{Polletta_2007} complemented with our own newly derived templates. The latter are obtained by combining a 1 Gyr old stellar population template from the P\'{e}gase.2 spectral evolution model \citep*{Fioc_1999} dominating the near-IR emission and empirical dust templates dominating the far-IR/sub-mm emission. Three dust templates are derived by (1) fitting the dust peak of 4C+41.17, a typical AGN dominated galaxy at high redshift; (2) the dust peak of the lensed ``eyelash" galaxy at $z  = 2.3$ \citep[SMM J2135-0102, ][]{Ivison_2010, Swinbank_2010}, a typical starburst galaxy at high redshift; and (3) source 11, which is very well sampled at far-IR/sub-mm wavelengths and for which the spectroscopic redshift is known \citep[$z_{\rm{spec}} = 1.18, $][]{Greve_2007}. In this way, not only templates derived from lower redshift galaxies, such as the SWIRE templates, are available to us but also templates derived from higher redshift galaxies.  For each of the three dust templates we create 15 composite templates with different ratios between the stellar emission in the near-IR and the dust emission in the far-IR. In order to get a matching wavelength coverage of the self-derived templates and the SWIRE templates we extend our templates by using greybody fitting results (see Section 4.2) for $\lambda > 1200\ \mu$m. We then extract the best-fitting templates from our 45 self-derived templates and the SWIRE library templates. Ultimately, a set of 9 different templates (see Table \ref{templates}) are used for the 11 sources for which we derive photometric redshifts. Templates 1 (Spiral C) and 3 (starburst) (both from the SWIRE library) are generated from the SED of these objects using the GRASIL code \citep{Silva_1998} and improved by using IR spectra from the PHT-S spectrometer on the \textit{Infrared Space Observatory} and from IRS on \textit{Spitzer} \citep{Houck_2004}. Template 2 is an empirical composite AGN+starburst template that fits IRAS 19254-7245. Template 4, 5, 6, 7, 8 and 9 are new, self-derived templates, with their properties described in Table \ref{templates}. The resulting $\chi^{2}$ distribution and the best $\chi^{2}$ are thus derived by considering all redshifts and all templates in the final set. Note that the final $\chi^2$ curve shows the minimum $\chi^{2}$ for the template set as a function of redshift and therefore is dependent on the template set used. \newline Because of varying spatial coverage of the multi-wavelength data, filters are ignored for `out-of-field' sources, but when a source is observed, but undetected, 3$\sigma$ upper limits are taken into account by {\tt hyperz}. We present the results of our photometric redshift estimates in Table \ref{photoz} and show the best-fitting SEDs in the Appendix. \newline As mentioned above, differing errors bars for the near-IR and far-IR observations introduce a bias in the fitting procedure giving the high signal-to-noise IRAC data more weight. However, by allowing a range of various ratios between the stellar (near-IR) and dust (far-IR) emission in the fitting templates we already make sure that the fits are not dependent on the IRAC data only but that also the relative contribution of the sources of emission is taken into account. To test this in more detail, we repeat the fitting procedure by relaxing the IRAC uncertainties to $20\%$. The best-fit redshifts are in agreement with the previously derived ones within the uncertainties. This shows that our results are not strongly biased by the IRAC data. \newline \citet{Greve_2007} derived photometric redshifts using the 1.6 $\mu$m rest-frame stellar `bump' in the observed IRAC data. They also estimate the redshift from the radio/sub-mm/mm color, but these only yield crude estimates, consistent with the redshift estimation from the stellar bump. Therefore, we only list $z_{\rm{bump}}$ in Table \ref{photoz}, which compares to our photometric redshifts and spectroscopic redshifts that exist for some of the sources. The uncertainties listed in Table \ref{photoz} only reflect the 1$\sigma$ formal uncertainties near the minimum of the $\chi^{2}$ distribution and may be severely underestimated. We describe the quality of the photometric redshifts for each source in the Appendix. Figure \ref{zvert} shows the spatial distribution of the sources, with white, open stars representing lower redshift objects ($z < 3$) and filled stars representing objects with $z >3$.
\begin{table*}
\caption{Summary of the templates used for deriving photometric redshifts. Template 1-3 are from the SWIRE template library \citep{Polletta_2007}. Template 4-9 are newly derived by combining the far-IR emission of 4C+41.17, the eyelash galaxy (SMM J$2135$-$0102$) and source 11, for which the spectroscopic redshift is known, with a 1 Gyr old stellar population from the P\'{e}gase.2 spectral evolution model \citep*{Fioc_1999}. The far-IR and stellar emission were normalized to their peak flux densities and combined with varying ratios, as indicated.}
\begin{center}
\begin{tabular}{c l}
\hline\hline
Template ID & Description \\
\hline
1&  Spiral C galaxy template, SWIRE template library \\
2 & Seyfert 2$+$Starburst/ULIRG template for IRAS 19254-7245, SWIRE template library \\
3 &  Starburst/ULIRG template for IRAS 20551-4250, SWIRE template library \\
4 & 4C+41.17 far-IR template $+$ old stellar population, stellar peak to dust peak ratio$: 300:1$\\
5 & 4C+41.17 far-IR template $+$ old stellar population, stellar peak to dust peak ratio$: 700:1$\\
6 & 4C+41.17 far-IR template $+$ old stellar population, stellar peak to dust peak ratio$:4500:1$\\
7 & SMM J2135-0102 far-IR template $+$ old stellar population, stellar peak to dust peak ratio$: 500:1$ \\
8 & source 11 far-IR template $+$ old stellar population, stellar peak to dust peakratio$: 700:1$ \\
9 & source 11 far-IR template $+$ old stellar population, stellar peak to dust peak ratio$: 1700:1$ \\
\hline
\end{tabular}
\end{center}
\label{templates}
\end{table*}
\begin{table}
\caption{Photometric redshifts, $z_{\rm{phot}}$, were derived with {\tt hyperz} for all sources with at least five detections. Stellar bump photometric redshifts, $z_{\rm{bump}}$, and spectroscopic redshifts, $z_{\rm{spec}}$ are from \citet{Greve_2007}. Template ID's are described in Table \ref{templates}.}
\begin{center}
\begin{tabular}{c c c c c}
\hline\hline
Source & $z_{\rm{phot}}$& template ID  & $z_{\rm{bump}}$& $z_{\rm{spec}}$\\
\hline
2 & 2.5$\pm 0.4$ &3& $\sim$ 1.8& \ldots\\
5 &2.4$\pm 0.2$  &3&$\sim$ 2.6& 2.672$\pm$ 0.001\\
7 &0.5$\pm 0.1$& 1 &\ldots &\ldots  \\
9 & 0.6$\pm 0.1$& 8 & \ldots&\ldots \\
11 & 1.2$\pm 0.2$& 9 & $<$ 1.3 &1.184 $\pm$ 0.002\\
13 & 2.2$\pm 0.4$& 5 &\ldots &\ldots  \\
16 &4.0$\pm 0.1$& 6 & \ldots& \ldots \\
4C+41.17 & 3.5$\pm 0.2$ & 2 & $\sim$ 4&3.792 $\pm$ 0.001\\
19 &  2.0$\pm 0.1$ & 4& $<$ 1.3 & 0.507 $\pm$ 0.020 \\
21 &  2.7$\pm 0.2$ & 7& $\sim$ 1.8 &\ldots  \\
29 &1.0$\pm 0.1$& 4 &\ldots & \ldots \\
\hline
\end{tabular}
\end{center}
\label{photoz}
\end{table}
\newline The average redshift for these 11 sources is 2.0 $ \pm$ 0.8. This agrees with the average redshift for SPIRE 250 $\mu$m selected sources, $z =1.8 \pm 0.2$, recently found by \citet{Mitchell_Wynne_2012}. \newline Most of the SPIRE-selected sources are found to be at $z < 2.5$, ruling out any physical connection with the radio galaxy and confirming that most of the far-IR sources in the vicinity of 4C+41.17 are likely foreground. Only one source, object 16, potentially lies at the same redshift as 4C+41.7. The $\chi^{2}$ distribution of this source shows a clear dip at $z = 3.8$. We therefore assume that object 16 and 4C+41.17 are at the same redshift, $z = 3.8$, and adopt this assumption for our subsequent analysis.
\begin{figure}
\parbox{84mm}{
\centering
\includegraphics[width = 84mm]{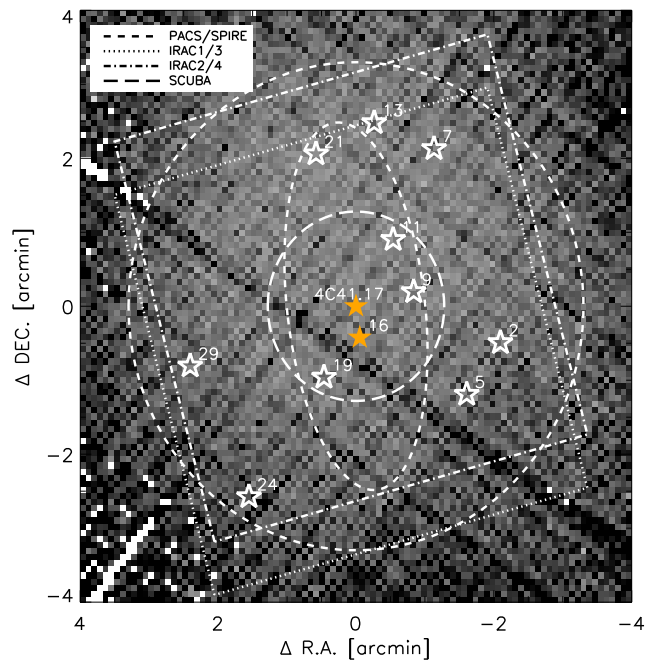}
}
\caption{Coverage map and spatial distribution of sources with derived photometric redshifts, centered on 4C+41.17. Dark pixels indicate regions with low coverage. White, open stars indicate sources that have  $z_{\rm{phot}}<$ 3, orange, filled stars show sources with $z_{\rm{phot}} >$ 3. 4C+41.17 and source 16 are very likely to be at the same redshift. The dashed circle ($r = 3.3^{\prime}$)/ellipse ($a=2.5^{\prime}$, $b = 1^{\prime}$) indicates the area for identifying targets in the SPIRE and PACS maps, respectively. The dotted box shows the coverage of IRAC1 and IRAC3, the dashed-dotted box shows the coverage of IRAC2 and IRAC4 and the long dashed circle shows the coverage of SCUBA. The MIPS and MAMBO images cover the whole SPIRE area and are therefore not illustrated here. The coverage/error in the region used for extraction is not homogeneous.}
\label{zvert}
\end{figure}

\subsection{\textit{Herschel} non-detections}

\citet{Ivison_2010} report five 850 $\mu$m detected sources, of which two are described as marginal detections; for the robust subset, we find viable \textit{Herschel} counterparts to two. We find the same detection rate for the robust 1200-$\mu$m sources reported by \citet{Greve_2007}. The fact that they are not detected in any other wavelength band may suggest some of them are just statistical fluctuations. This would be especially true for those that are only marginally detected. The 1200 $\mu$m sources, however, are all observed at a significance $\geqslant$ 5$\sigma$. If the sources are real they are likely very dust obscured sources belonging to the high redshift ($z > 4$) tail of sub-mm bright star-forming galaxies \citep{Swinbank_2012, Walter_2012}. None of these sources has significant MIPS detections or unambiguous IRAC detections and we are therefore not able to estimate a likely redshift range.  

\subsection{Far-IR Luminosities, Star-Formation Rates and Limits}

For sources with more than 4 detections in the the SPIRE, MAMBO and SCUBA bands we derive dust temperatures, far-IR luminosities and star formation rates (SFR). Sources 2, 5, 11, 16, 4C+41.17 and 21 were fitted with a grey-body law of the form: $S_{\nu} \propto \nu^{\beta}B_{\nu}(T) = \frac{\nu^{\beta+3}}{(e^{h\nu/kT_{d}}-1)}$ where $S_{\nu}$ is the flux density at the rest-frame frequency $\nu$, $\beta$ the grain emissivity index and $T_{d}$ the dust temperature. Dust temperatures for an interstellar medium only heated by star-formation in expected to range between $\sim$ 20--60 K; $\beta$ can range between 1--2.5  \citep{Casey_2012}. Far-IR luminosities were derived by integrating their SED over the wavelength range 40-1000 $\mu$m and applying the relation $L_{\rm{FIR}} = 4\pi D_{L}^{2}F_{\rm{FIR}}$ where $D_{L}$ is the luminosity distance computed from their photometric redshifts. Where spectroscopic redshifts were available those were applied. Source 16 was assumed to be at the redshift of the radio galaxy ($z = 3.8$). We then estimated the star formation rates by using $SFR \ [M_{\astrosun}] = L_{\rm{FIR}}/5.8\times 10^{9}$ L$_{\astrosun}$ \citep{Kennicutt_1998}. The results are listed in Table \ref{sfrs}. \newline Given the shallowness of the SPIRE images, at $z=3.8$, we are only sensitive to the most massive starbursts. Assuming a dust emission from the starburst with $\beta =  1.5$ and $T_{d} = 45 \  \rm{K}$, we find that at $z = 3.8$ we can only detect galaxies with a SFR $\gtrsim 2600$ M$_{\astrosun}$yr$^{-1}$. We therefore can only report on the presence of strongly starbursting galaxies in the field of 4C+41.17. 
\begin{table}
\caption{Derived dust temperatures ($T_d$), grain emissivity indices ($\beta$), far-IR luminosities ($L_{\rm{FIR}}$) and star formation rates (SFRs) for sources with at least 4 detections in SPIRE, MAMBO and SCUBA. Spectroscopic redshifts where used where available and are marked in italics. For source 16, we assumed the redshift of 4C+41.17 ($z=3.8$).}
\begin{center}
\begin{tabular}{c c c c c c}
\hline\hline
Source & $z$ & $T_{d}$ & $\beta$ & $L_{\rm{FIR}}$ & SFR \\
 & & [K] & & [$10^{13}$ L$_{\astrosun}$] & [M$_{\astrosun}$yr$^{-1}$] \\
\hline
\hline
       2    & 2.6 & 38 &  1.6  & 0.7   & 1200 \\
       5 &    \textit{2.7} & 40 & 1.5 &0.3  &     500 \\
      11 &  \textit{1.2} & 26 & 1.4 & 0.1  &   200    \\
      16  &  3.8 & 48  & 1.7 &1.8   &    3100  \\
      4C+41.17  & \textit{3.8} & 52  & 1.6 &1.6   &    2800 \\
      21 & 2.7 & 32   & 1.7 &0.4   &    700 \\
      \hline
\end{tabular}
\end{center}
\label{sfrs}
\end{table}

\subsection{Number Density}
In order compare the source density to the \textit{Herschel} wide field surveys \textit{Herschel}-ATLAS \citep{Eales_2012} and HerMES \citep{Oliver_2012}, we restrict this analysis to a flux density limit at which our catalogs are complete. We estimate the incompleteness by placing artificial sources in our images and applying the source extraction algorithm on the modified images. The number of successful recoveries then provides us with an estimate of the incompleteness for various flux density bins. A completeness of $\sim$ 95\% is reached at $\sim$ 35mJy (corresponding to a SFR $\sim 2600$ M$_{\astrosun}$yr$^{-1}$ at $z = 3.8$) for our 250 $\mu$m catalog. We find 8 sources above that flux density limit in the extraction area of 34.2 arcmin$^{2}$, resulting in a surface density of $\sim$ 0.23 $\pm$ 0.08 arcmin$^{-2}$ for the field of 4C+41.17. The SPIRE blank field number counts at our flux density limit found by \citet{Clements_2010} are $N(S_{\rm{250 \mu m}}> 36\  \rm{mJy}) \sim 0.121 \pm 0.002\ \rm{arcmin}^{-2}$ suggesting a marginally significant overdensity of a factor $\sim$ 2. On the other hand, \citet{Oliver_2010} find blank field number counts of $N(S_{\rm{250 \mu m}}> 30\  \rm{mJy}) \sim 0.18 \pm 0.15\ \rm{arcmin}^{-2}$. We find 10 sources with flux densities above 30mJy corresponding to a surface density of $0.3\pm 0.1$ arcmin$^{-2}$. As our catalogs are not yet complete at that limit this result has to be treated with caution but still hints to a slight overdensity for the field of 4C+41.17. 
\newline \citet{Mayo_2012} find a density of MIPS 24 $\mu$m sources which is $\sim$ 2 times higher than the blank field mean density, which agrees with our \textit{Herschel} observations. Nevertheless, this field was not counted as significantly overdense as it is still less than $3\sigma$ above the mean. Compared to the mean density of HzRG fields analyzed by \citet{Mayo_2012}, the field around 4C+41.17 is typical as compared to other radio galaxies in terms of density ($\delta_{\rm{4C+41.17}} = 2.1\ \rm{arcmin}^{-2}$, compared to $\langle\delta_{\rm{HzRG}}\rangle \sim 2.2\ \rm{arcmin}^{-2}$). \newline Selecting color-selected high-redshift IRAC sources in the fields of HzRGs, \citet{Galametz_2012} finds 4C+41.17 to be overdense with a compact clump of IRAC sources identified $\sim$ 1$^{\prime}$ south of the radio galaxy. However, the IRAC color criterion applied simply identifies sources at $z > 1.2$. Given that we find an excess of galaxies at $z \sim 2.5$ (see Figure \ref{zvert}), the clump detected by \citet{Galametz_2012} is likely a foreground structure, as also suggested by \citet{Greve_2007}.
\newline Considering all data at hand, we therefore find no indications for a remarkable overdensity in the field of 4C+41.17.\newline
In the 250 $\mu$m image, source 16 is found $\sim$ 25 arcsec from the 250 $\mu$m position of 4C+41.17 ($\sim$ 180 kpc at $z = 3.8$). The probability of finding such a bright far-IR source (42 mJy at 250 $\mu$m, 48 mJy at 350 $\mu$m) at this distance to the HzRG is 4\% and 8\% for 250 $\mu$m and 350 $\mu$m, respectively \citep{Oliver_2010}. This probability is not remarkable when considering the whole sky but is very special when the evidence points to the two sources lying at the same redshift. The probability of finding a 250 $\mu$m selected source at $z=3.8$ is also only $\sim$ 5\% \citep{Mitchell_Wynne_2012}. The probability of finding a source of that flux density at 25$^{\prime\prime}$ distance at $z=3.8$ will therefore be $\ll 4$\% and is a very unlikely event.

\section{Summary and Conclusions}

Using \textit{Herschel} PACS and SPIRE observations combined with \textit{Spitzer} mid-IR observations, we have carried out a multi-wavelength study of the environment of 4C+41.17, a powerful radio galaxy at $z = 3.8$. This pilot study for the HeRG\'E project clearly shows that far-IR observations combined with shorter wavelength observations improve our ability to securely distinguish overdensities found by different selection criteria \citep[e.g.][]{Galametz_2012, Mayo_2012} from truly clustered structures. The field of 4C+41.17 has long been thought to host a galaxy cluster associated with the radio galaxy \citep{Ivison_2000}. \citet{Greve_2007} already concluded from stellar bump photometric redshifts that most of the sources might belong to a foreground structure. Only source 16 \citep[J065051.4 in ][]{Greve_2007} appeared possibly associated with 4C+41.17. In this work we find strong indications that these two sources lie at the same redshift and thus that there is a physical association between them. \citet{Ivison_2000} and \citet{Greve_2007} find that the radio galaxy and source 16 appear to be connected by a faint bridge in both the SCUBA and MAMBO map increasing the likelihood that this source is part of the same system as 4C+41.17. Source 16 makes the environment of 4C+41.17 special as the probability of finding such a bright source that close ($\sim$ 25 arcsec distant, 180 kpc at $z = 3.8$) is only $\sim$ 5\%. \newline However, close-by companion sources might actually be a common feature for HzRGs. \citet{Ivison_2008} finds two clumps of emission 3.3 arcsec distant from the HzRG 4C+60.07 that are most likely merging with the $z = 3.8$ radio galaxy. \citet{Ivison_2012} finds a bright sub-mm galaxy near the radio galaxy 6C 1909+72 that is most likely sharing the same node or filament of the cosmic web. Also, \citet{Nesvabda_2009} find two CO-emission line components at a distance of $\sim$ 80 kpc from the HzRG TXS0828+193 ($z = 2.6$) which may be associated with a gas-rich, low-mass satellite galaxy. Although these companions are found much closer to the HzRG than source 16 is to 4C+41.17, these observations suggest that companion sources around HzRGs may be a common feature \citep[see also][]{Ivison_2012}. We find that most of the \textit{Herschel} far-IR sources in the vicinity of 4C+41.17 are foreground sources. However, this does not rule out the presence of a cluster around 4C+41.17 as our observations are only sensitive to galaxies with SFRs $\gtrsim 2600$ M$_{\astrosun}$yr$^{-1}$. \newline Caution is needed when identifying overdensities from a single wavelength data set. With  IRAC and MIPS data available for all sources being observed by the HeRG\'E project we will be able to identify likely protocluster candidates around the HzRGs. However, $850\ \mu$m data are required to constrain the Rayleigh-Jeans part of the SED. We have therefore started a systematic SCUBA-2 follow up campaign to map the full SPIRE area of the HeRG\'E fields.

\section*{Acknowledgments}

T. R. Greve acknowledges support from the UK Science and Technologies Facilities Council. N. Seymour is the recipient of an Australian Research Council Future Fellowship. The work of D. Stern was carried out at Jet Propulsion Laboratory, California Institute of Technology, under a contract with NASA. The \textit{Herschel} spacecraft was designed, built, tested, and launched under a contract to ESA managed by the \textit{Herschel}/\textit{Planck} Project team by an industrial consortium under the overall responsibility of the prime contractor Thales Alenia Space (Cannes), and including Astrium (Friedrichshafen) responsible for the payload module and for system testing at spacecraft level, Thales Alenia Space (Turin) responsible for the service module, and Astrium (Toulouse) responsible for the telescope, with in excess of a hundred subcontractors.

\appendix
\section[]{Notes on individual sources}
In the Appendix  we give more details on all sources with photometric redshifts derived in this work. For each source $60^{\prime\prime}\times60^{\prime\prime}$ IRAC and MIPS grey-scale postage stamps are shown in the first row. The second row shows PACS and SPIRE grey scale postage stamps, $100^{\prime\prime}$ on a side. The postage stamps are centered on the 250 $\mu$m centroid, indicated by the orange square. The blue circle in the upper row shows the $10^{\prime\prime}$ search radius for the cross correlation analysis, the smaller, red circle in all stamps shows the apertures for the IRAC, MIPS, PACS and SPIRE images. The green diamond indicates the MAMBO position (in case of no MAMBO position, the SCUBA position) from \citet{Greve_2007}. We also show the SEDs and the minimum $\chi^{2}$ as a function of redshift from {\tt hyperz}. Black data points are measured values, green arrows upper limits and red dashed lines the best-fitting redshifted template. Detailed notes for each source are given below.

\begin{figure*}
 \parbox[height=20mm]{20mm}{
    \centering 
    \includegraphics[width=20mm]{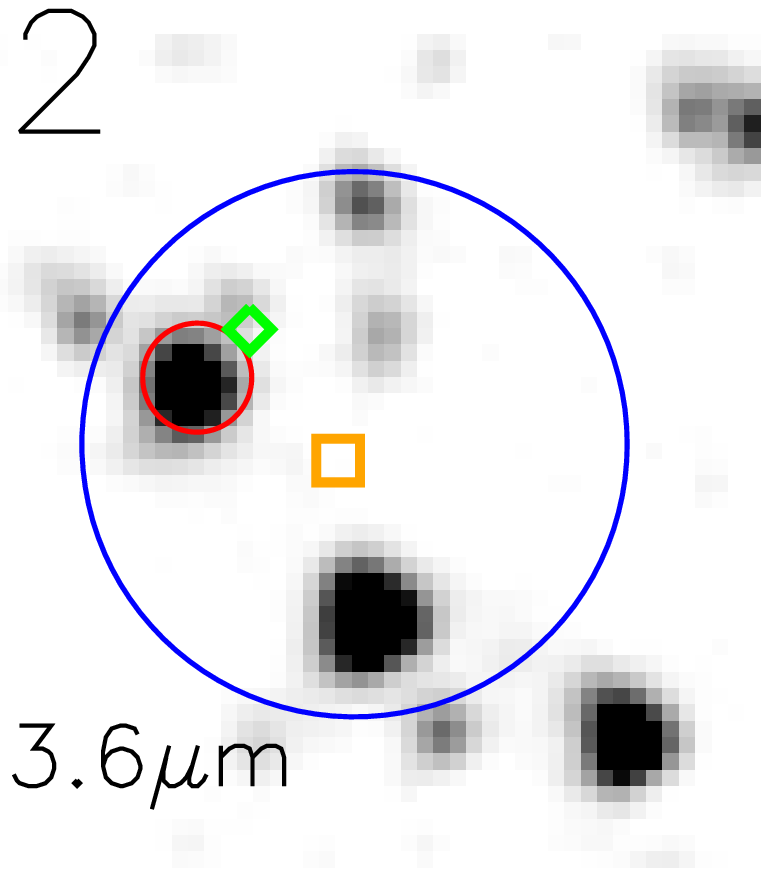}
  }
  \parbox[height=20mm]{20mm}{
    \centering
    \includegraphics[width=20mm]{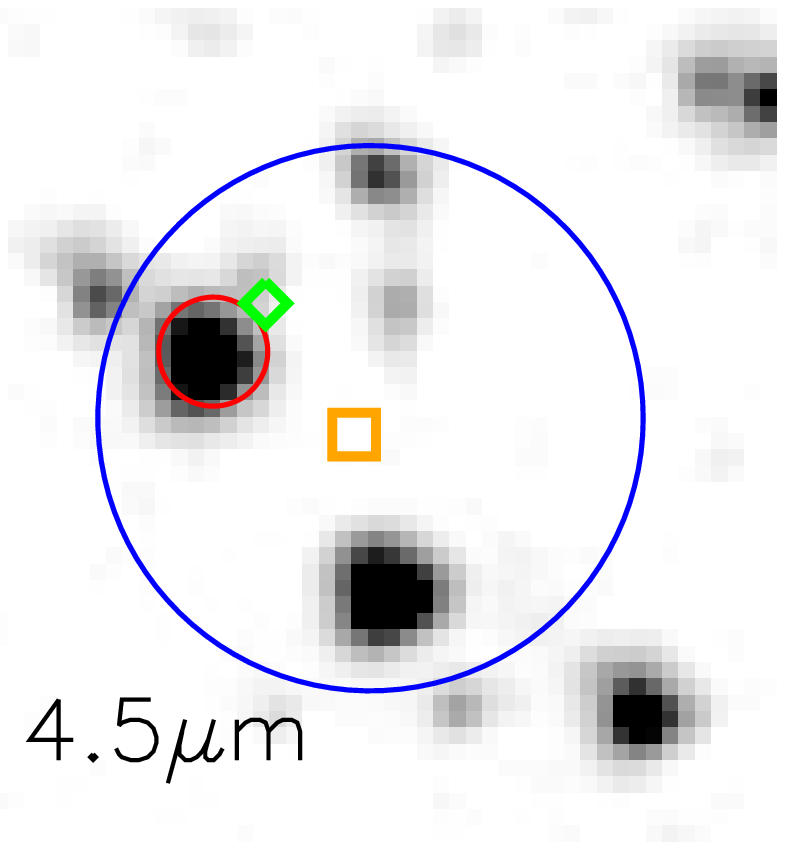}
  }
   \parbox[height=20mm]{20mm}{
    \centering
    \includegraphics[width=20mm]{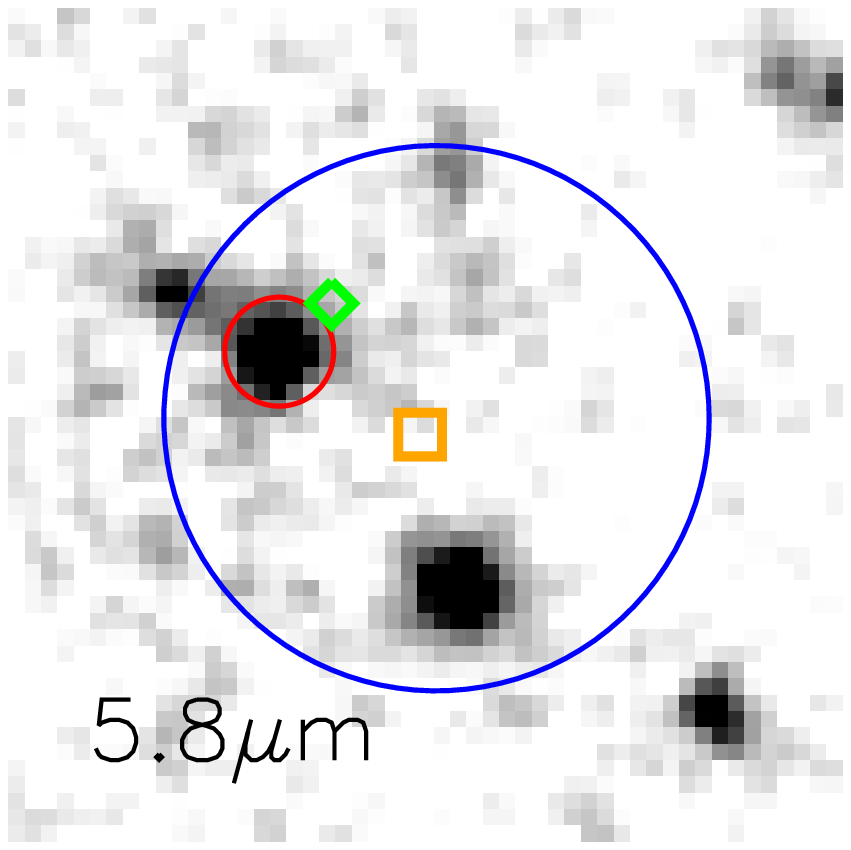}
   }
\parbox[height=20mm]{20mm}{
   \centering
    \includegraphics[width=20mm]{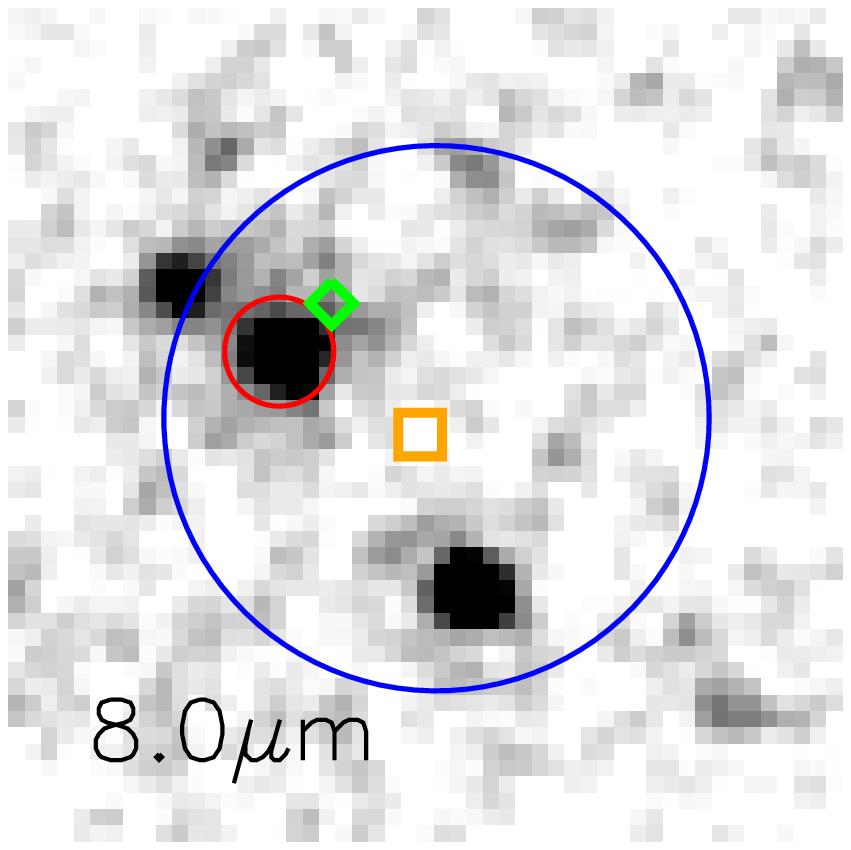}
   }
\parbox[height=20mm]{20mm}{
    \centering
    \includegraphics[width=20mm]{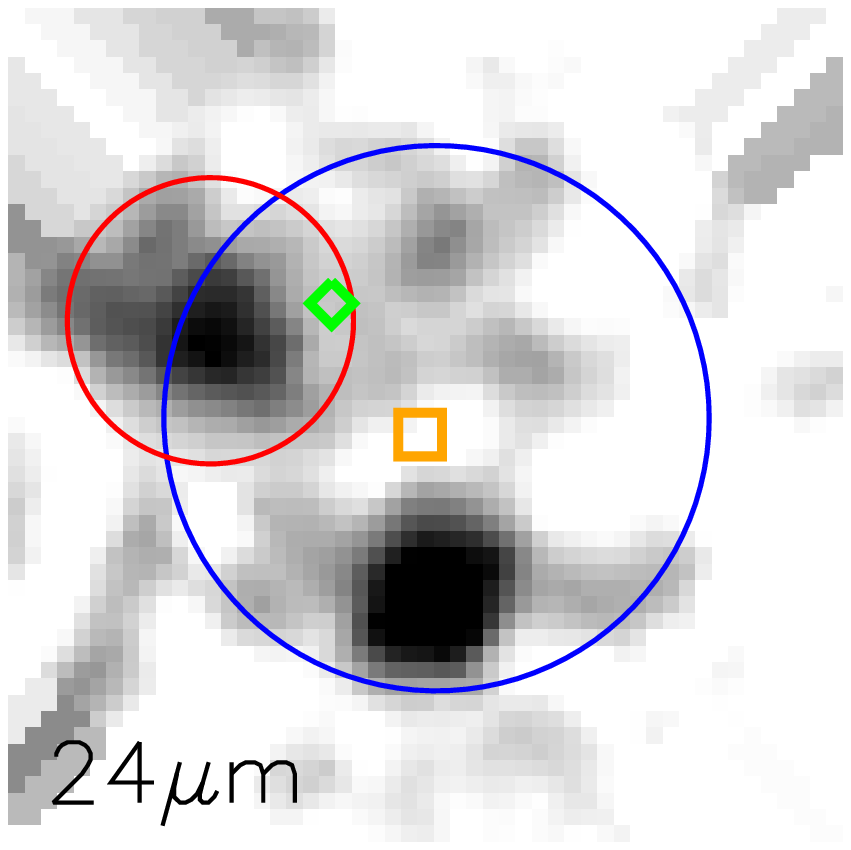}
   }

\parbox[height=20mm]{20mm}{
    \centering\hspace{20mm}
   } 
   \parbox[height=20mm]{20mm}{
    \centering\hspace{20mm}
   } 
   \parbox[height=20mm]{20mm}{
    \centering
    \includegraphics[width=20mm]{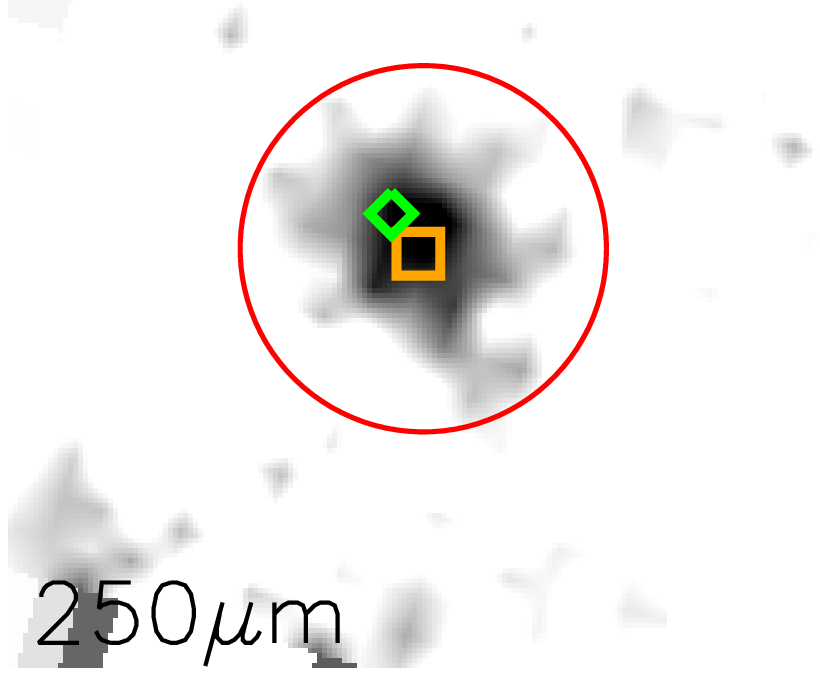}
   } 
   \parbox[height=20mm]{20mm}{
    \centering
    \includegraphics[width=20mm]{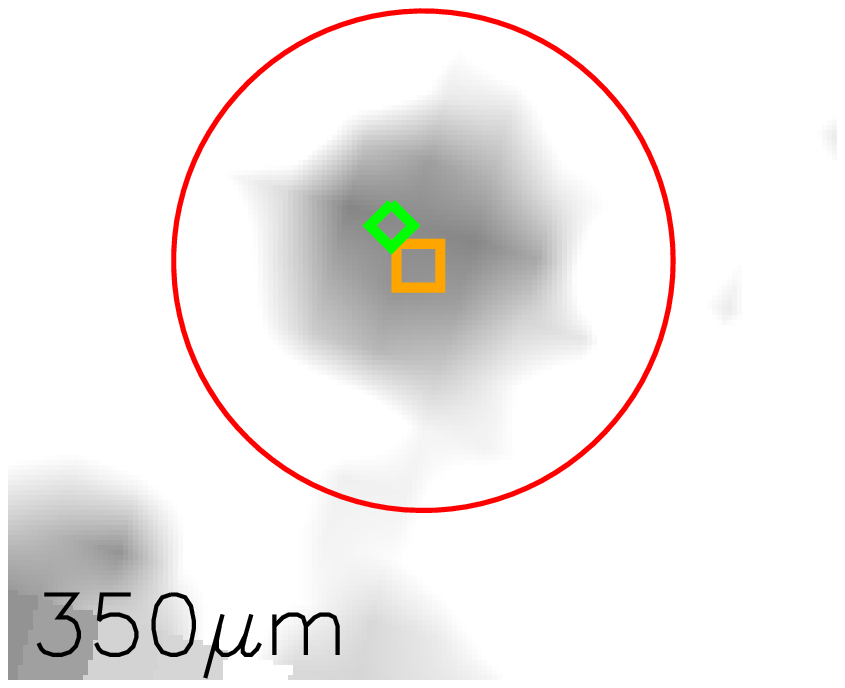}
   } 
   \parbox[height=20mm]{20mm}{
    \centering
    \includegraphics[height=20mm]{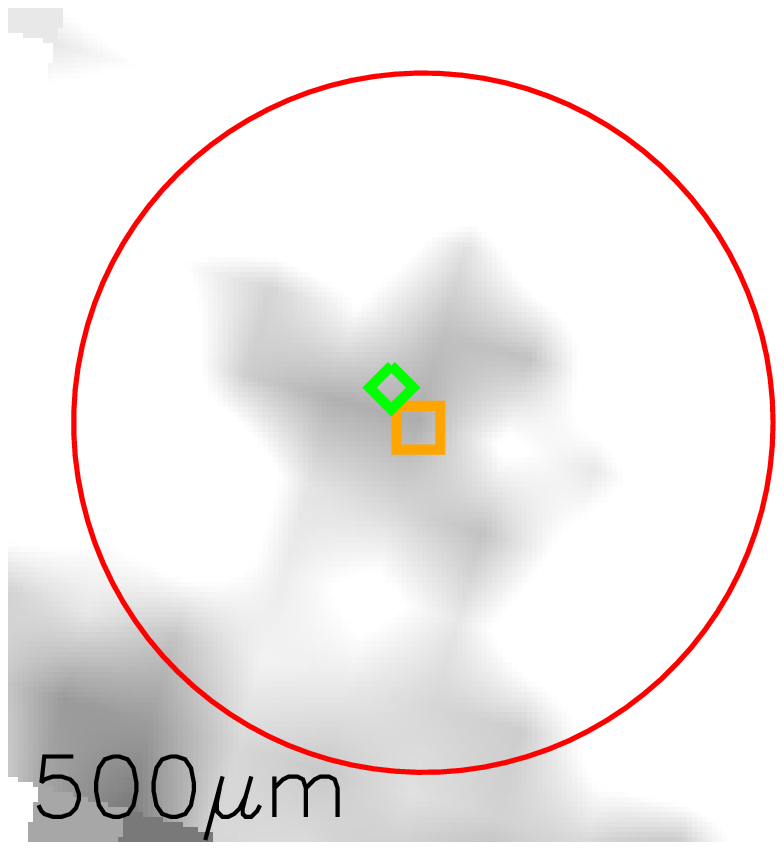}
   } 
\end{figure*}
 \begin{figure*}
    \parbox{70mm}{
    \centering
    \includegraphics[width=70mm]{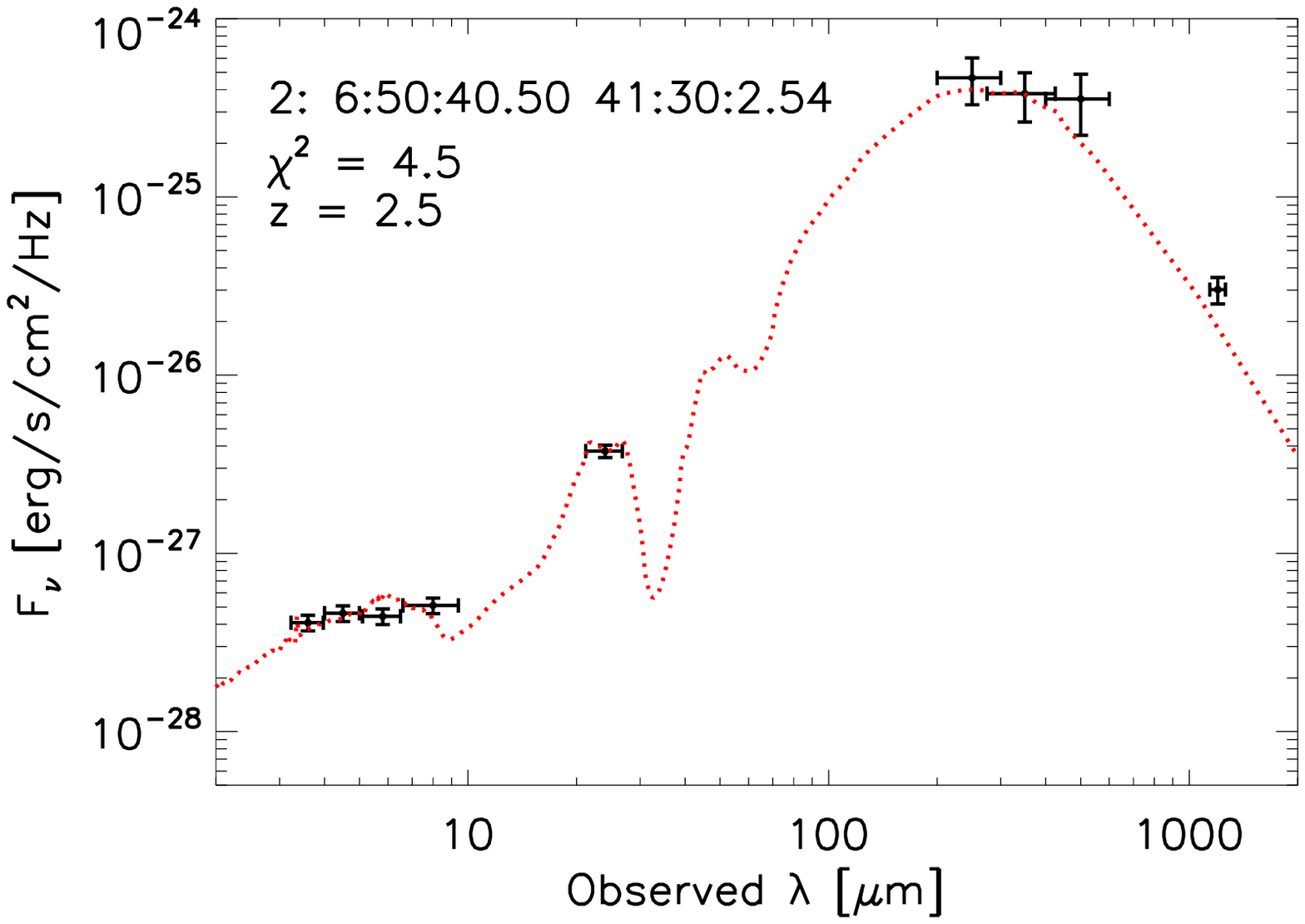}
  } 
   \parbox{70mm}{
    \centering
    \includegraphics[width=70mm]{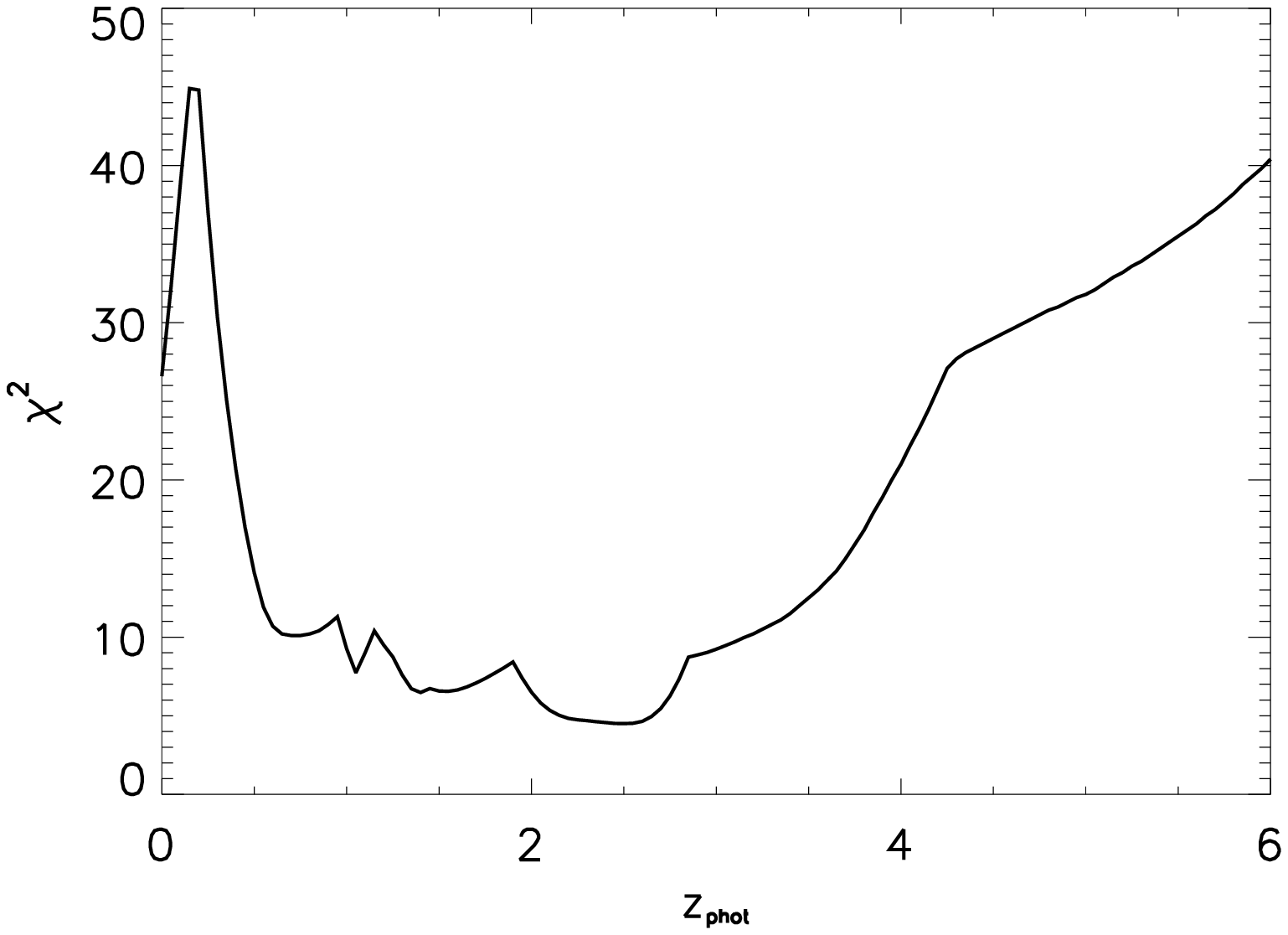}
  }
  \caption{This source is found to be at $z_{\rm{phot}}$ $=$ 2.5 and best fit with a starburst template (I20551) The dust peak is very well observed and well fit. The $\chi^{2}$ distribution shows a clear dip at this redshift, placing this source foreground to the radio galaxy.}
\end{figure*}
\begin{figure*}
\parbox[height=20mm]{20mm}{
    \centering
    \includegraphics[width=20mm]{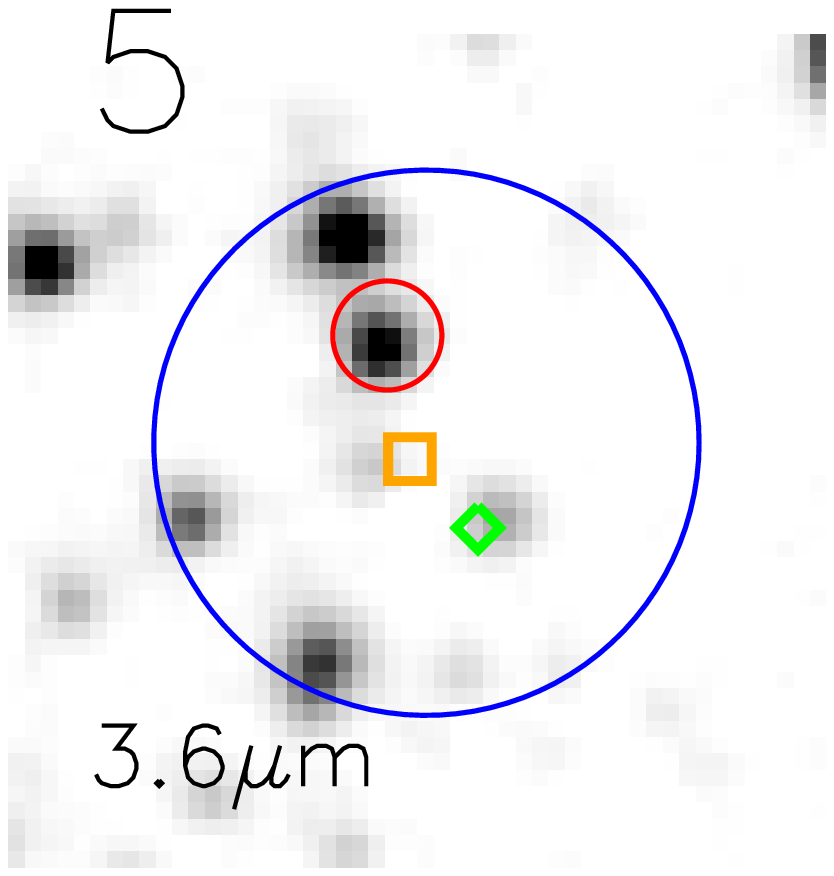}
  }
  \parbox[height=20mm]{20mm}{
    \centering
    \includegraphics[width=20mm]{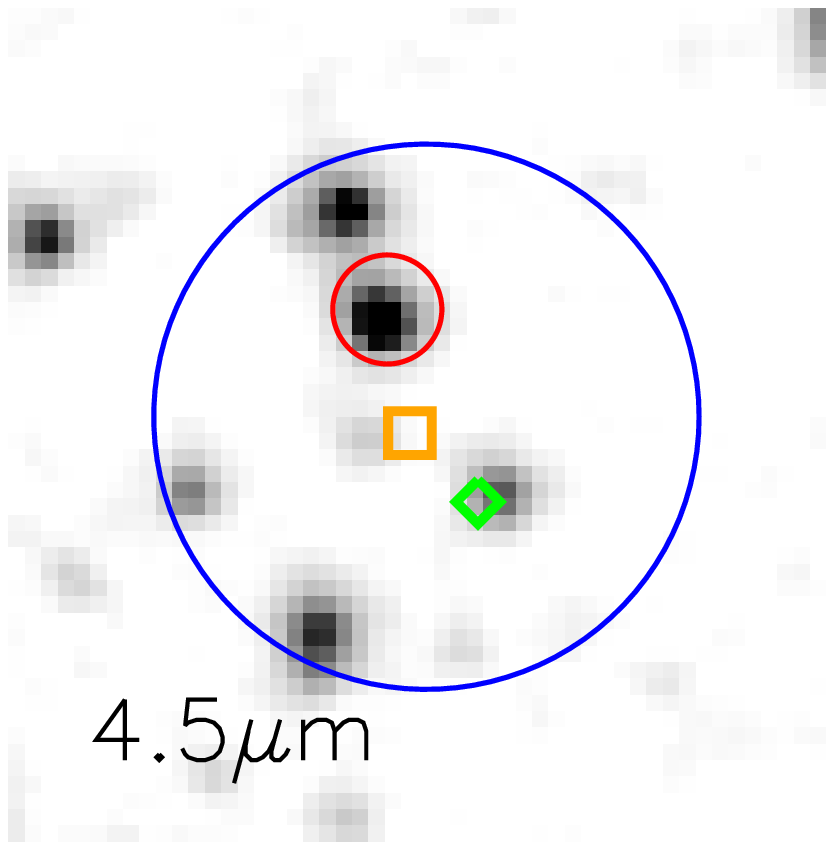}
  }
   \parbox[height=20mm]{20mm}{
    \centering
    \includegraphics[width=20mm]{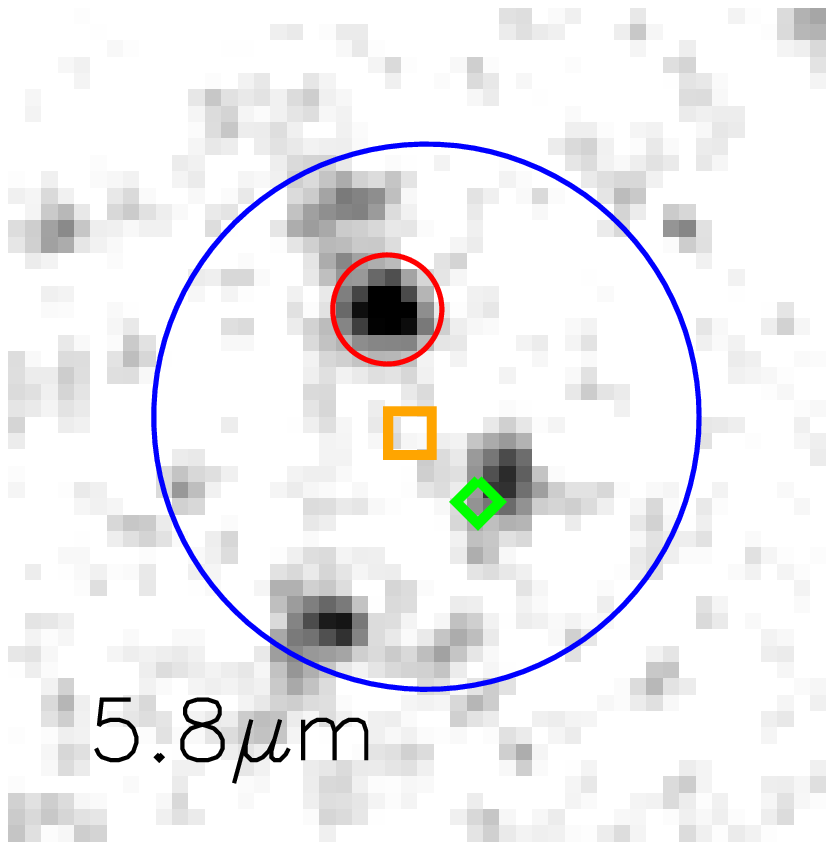}
   }
\parbox[height=20mm]{20mm}{
   \centering
    \includegraphics[width=20mm]{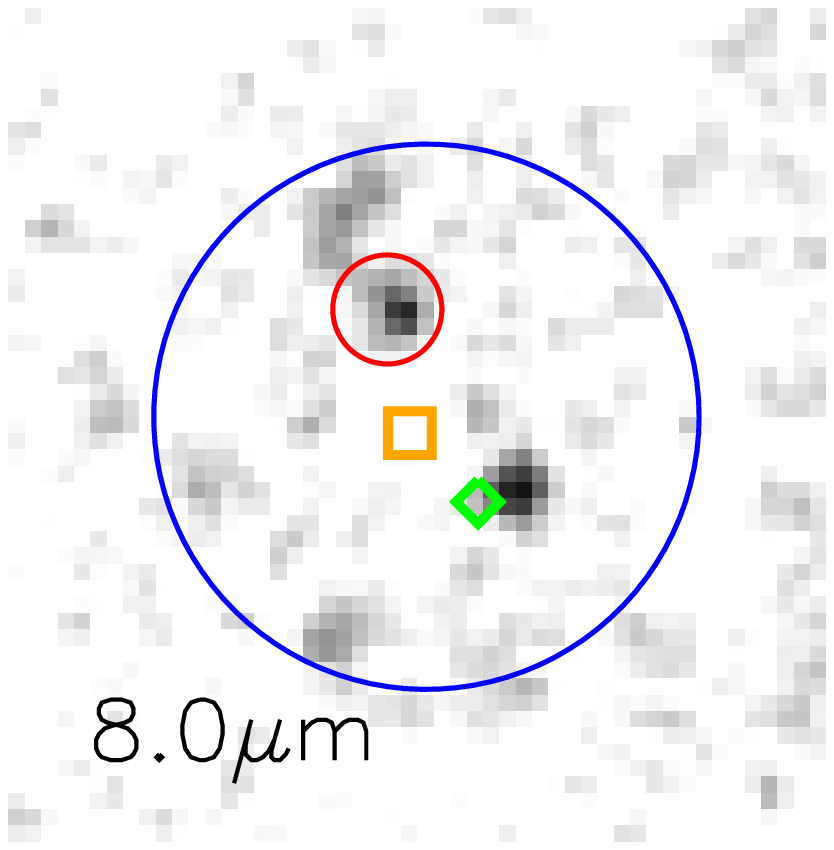}
   }
\parbox[height=20mm]{20mm}{
    \centering
    \includegraphics[width=20mm]{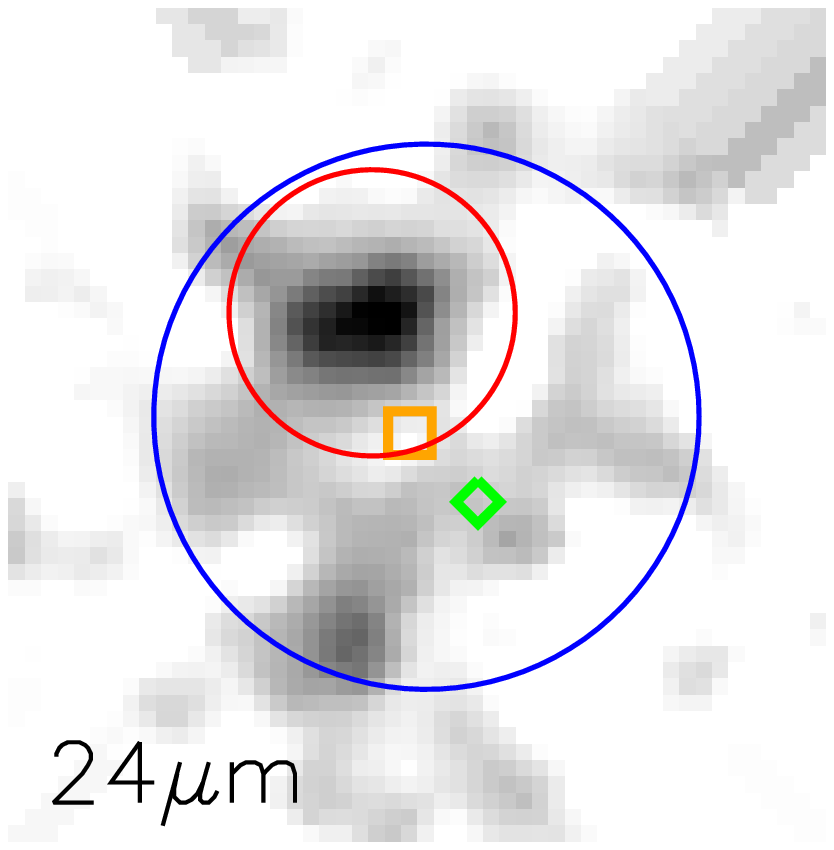}
}

\parbox[height=20mm]{20mm}{
    \centering\hspace{20mm}
   } 
\parbox[height=20mm]{20mm}{
    \centering\hspace{20mm}
   } 
\parbox[height=20mm]{20mm}{
    \centering
    \includegraphics[width=20mm]{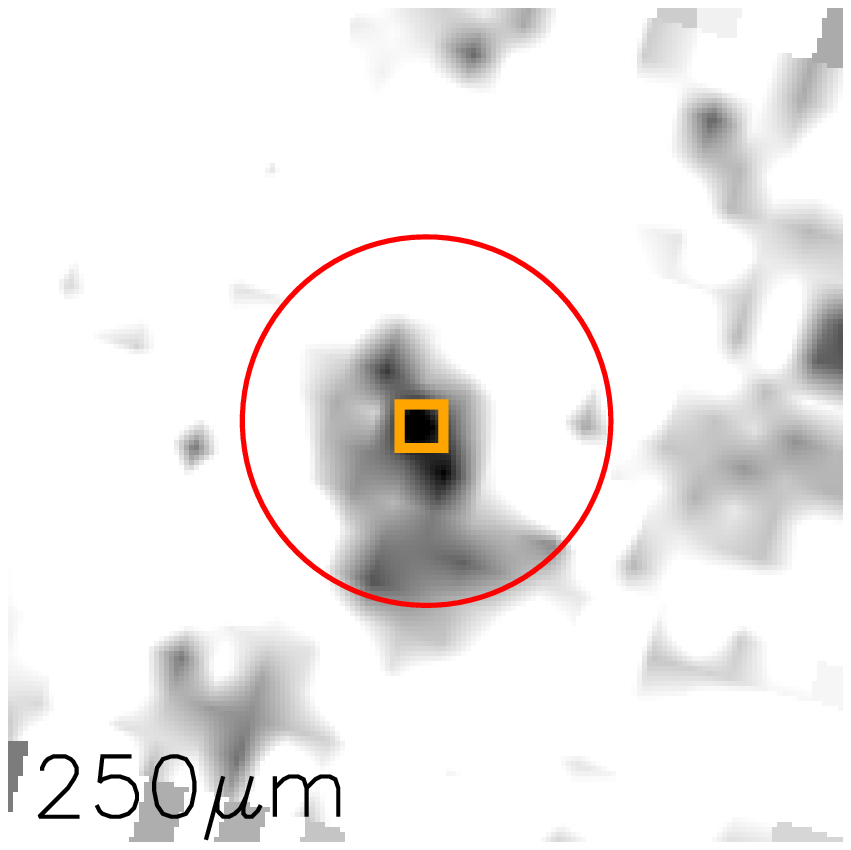}
   } 
\parbox[height=20mm]{20mm}{
    \centering
    \includegraphics[width=20mm]{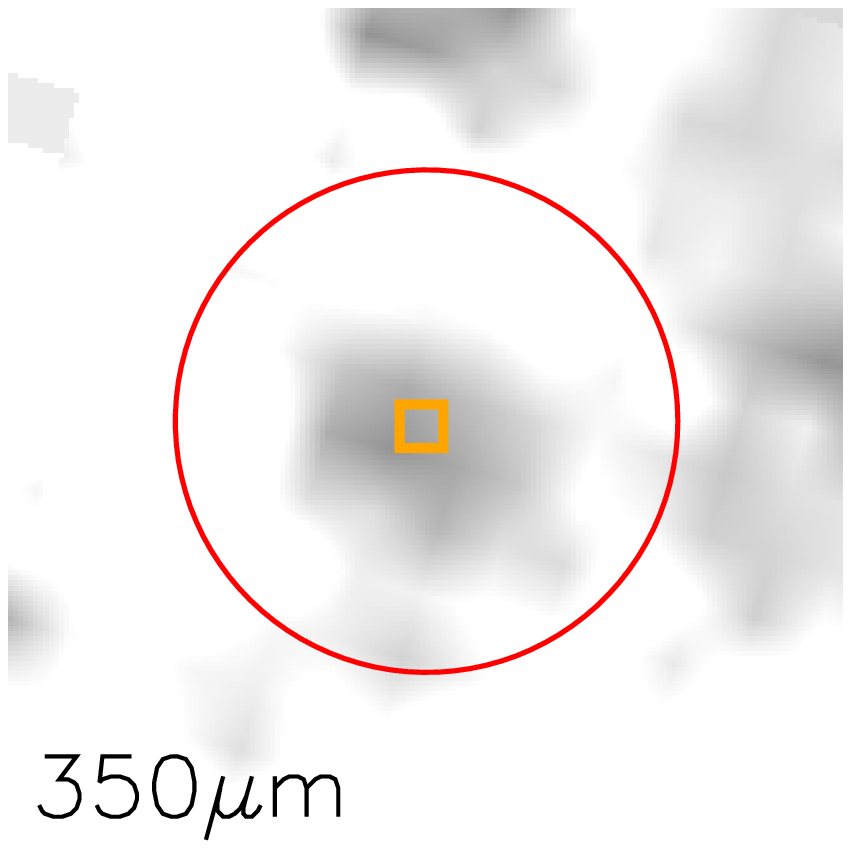}
   } 
\parbox[height=20mm]{20mm}{
    \centering
    \includegraphics[width=20mm]{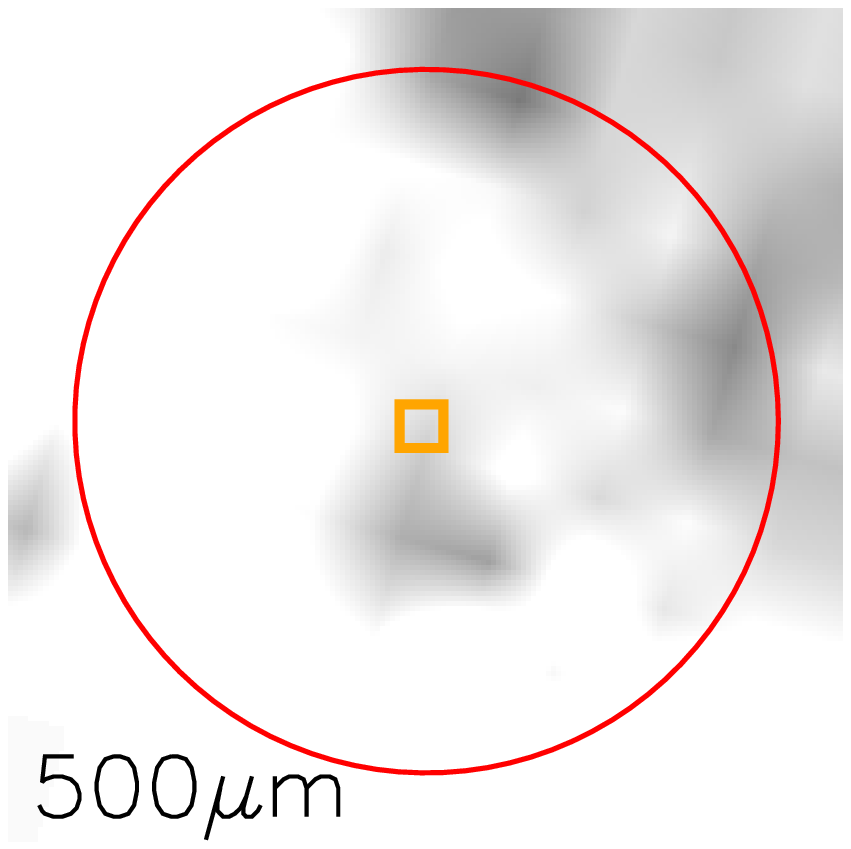}
   } 
   \end{figure*}
 \begin{figure*}
    \parbox{70mm}{
    \centering
    \includegraphics[width=70mm]{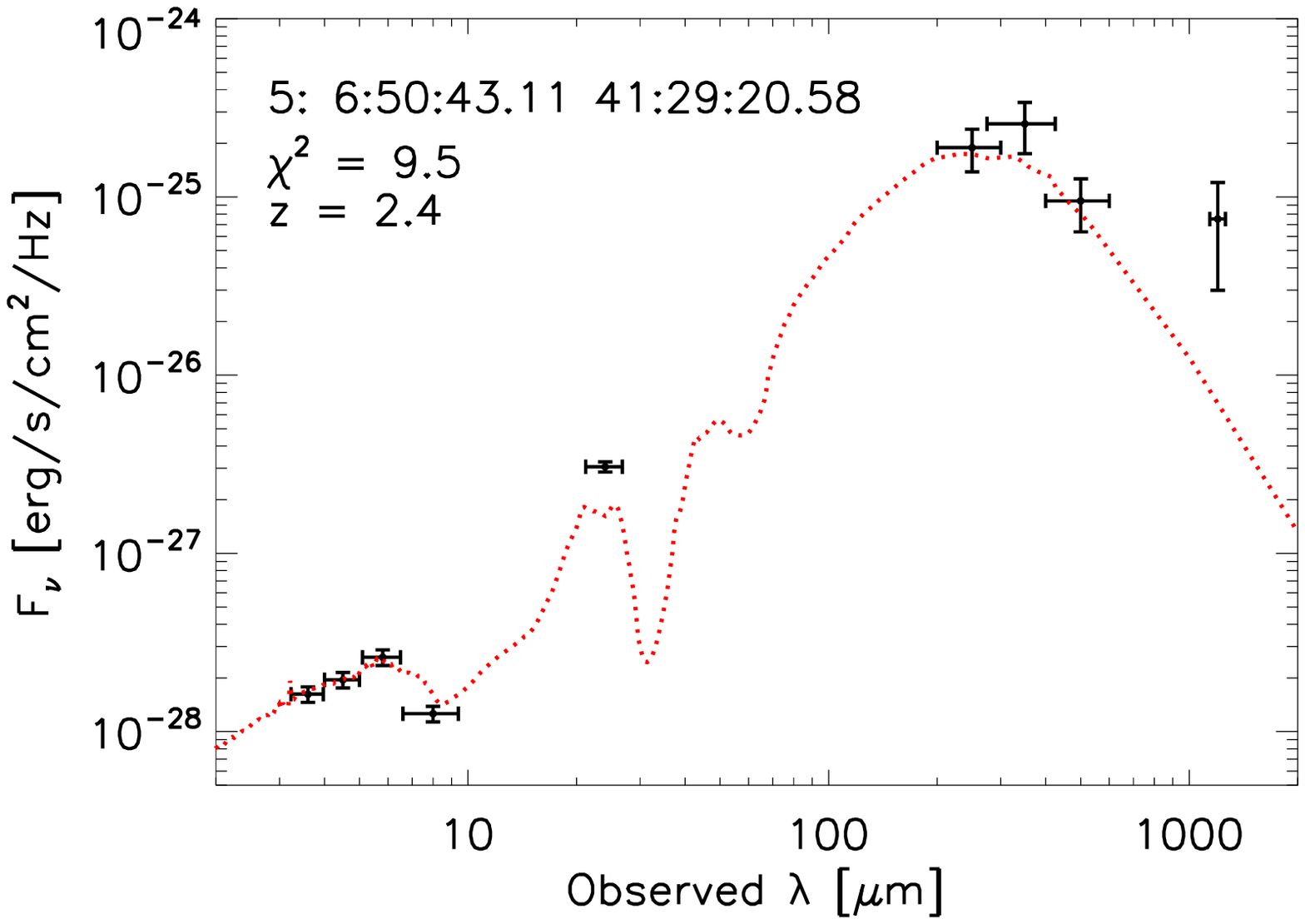}
  }
    \parbox{70mm}{
    \centering
    \includegraphics[width=70mm]{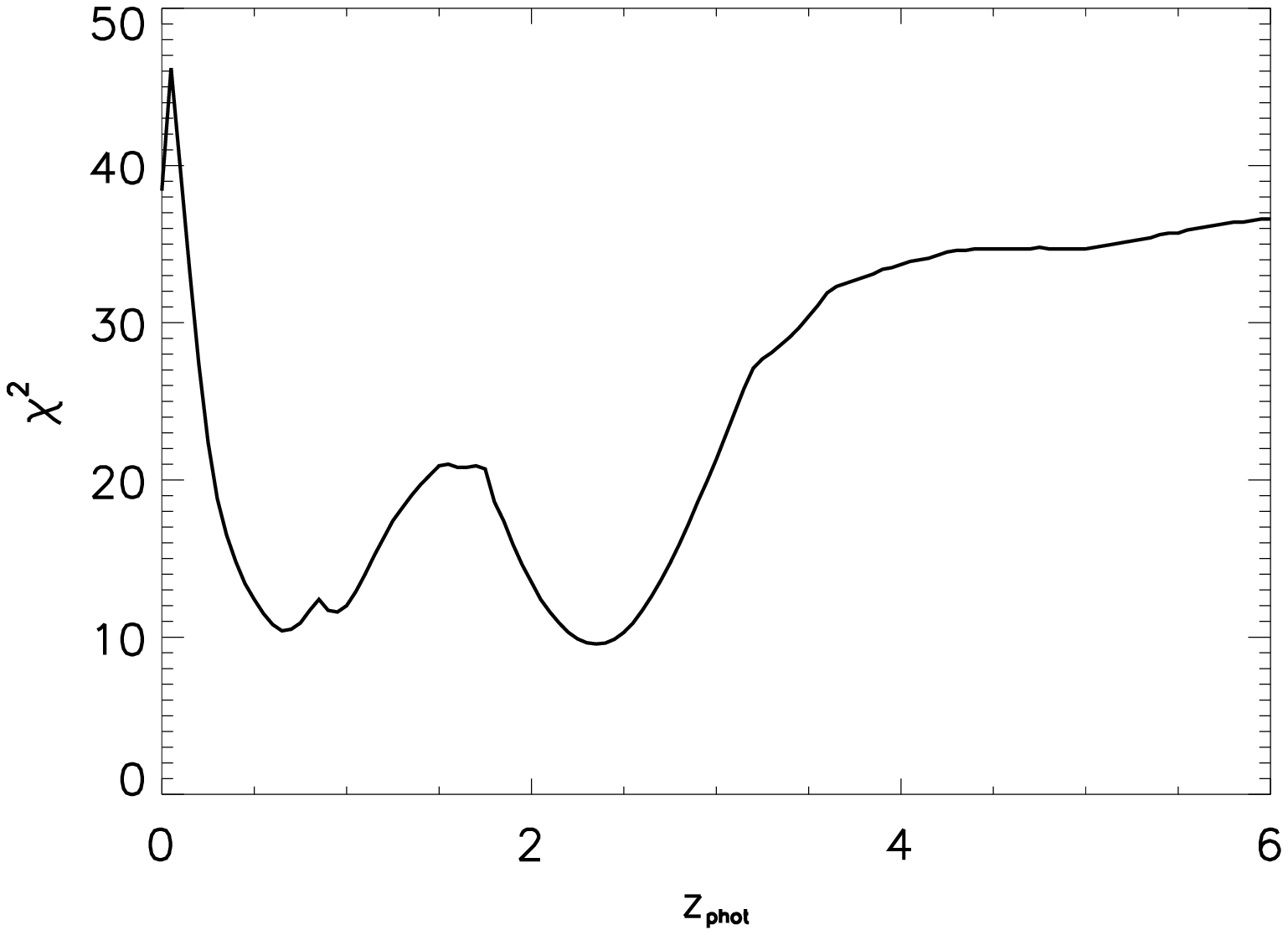}
  }
  \caption{The IRAC photometry of this source shows a very prominent stellar bump that is well fit by the starburst template, leaving, no doubt on the low redshift ($z_{\rm{pho}t}$ $\sim$ 2.4) of this source. This is also confirmed by its spectroscopic redshift ($z_{\rm{spec}}$ = 2.672) consistent with our photometric redshift. \citet{Greve_2007} finds a very extended Ly$\alpha$ halo extending 50 kpc from this source.}
\end{figure*}
\begin{figure*}
\parbox[height=20mm]{20mm}{
    \centering
    \includegraphics[width=20mm]{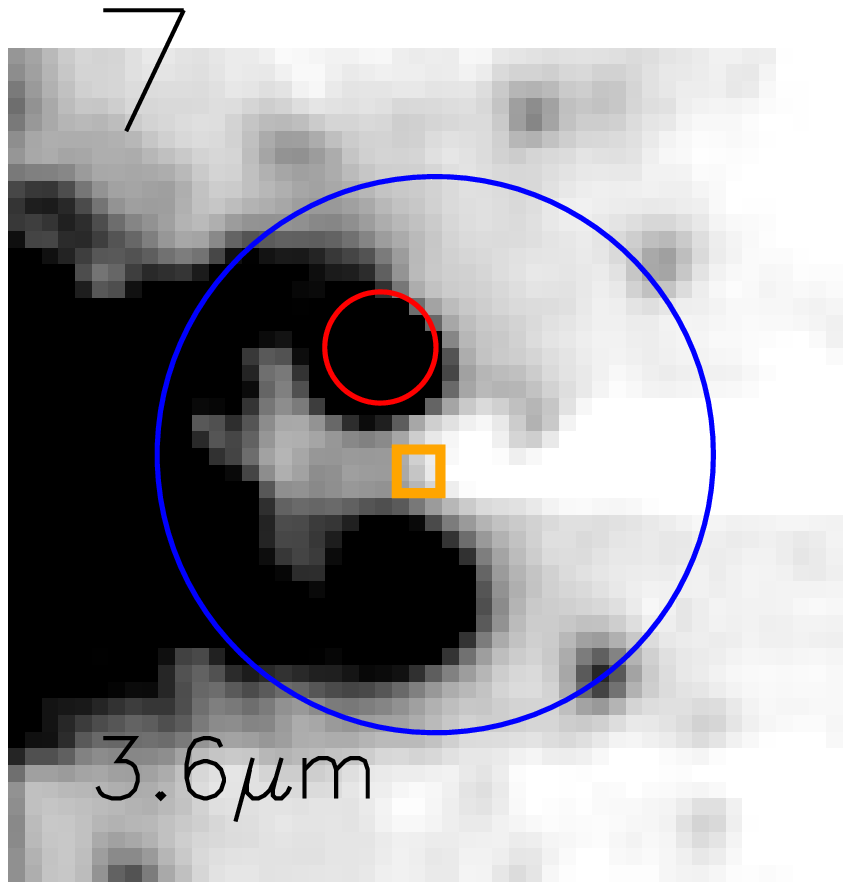}
  }
  \parbox[height=20mm]{20mm}{
    \centering
    \includegraphics[width=20mm]{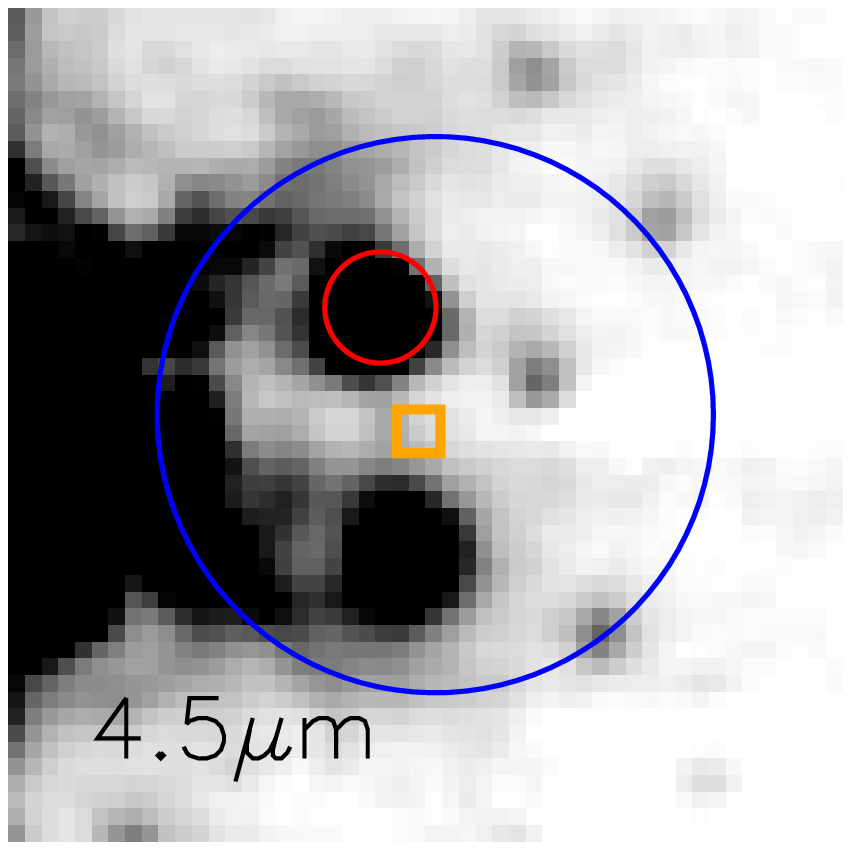}
  }
   \parbox[height=20mm]{20mm}{
    \centering
    \includegraphics[width=20mm]{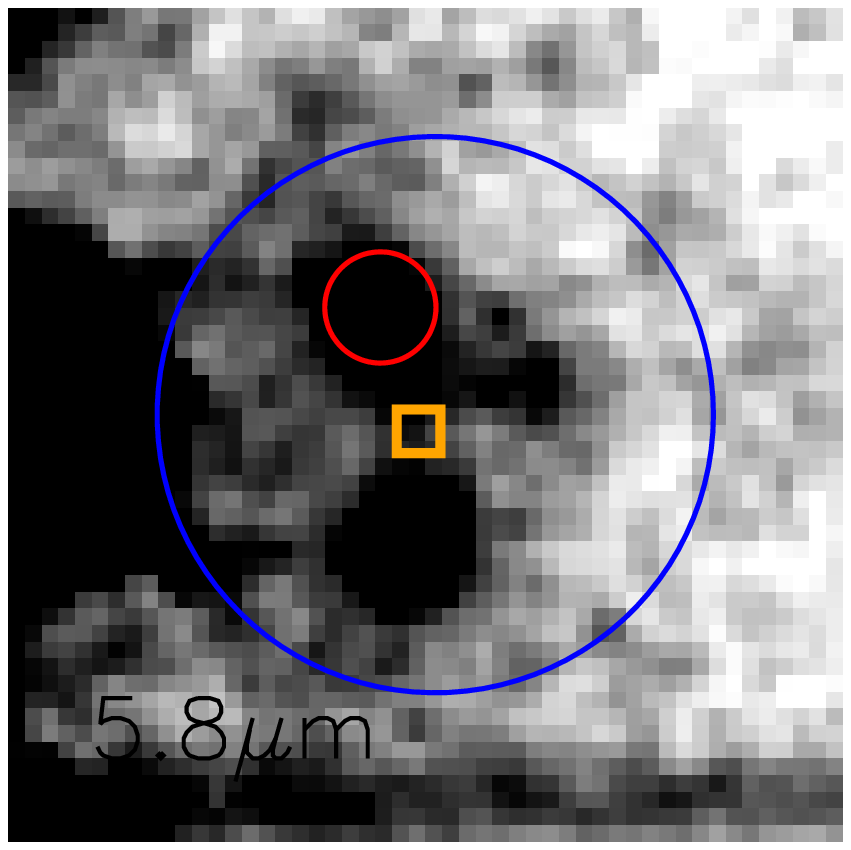}
   }
\parbox[height=20mm]{20mm}{
   \centering
    \includegraphics[width=20mm]{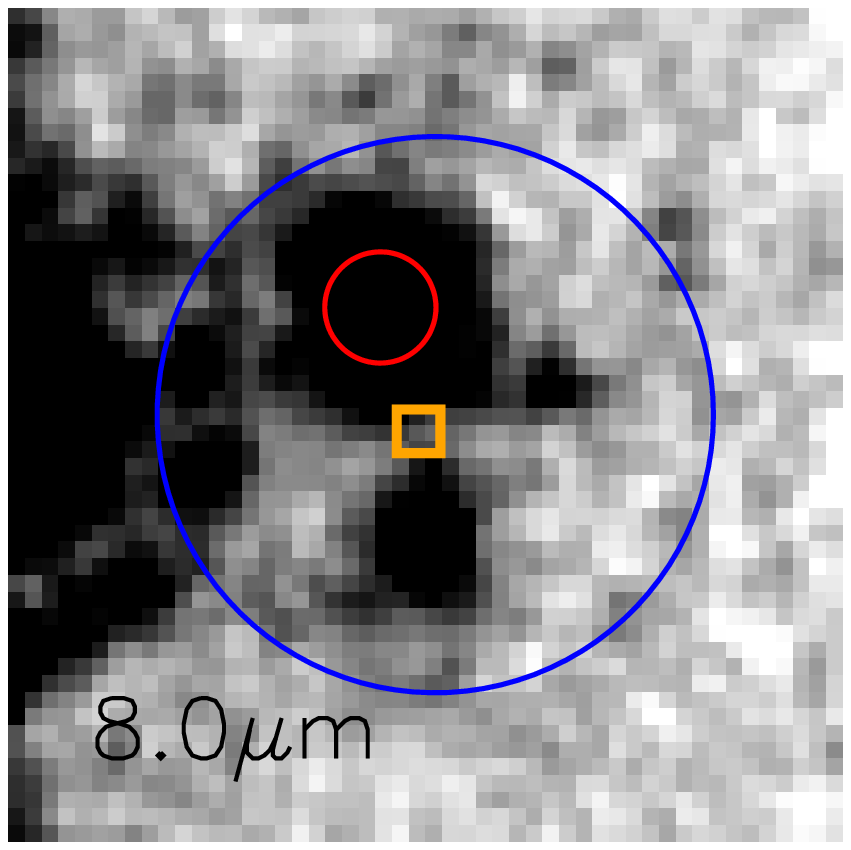}
   }
\parbox[height=20mm]{20mm}{
    \centering
    \includegraphics[width=20mm]{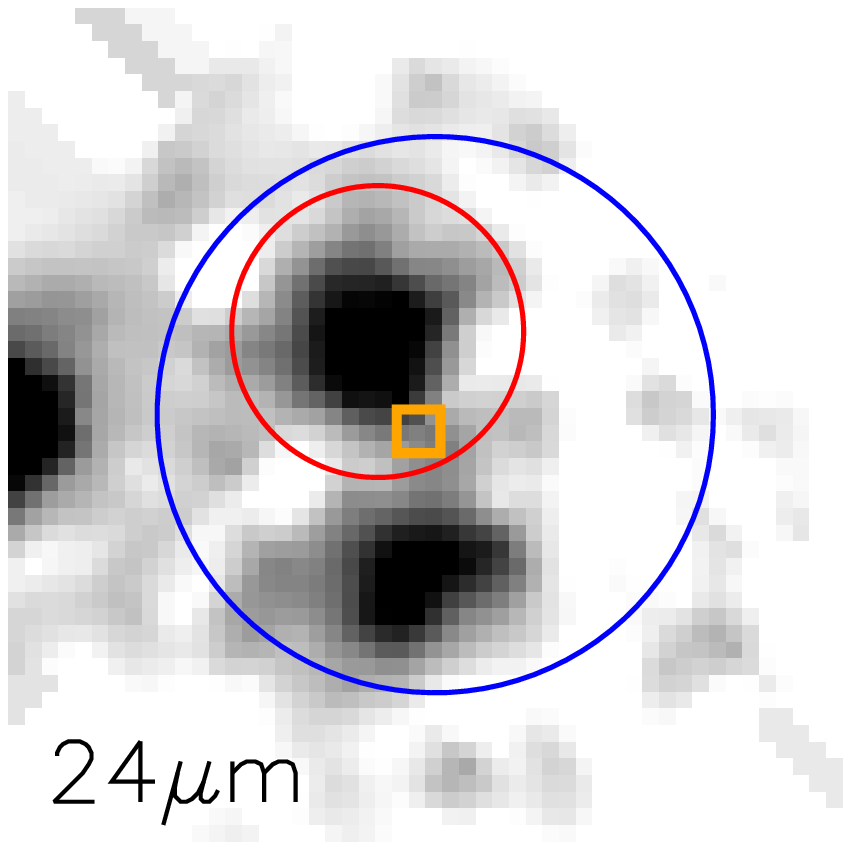}
   }
   
   \parbox[height=20mm]{20mm}{
    \centering\hspace{20mm}
   } 
   \parbox[height=20mm]{20mm}{
    \centering\hspace{20mm}
   } 
   \parbox[height=20mm]{20mm}{
    \centering
    \includegraphics[width=20mm]{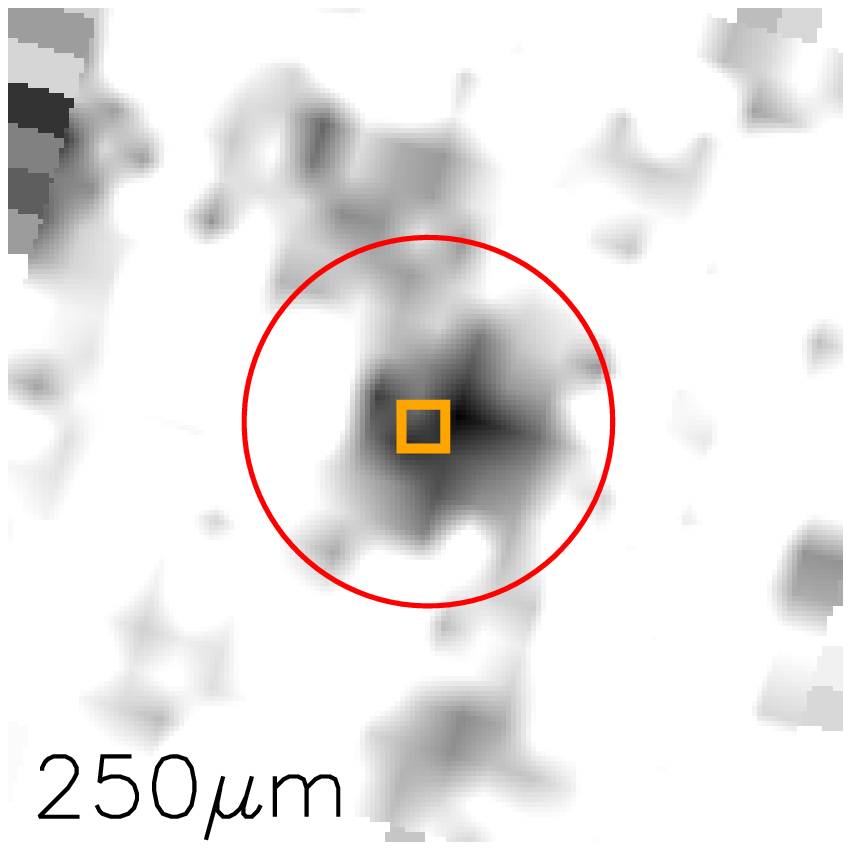}
   } 
   \parbox[height=20mm]{20mm}{
    \centering
    \includegraphics[width=20mm]{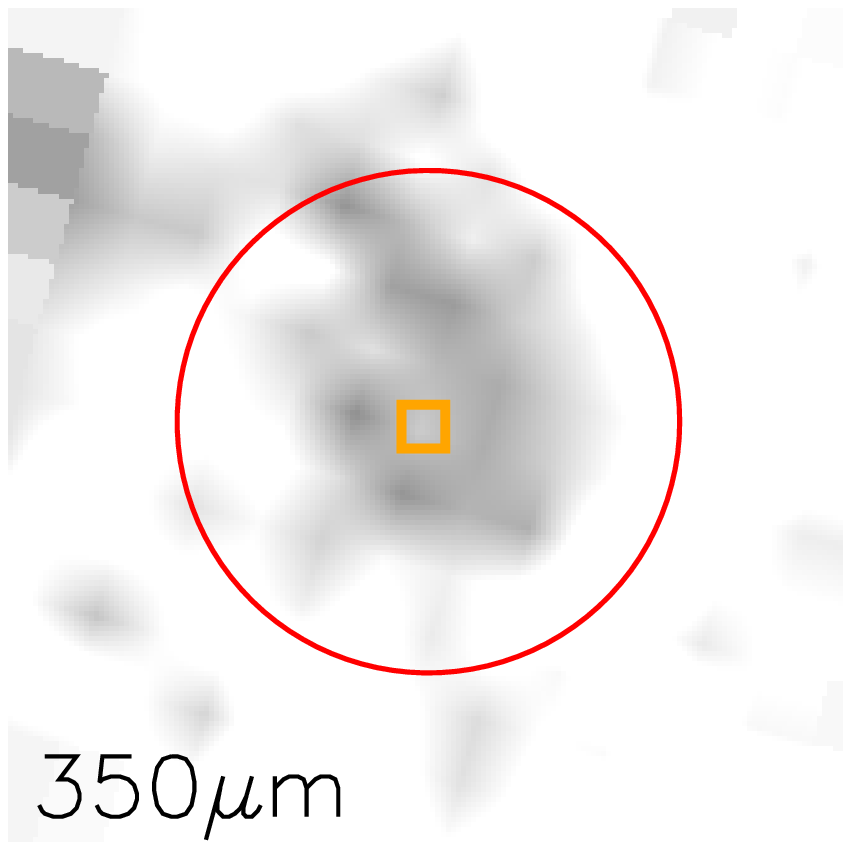}
   } 
   \parbox[height=20mm]{20mm}{
    \centering\hspace{20mm}
   } 
\end{figure*}
\begin{figure*}
    \parbox{70mm}{
    \centering
    \includegraphics[width=70mm]{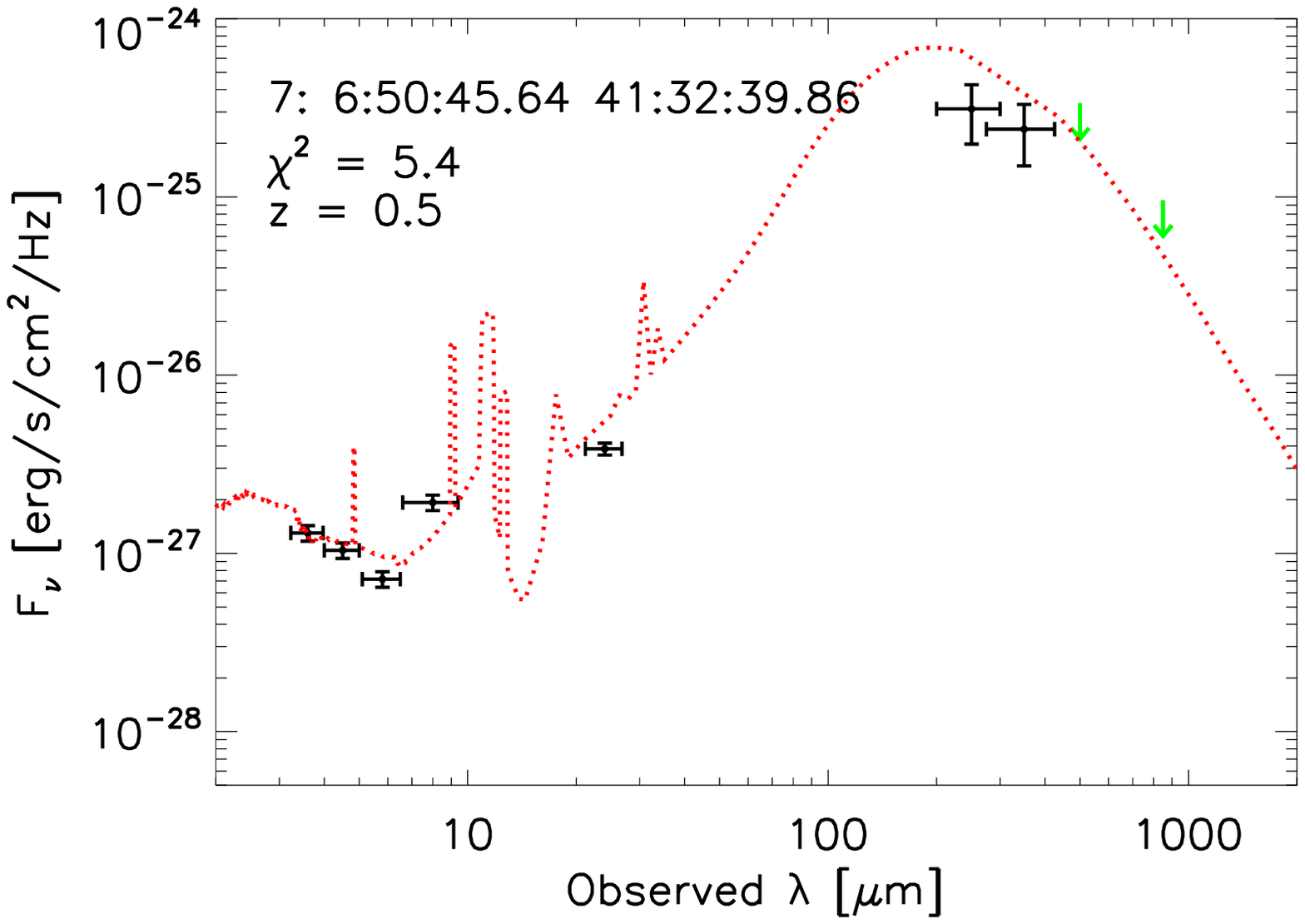}
  }
    \parbox{70mm}{
    \centering
    \includegraphics[width=70mm]{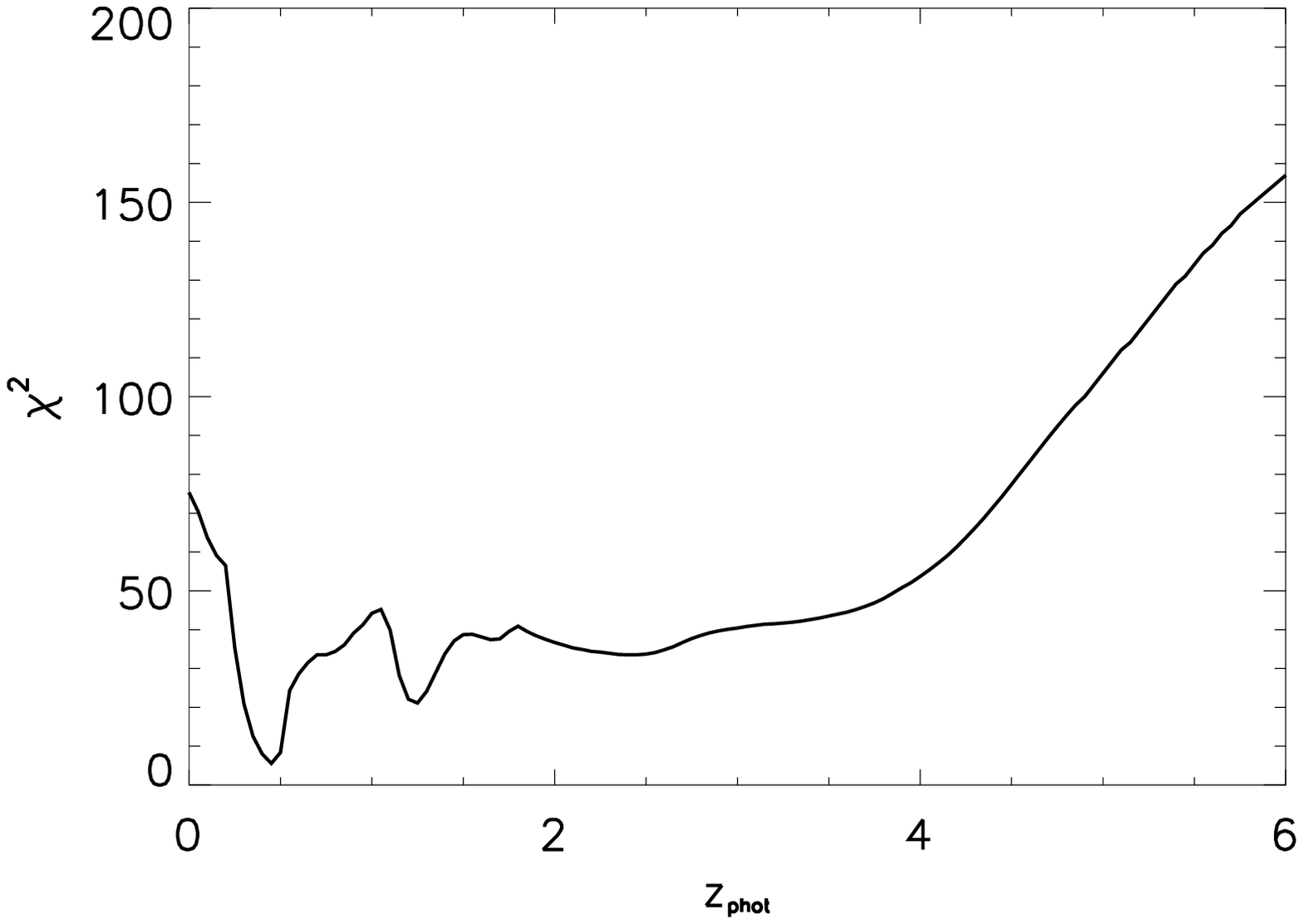}
      }
\caption{Although the far-IR photometric observations are not well fit by {\tt hyperz}, the upper limit at 850$\mu$m is very constraining and places the dust peak at $\sim$ 200 $\mu$m. The IRAC and MIPS photometric points can only be fit with a spiral template. The $\chi^{2}$ distribution places the source unambiguously at low redshift ($z_{\rm{phot}}$ = 0.5). }
\end{figure*}
\begin{figure*}
\parbox[height=20mm]{20mm}{
    \centering
    \includegraphics[width=20mm]{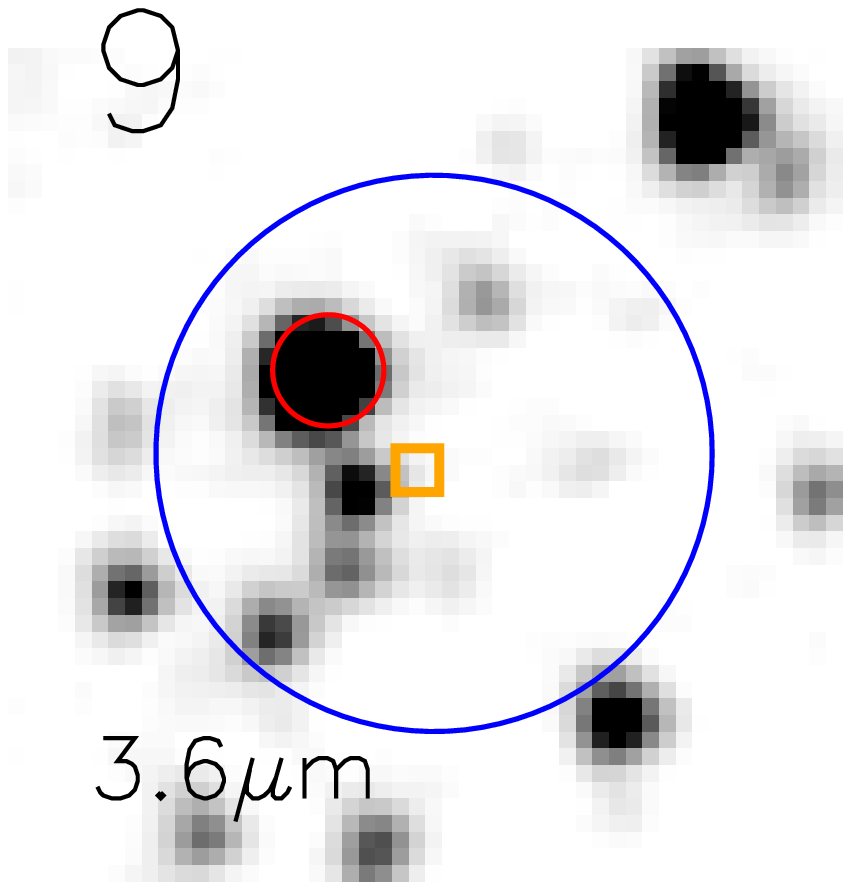}
  }
  \parbox[height=20mm]{20mm}{
    \centering
    \includegraphics[width=20mm]{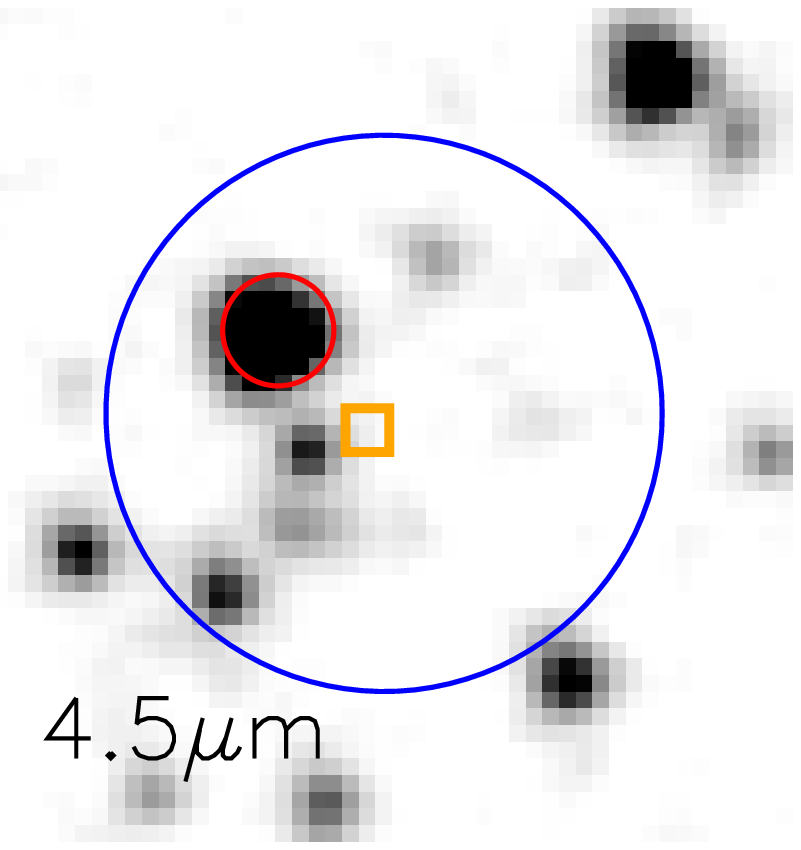}
   }
   \parbox[height=20mm]{20mm}{
    \centering
    \includegraphics[width=20mm]{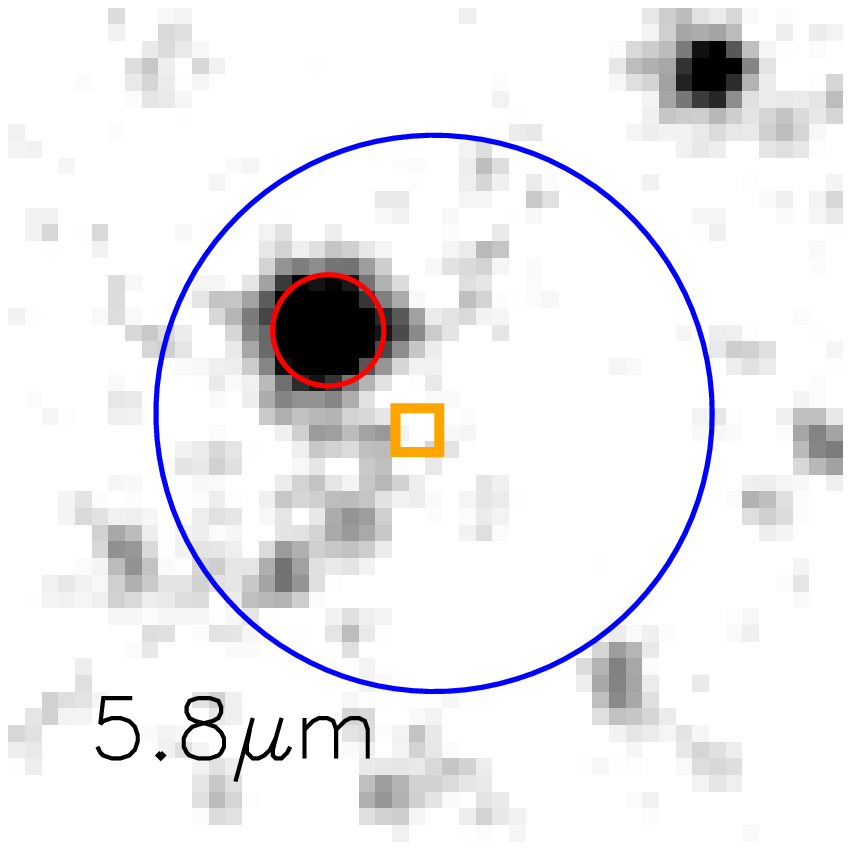}
   }
   \parbox[height=20mm]{20mm}{
    \centering
    \includegraphics[width=20mm]{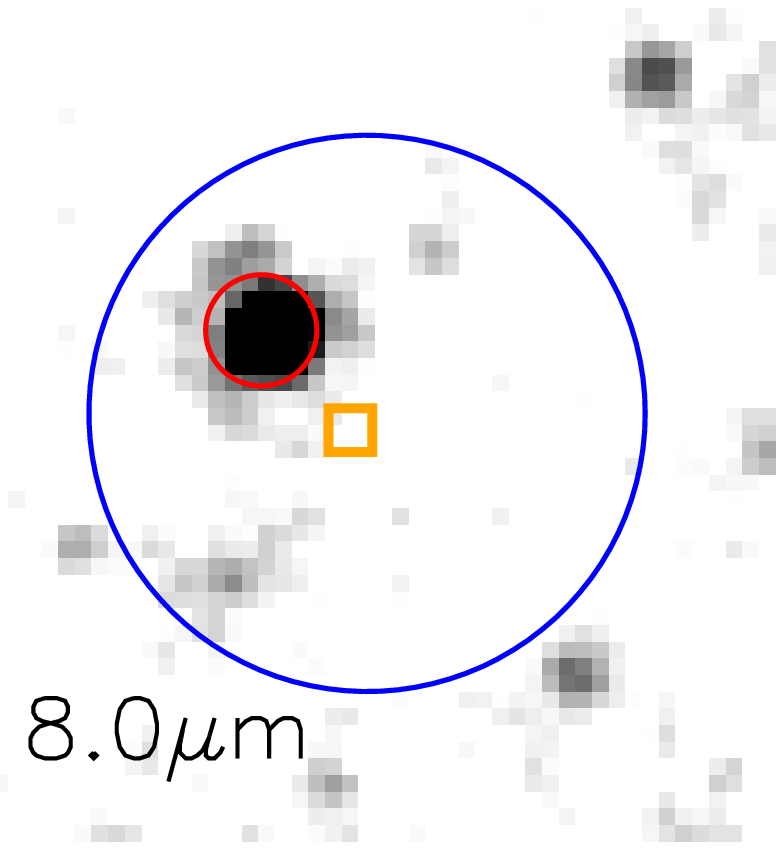}
   }
\parbox[height=20mm]{20mm}{
    \centering
    \includegraphics[width=20mm]{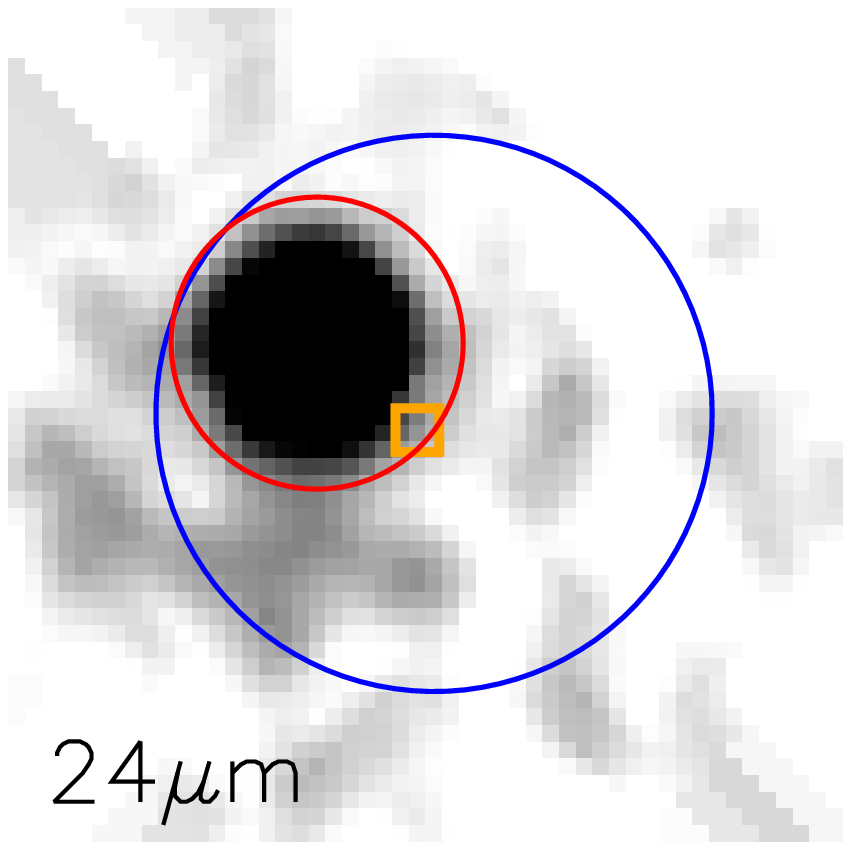}
   }
   
  \parbox[height=20mm]{20mm}{
    \centering
    \includegraphics[width=20mm]{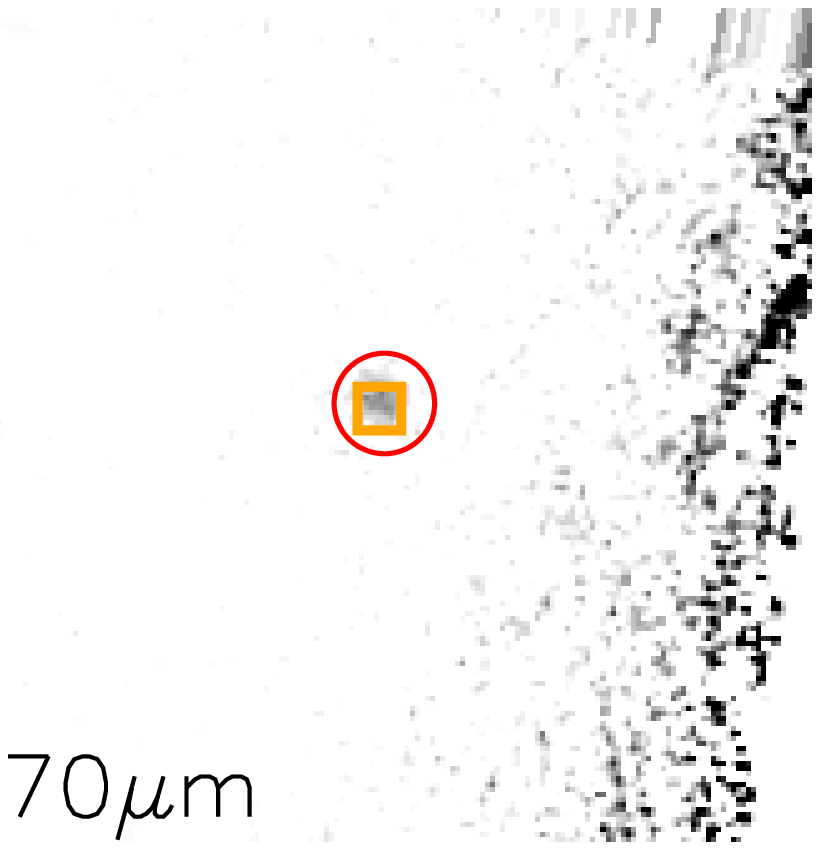}
   }  
  \parbox[height=20mm]{20mm}{
    \centering
    \includegraphics[width=20mm]{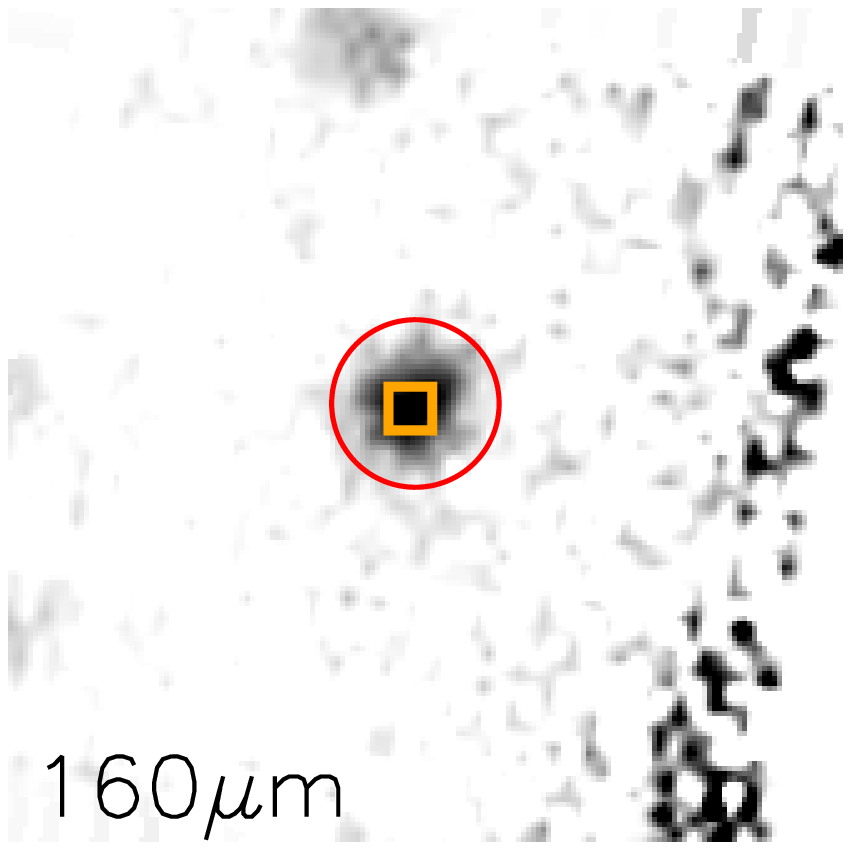}
   }  
  \parbox[height=20mm]{20mm}{
    \centering
    \includegraphics[width=20mm]{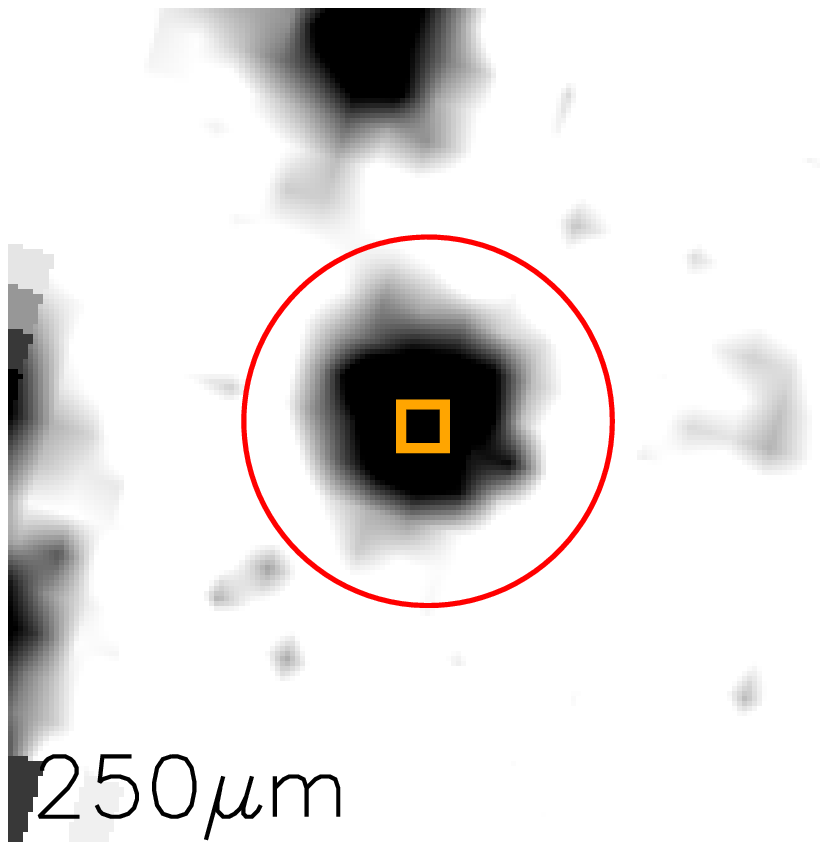}
   }  
  \parbox[height=20mm]{20mm}{
    \centering
    \includegraphics[height = 20mm]{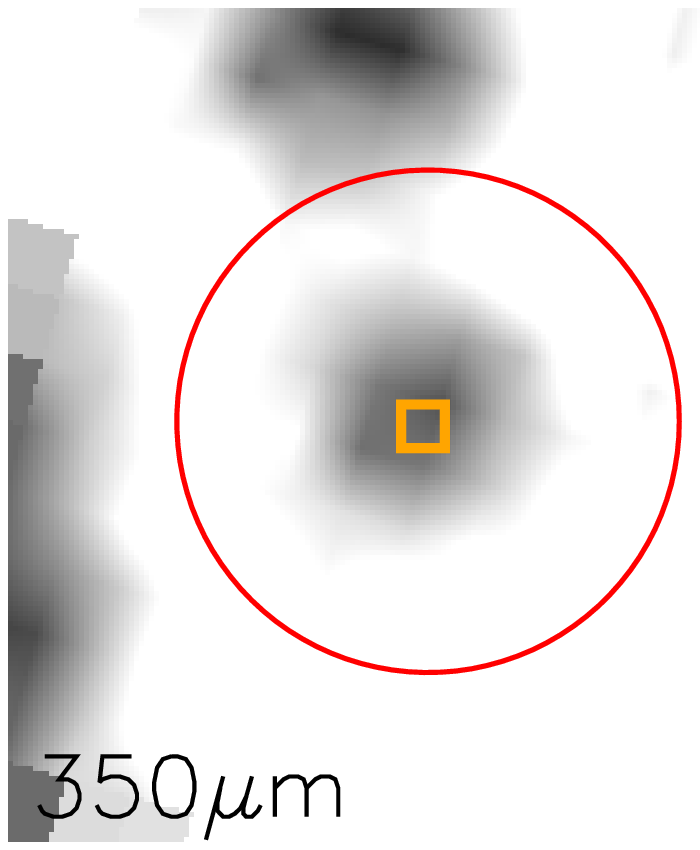}
   }  
  \parbox[height=20mm]{20mm}{
    \centering\hspace{20mm}
   }  
\end{figure*}
\begin{figure*}
   \parbox{70mm}{
    \centering
    \includegraphics[width=70mm]{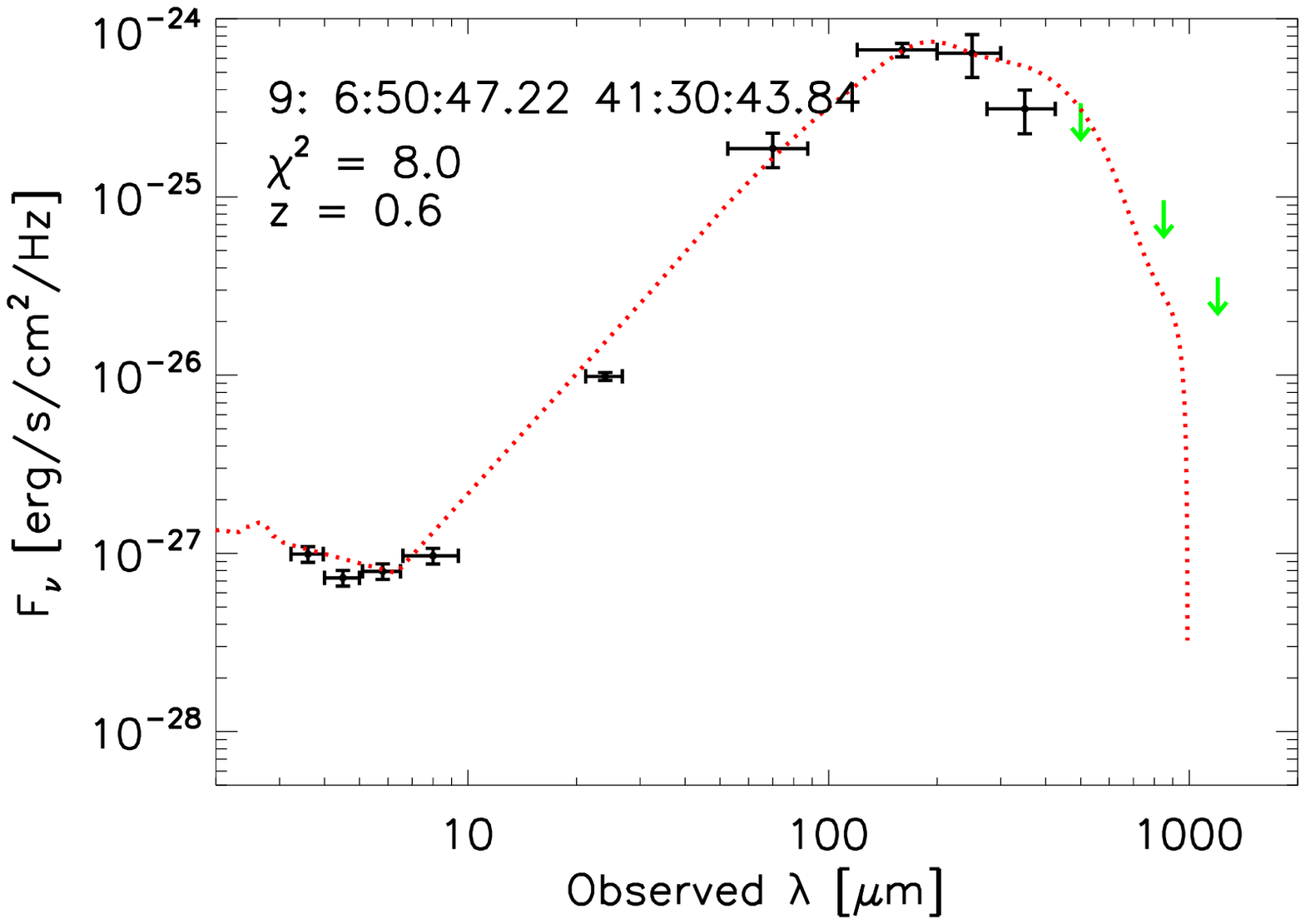}
  }   \parbox{70mm}{
    \centering
    \includegraphics[width=70mm]{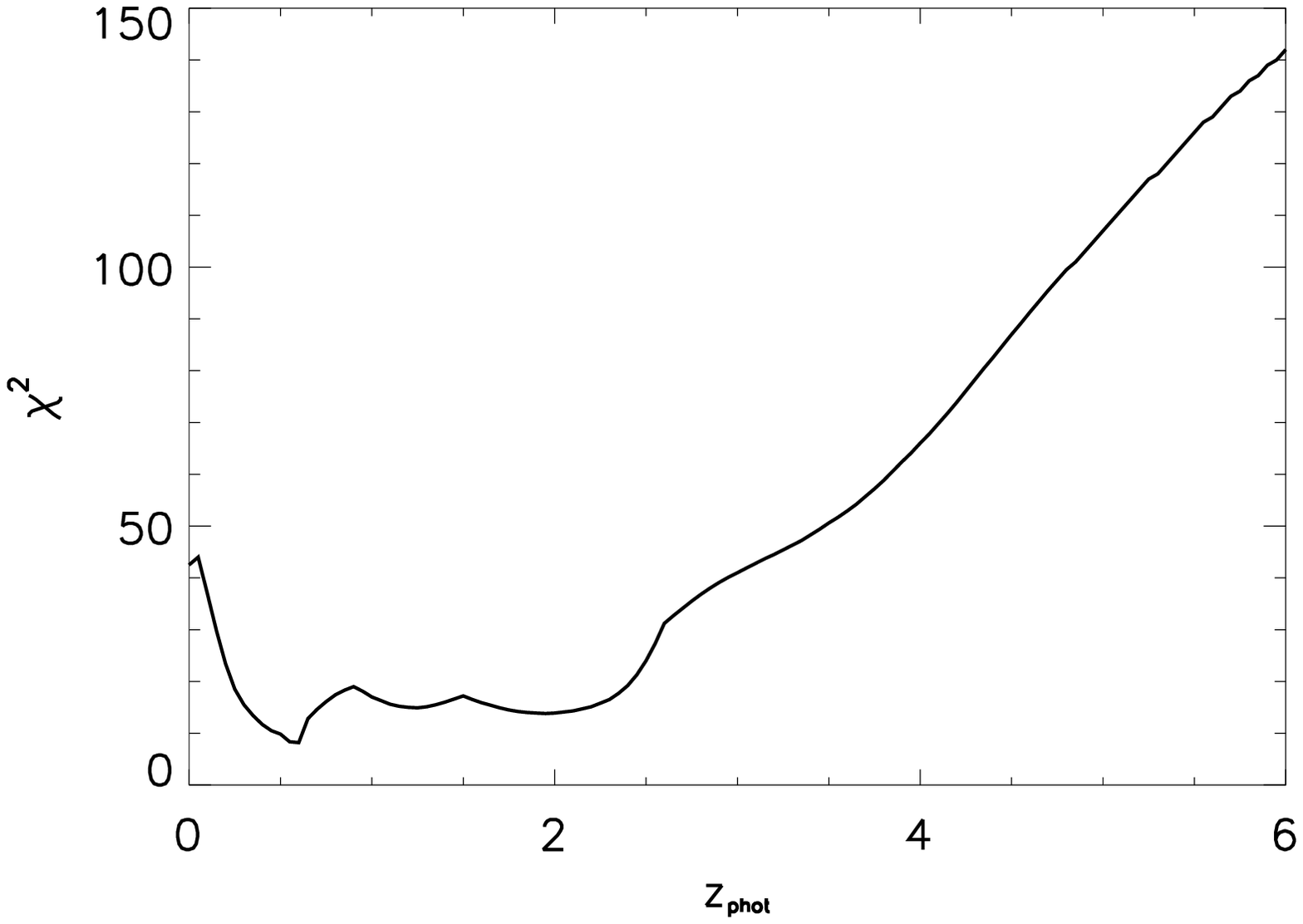}
  }
  \caption{The {\tt hyperz} fit (source 11 dust template + old stellar population) fits the far-IR emission very well. The IRAC observations show the long wavelength tail of the stellar bump (at 1.6 $\mu$m restframe wavelength) indicating a low redshift. We find no secondary prominent dips in the $\chi^{2}$ distribution and this source is thus found to be at lower redshift ($z_{\rm{phot}}$ = 0.6) than the radio galaxy.}
\end{figure*}
\begin{figure*}
\parbox[height=20mm]{20mm}{
    \centering
    \includegraphics[width=20mm]{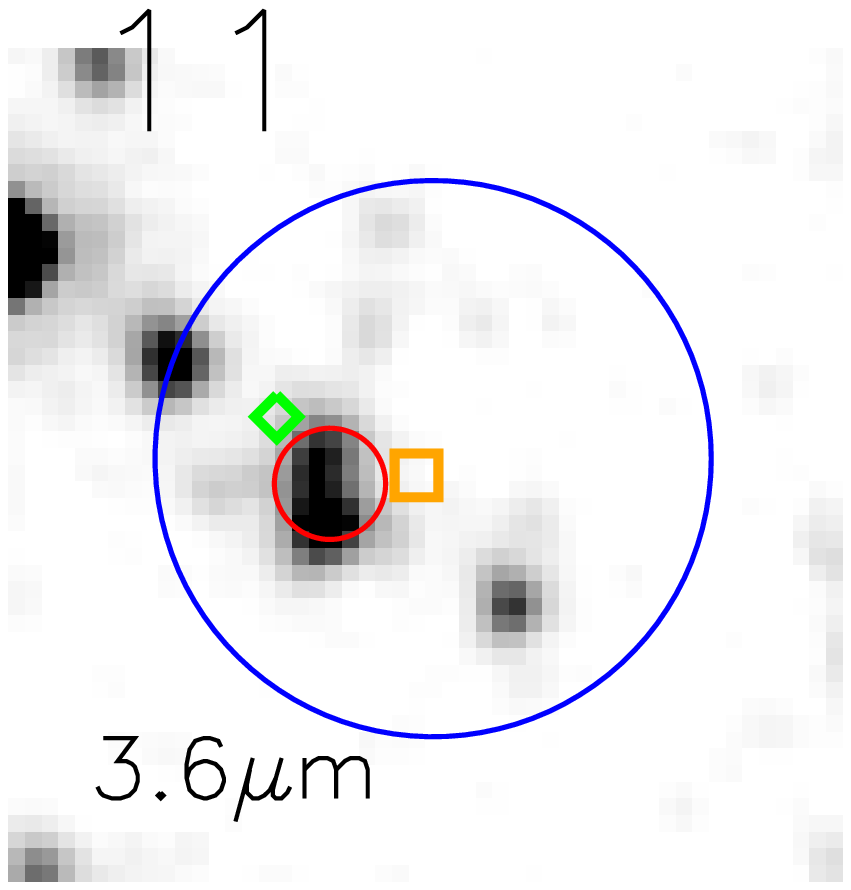}
  }
  \parbox[height=20mm]{20mm}{
    \centering
    \includegraphics[width=20mm]{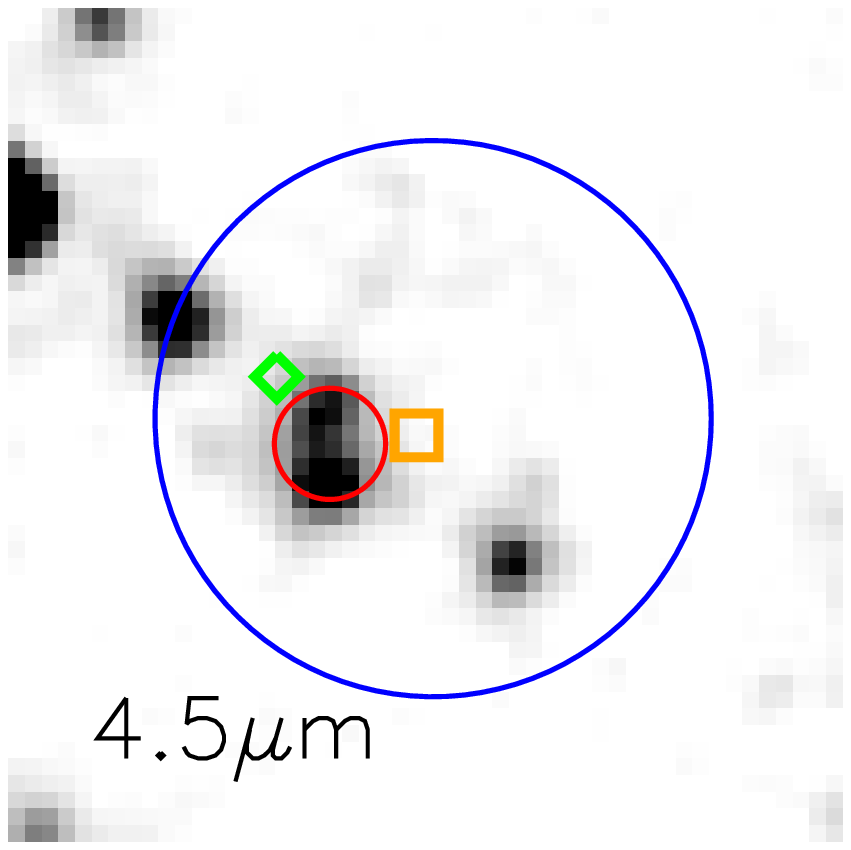}
   }
   \parbox[height=20mm]{20mm}{
    \centering
    \includegraphics[width=20mm]{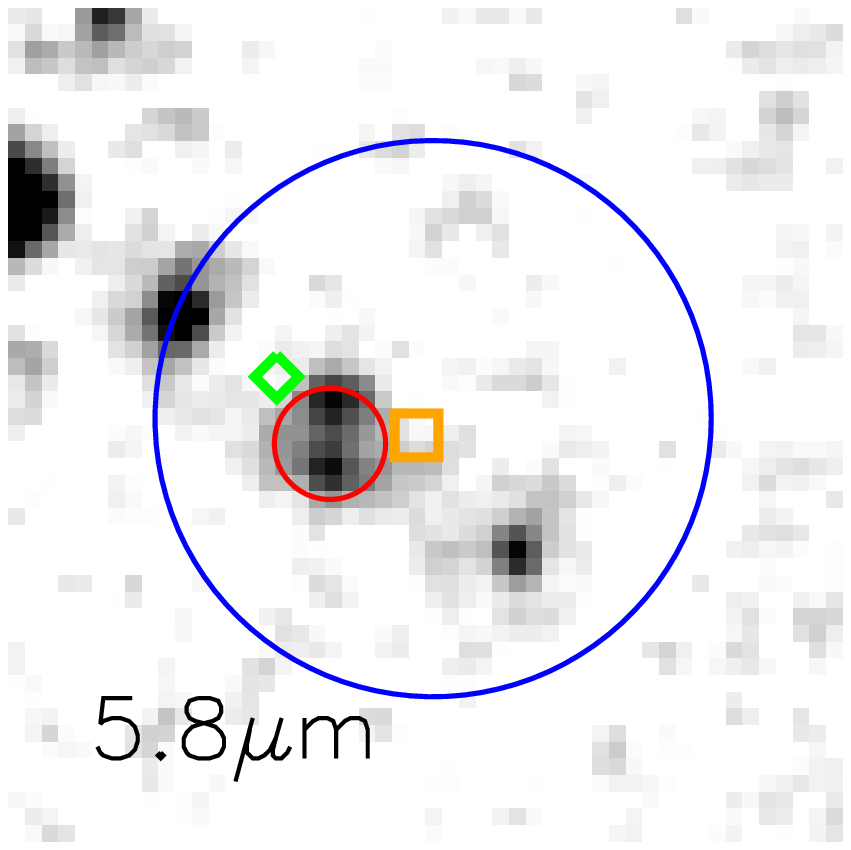}
   }
\parbox[height=20mm]{20mm}{
    \centering
    \includegraphics[width=20mm]{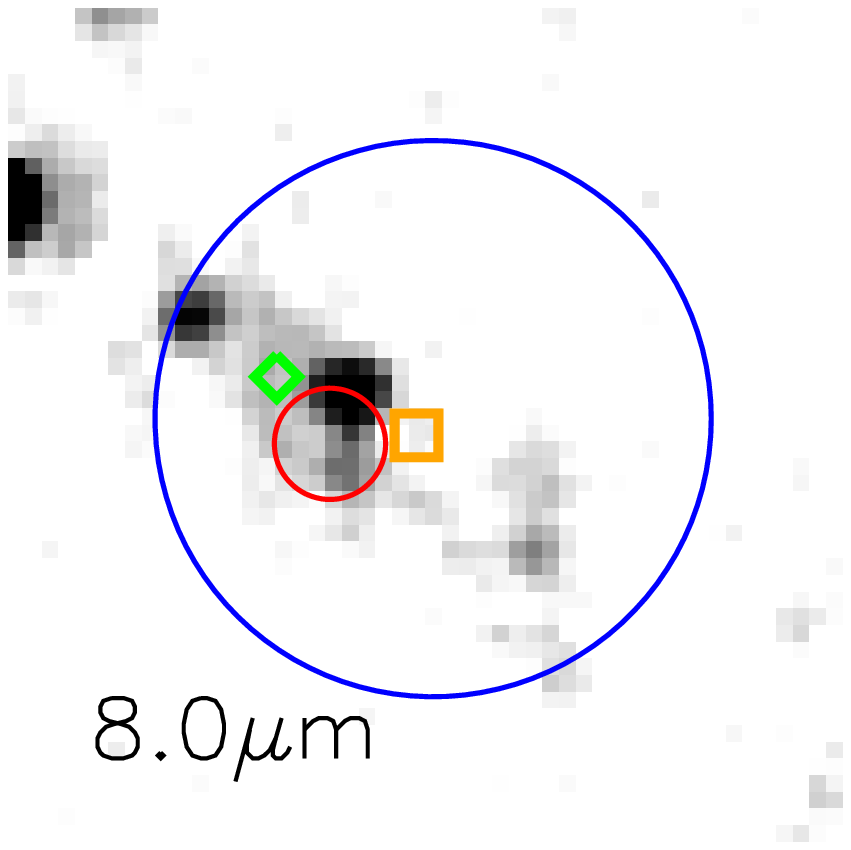}
   }
\parbox[height=20mm]{20mm}{
    \centering
    \includegraphics[width=20mm]{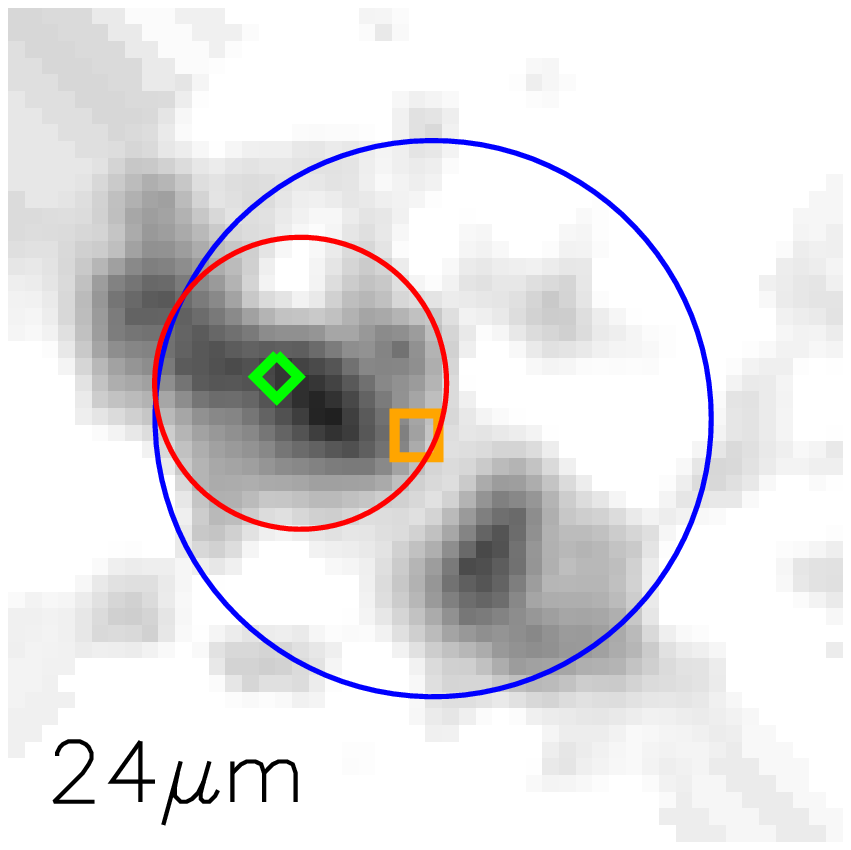}
   }
   
   \parbox[height=20mm]{20mm}{
    \centering\hspace{20mm}
   } 
   \parbox[height=20mm]{20mm}{
    \centering
    \includegraphics[width=20mm]{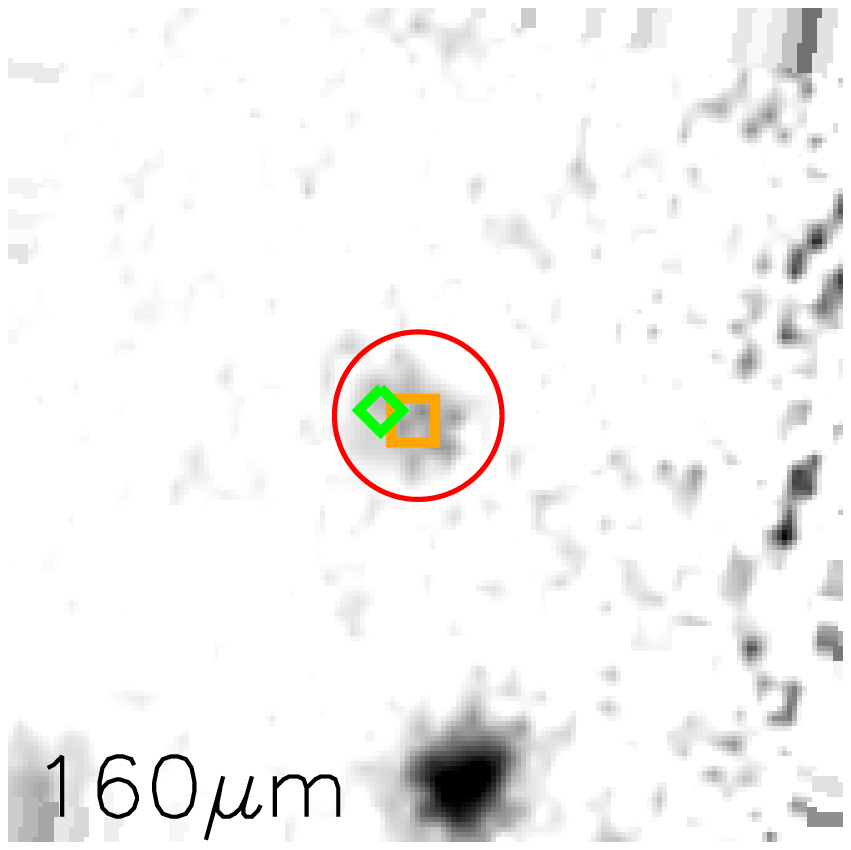}
   } 
  \parbox[height=20mm]{20mm}{
    \centering
    \includegraphics[width=20mm]{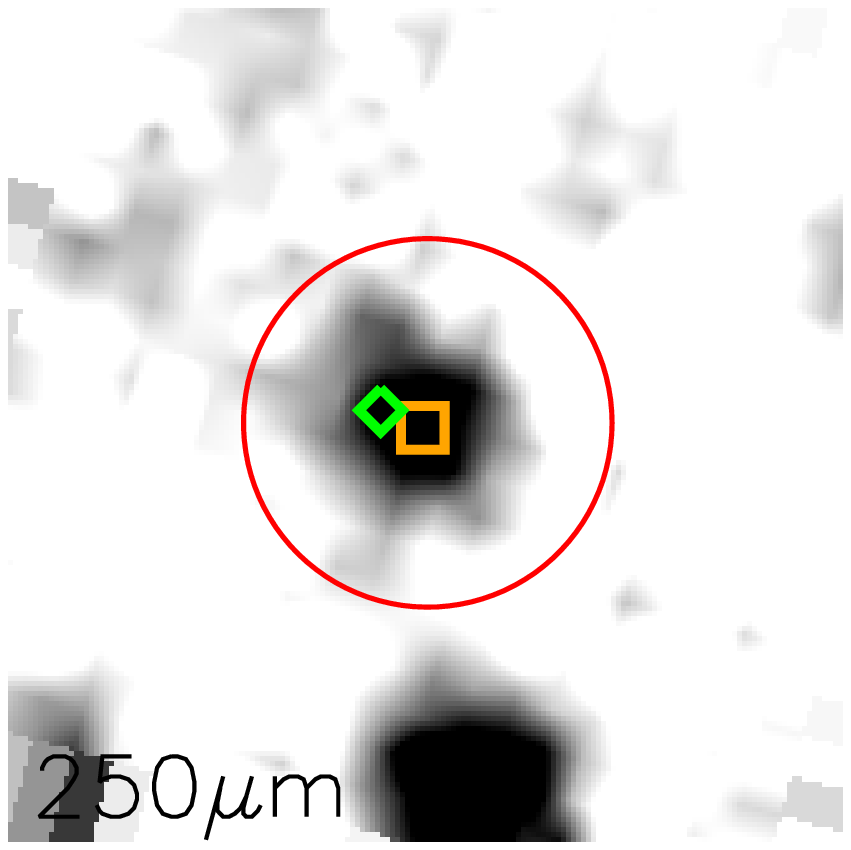}
   } 
  \parbox[height=20mm]{20mm}{
    \centering
    \includegraphics[width=20mm]{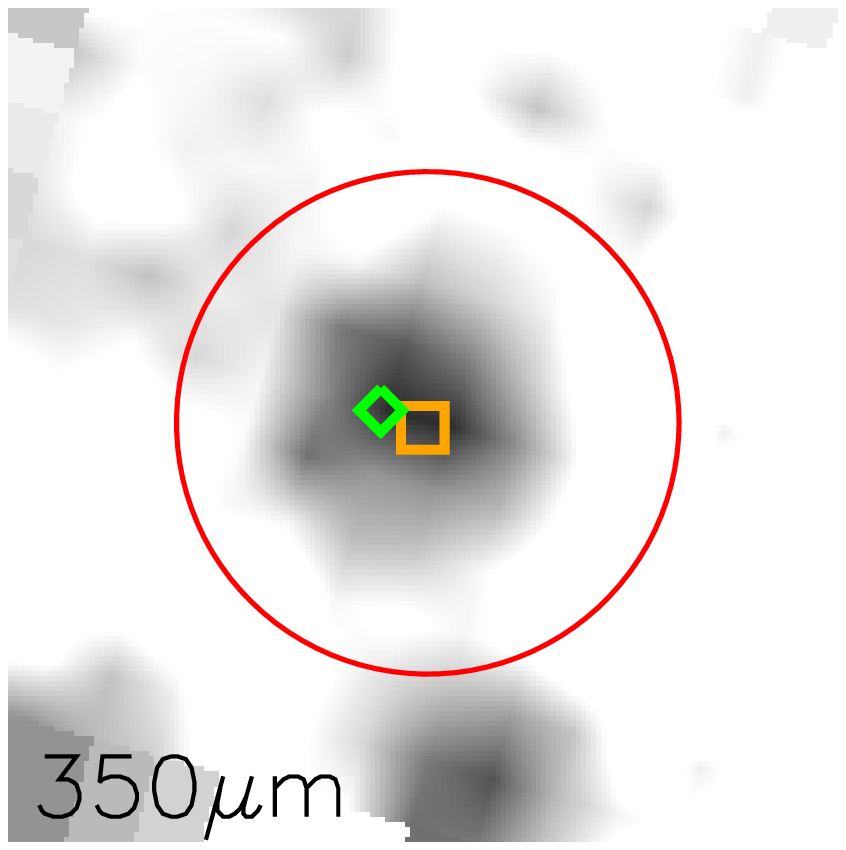}
   } 
  \parbox[height=20mm]{20mm}{
    \centering
    \includegraphics[width=20mm]{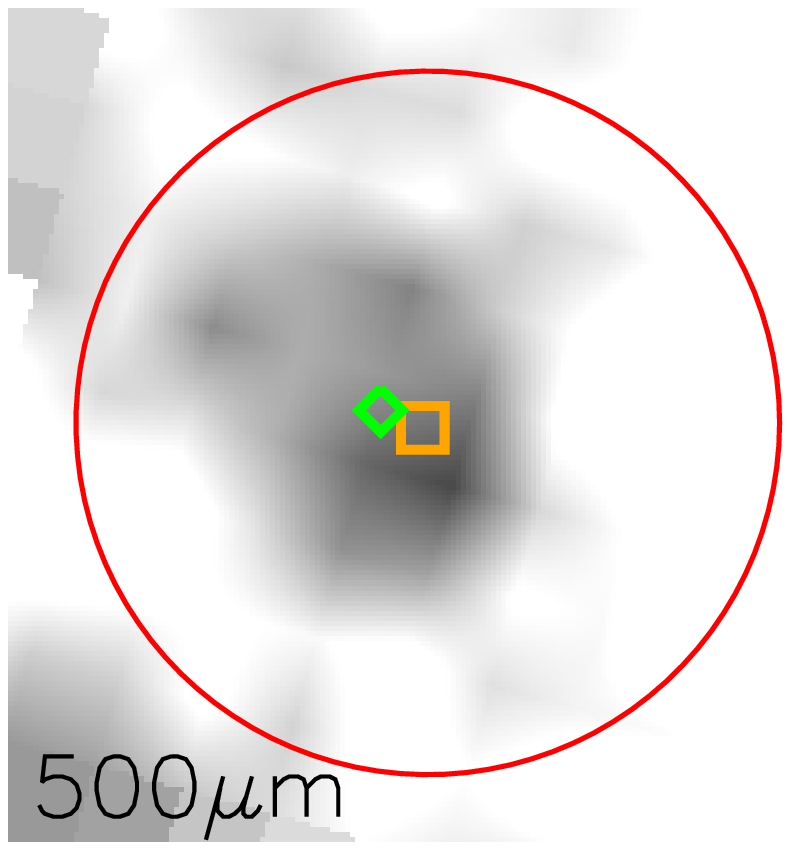}
   }   
\end{figure*}
\begin{figure*}
\parbox{70mm}{
    \centering
    \includegraphics[width=70mm]{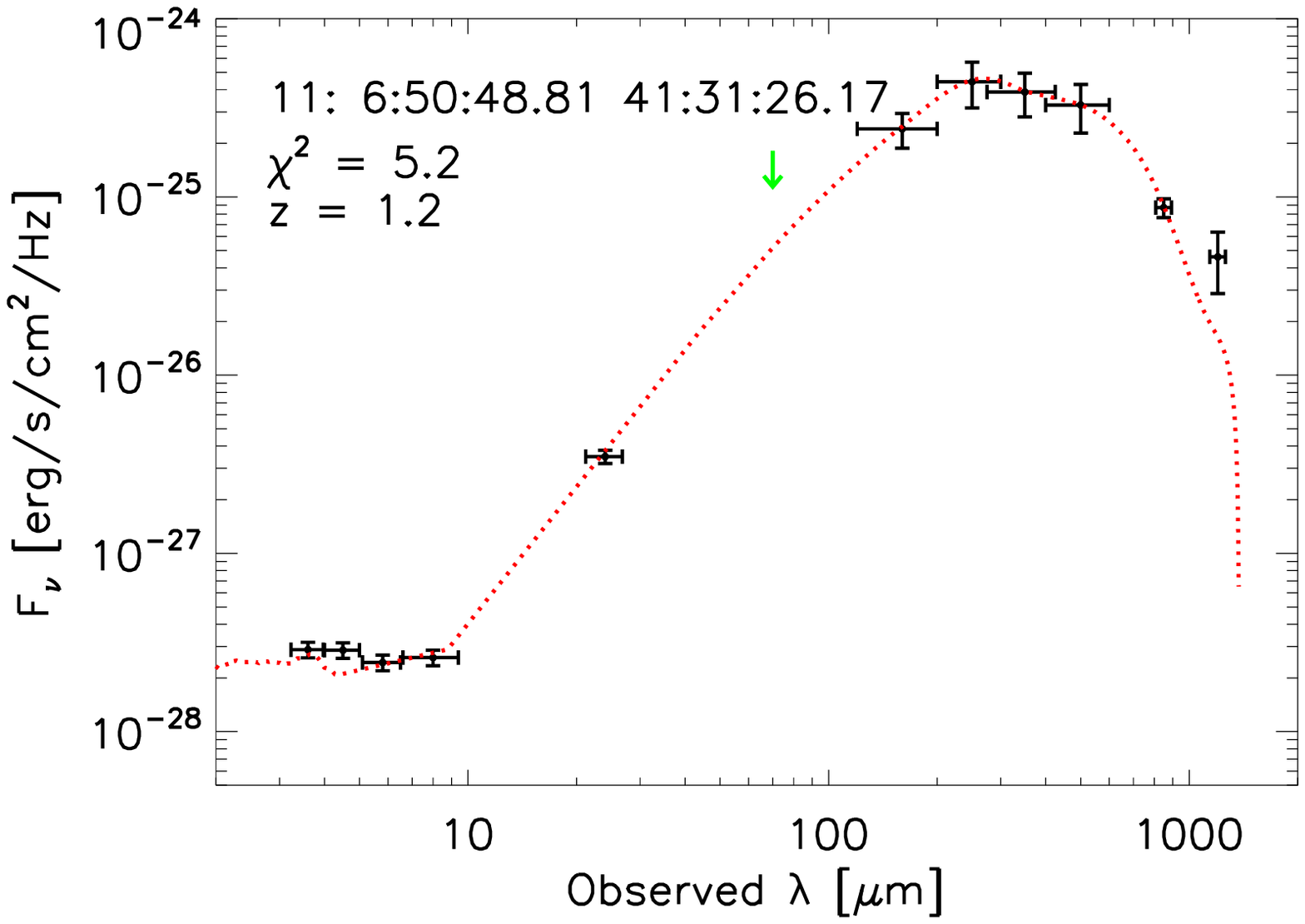}
  }
  \parbox{70mm}{
    \centering
    \includegraphics[width=70mm]{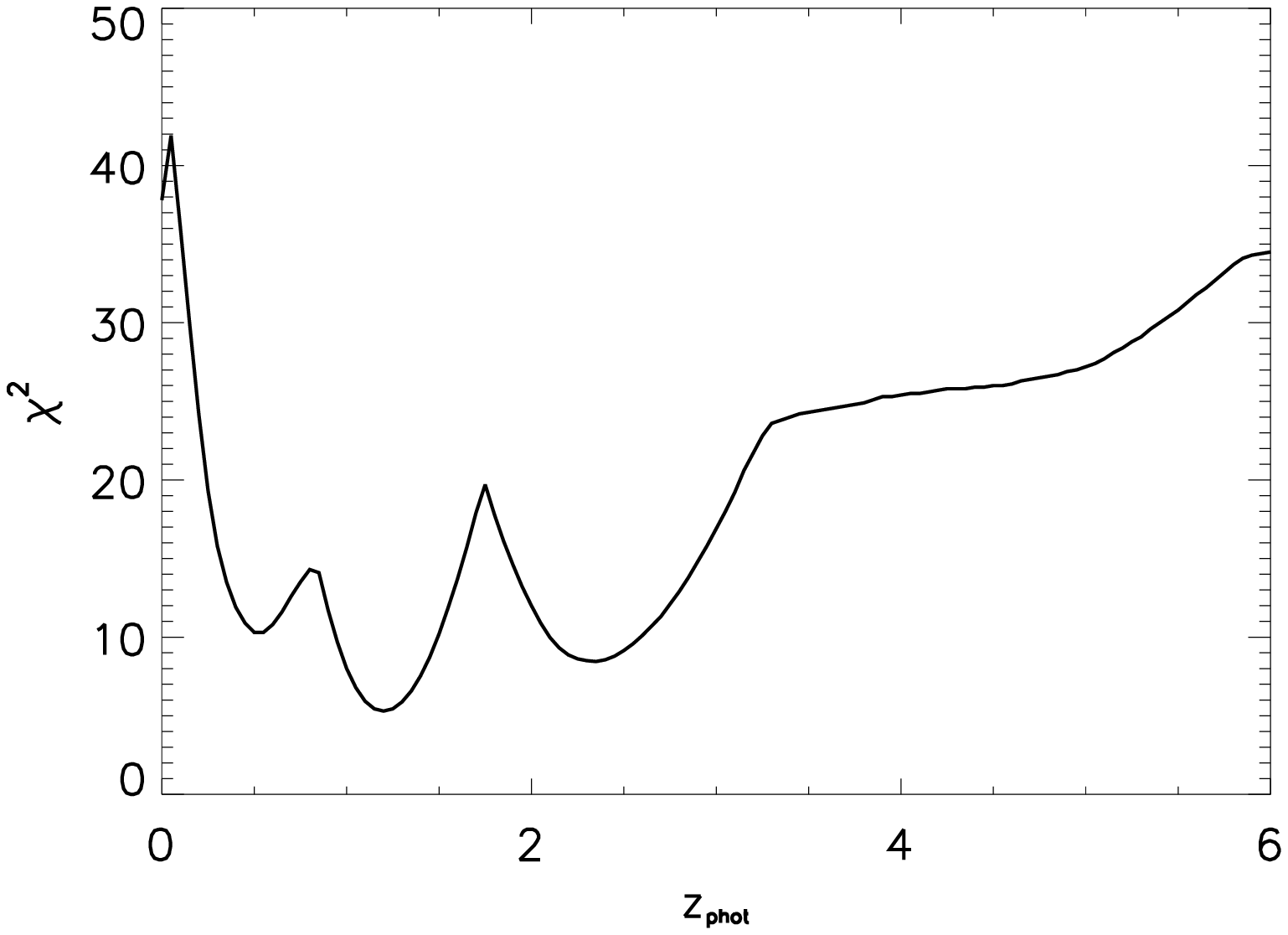}
  }
  \caption{This source is detected in all far-IR bands (160-1200 $\mu$m). The best $\chi^{2}$ is found with template 9 (this source + old stellar population) and gives a redshift of 1.2, matching the spectroscopic redshift of 1.184.}
\end{figure*}  
\begin{figure*}
\parbox[height=20mm]{20mm}{
   \centering
    13 \\ 6:50:50.28 41:33:0.93
  }
  \parbox[height=20mm]{20mm}{
    \centering
    \includegraphics[width=20mm]{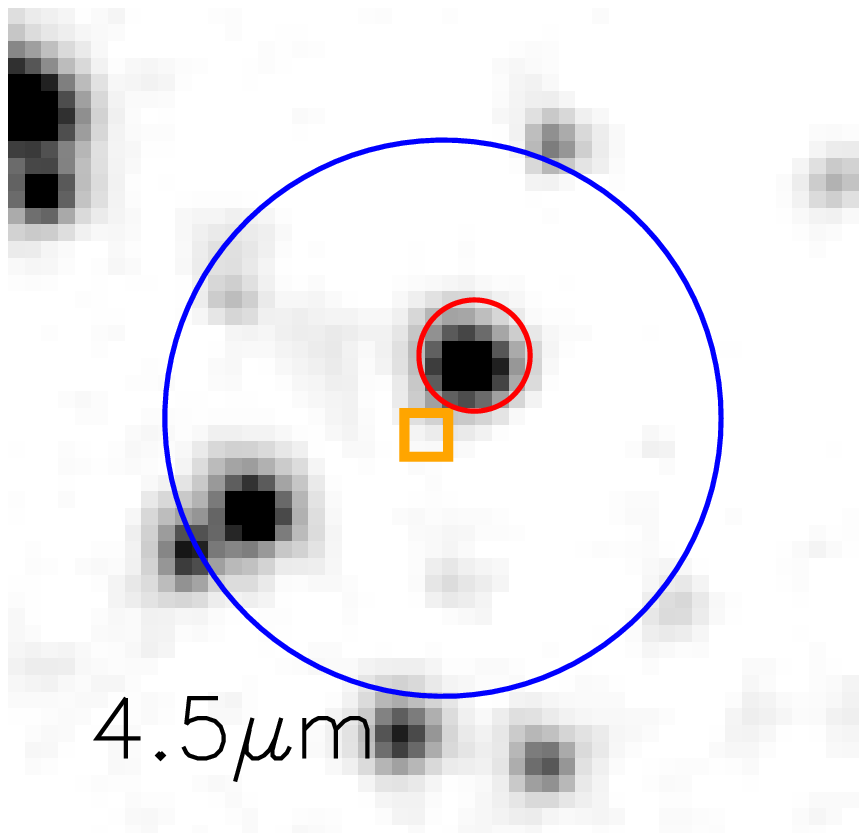}
  }
   \parbox[height=20mm]{20mm}{
    \centering\hspace{20mm}
   }
\parbox[height=20mm]{20mm}{
   \centering
    \includegraphics[width=20mm]{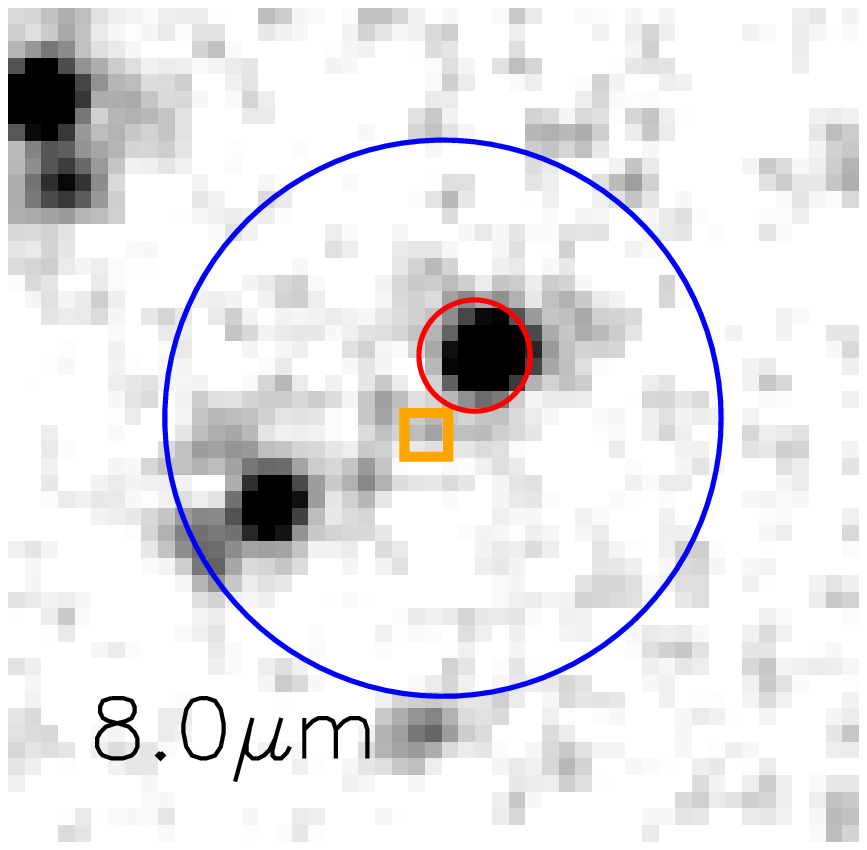}
   }
\parbox[height=20mm]{20mm}{
    \centering
    \includegraphics[width=20mm]{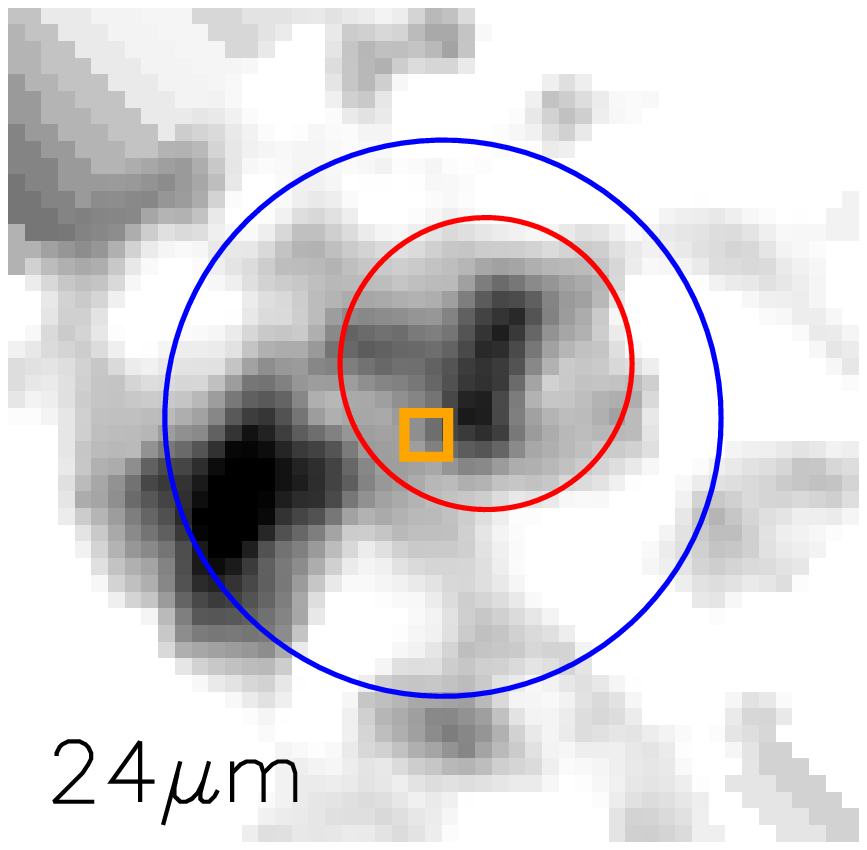}
   }
   
   \parbox[height=20mm]{20mm}{
    \centering\hspace{20mm}
   } 
  \parbox[height=20mm]{20mm}{
    \centering\hspace{20mm}
   } 
  \parbox[height=20mm]{20mm}{
    \centering
    \includegraphics[width=20mm]{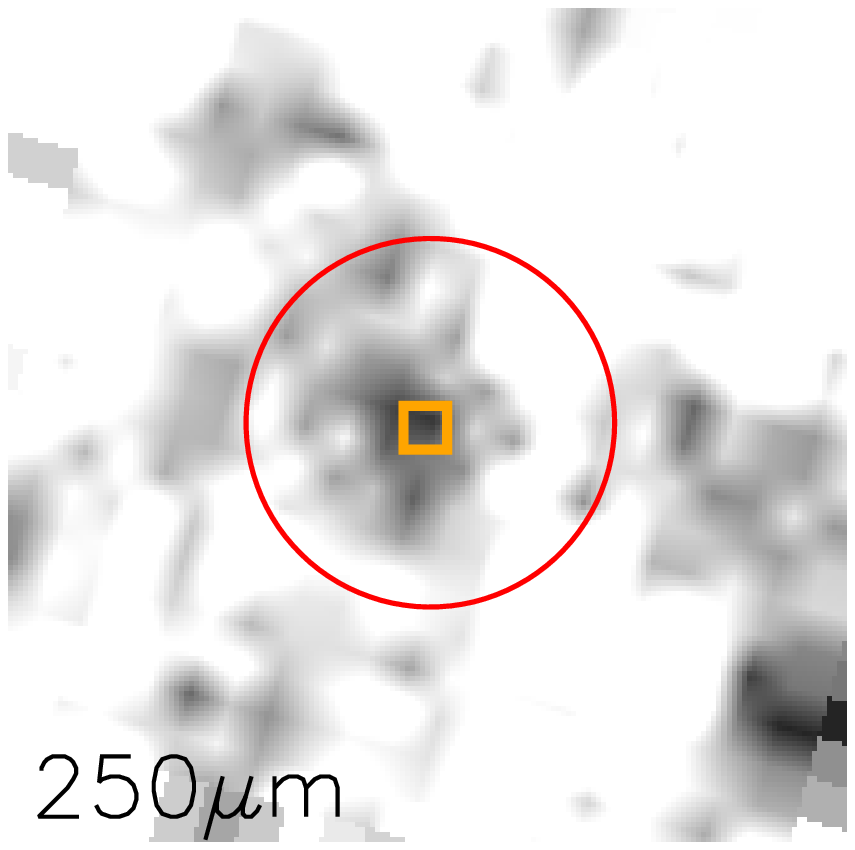}
   } 
  \parbox[height=20mm]{20mm}{
    \centering
    \includegraphics[height=20mm]{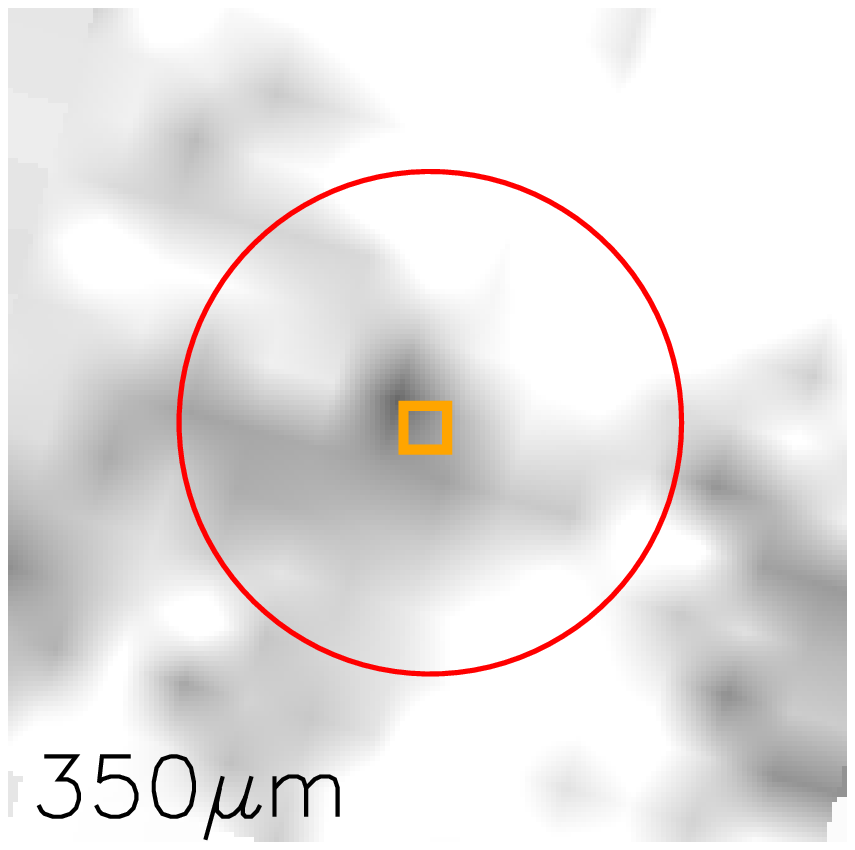}
   } 
  \parbox[height=20mm]{20mm}{
    \centering\hspace{20mm}
   } 
\end{figure*}
\begin{figure*}
 \parbox{70mm}{
    \centering
    \includegraphics[width=70mm]{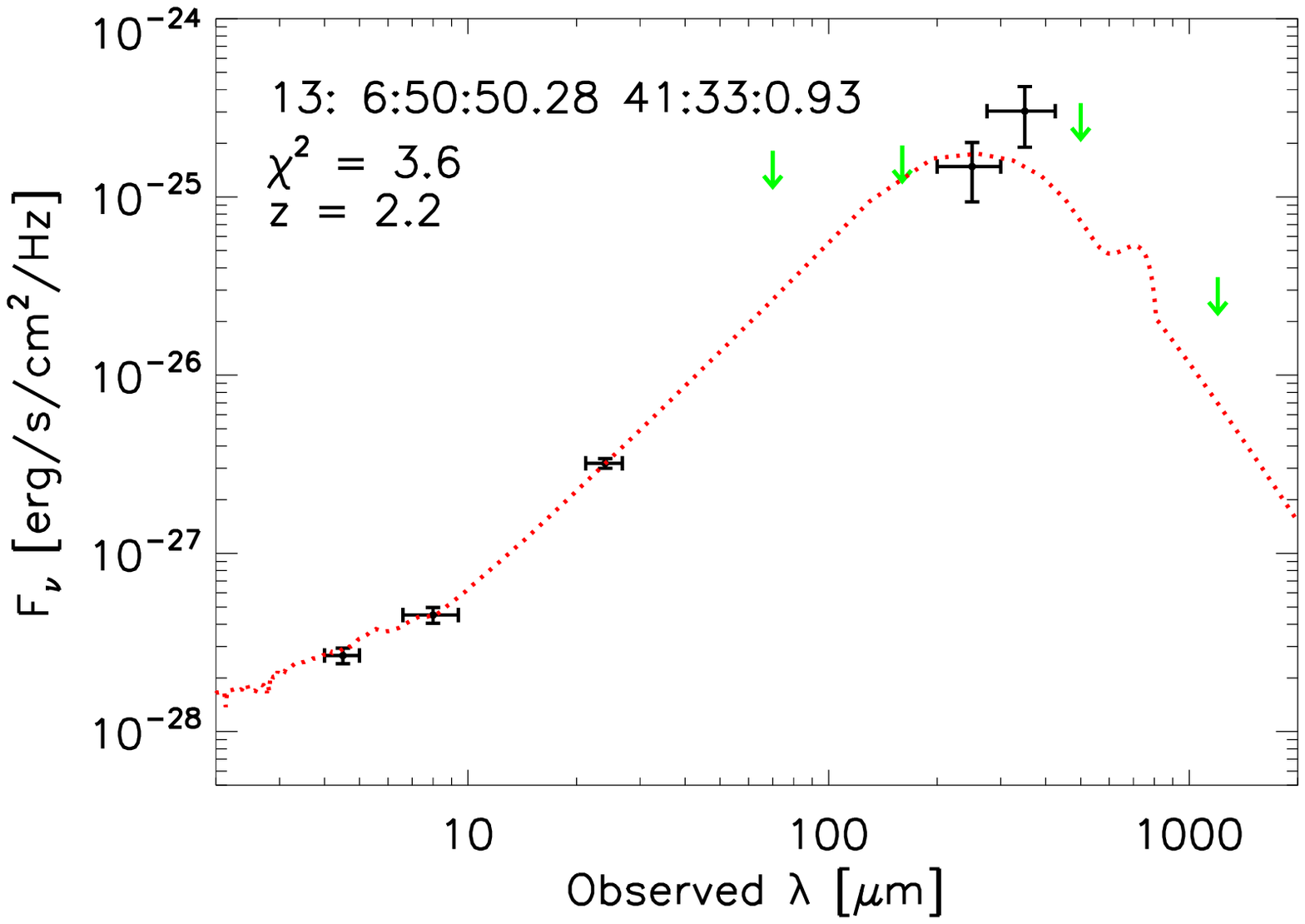}
  }
 \parbox{70mm}{
    \centering
    \includegraphics[width=70mm]{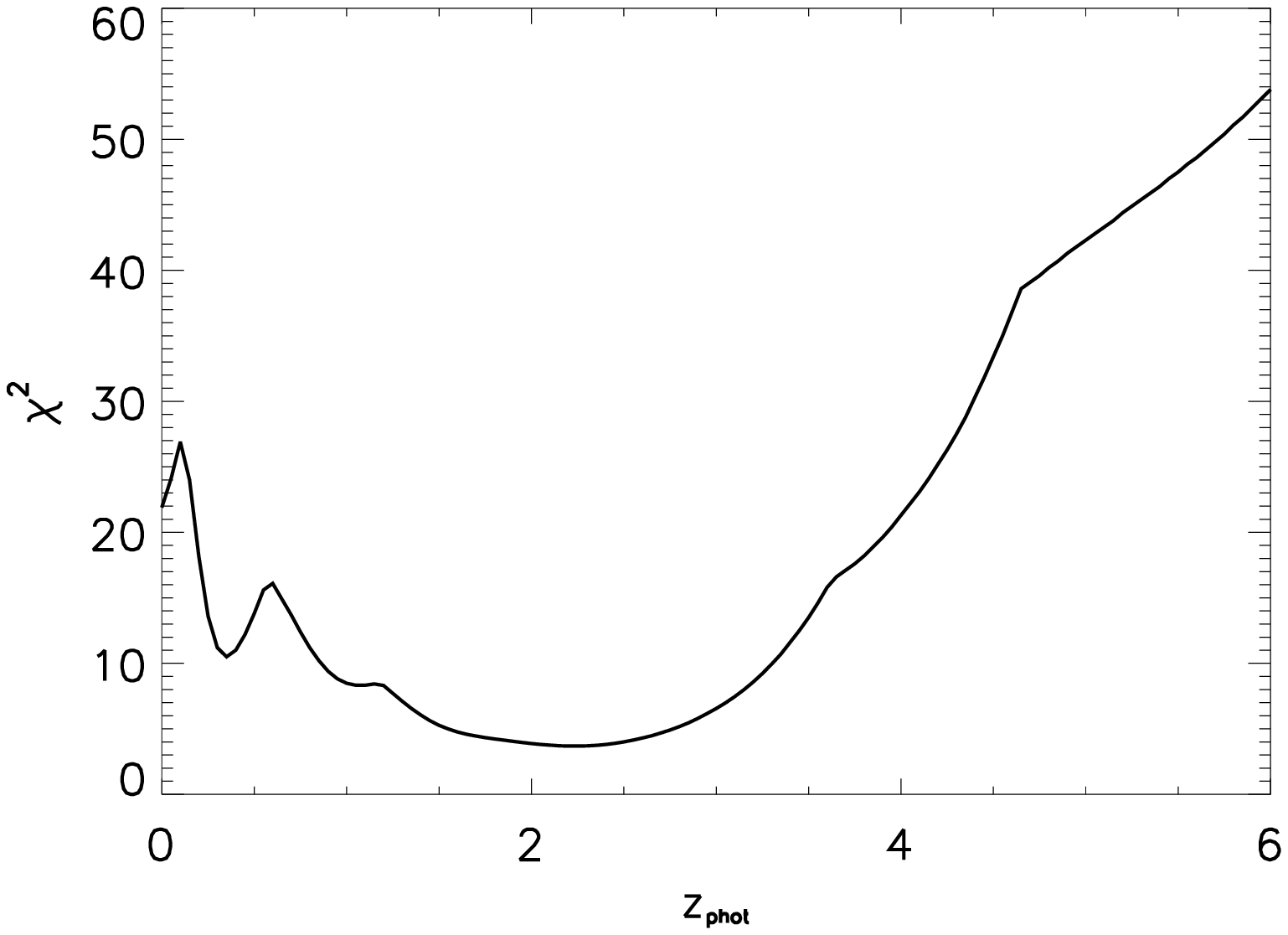}
  }
  \caption{Only five photometric points are available for fitting the SED of this source. The increasing emission towards 350 $\mu$m is not well fit and the best $\chi^{2}$ solution gives a redshift of 2.2. Blending of other sources in close proximity may cause this increasing flux at 350 $\mu$m.}
  \end{figure*}
  \begin{figure*}
  \parbox[height=20mm]{20mm}{
    \centering
    \includegraphics[width=20mm]{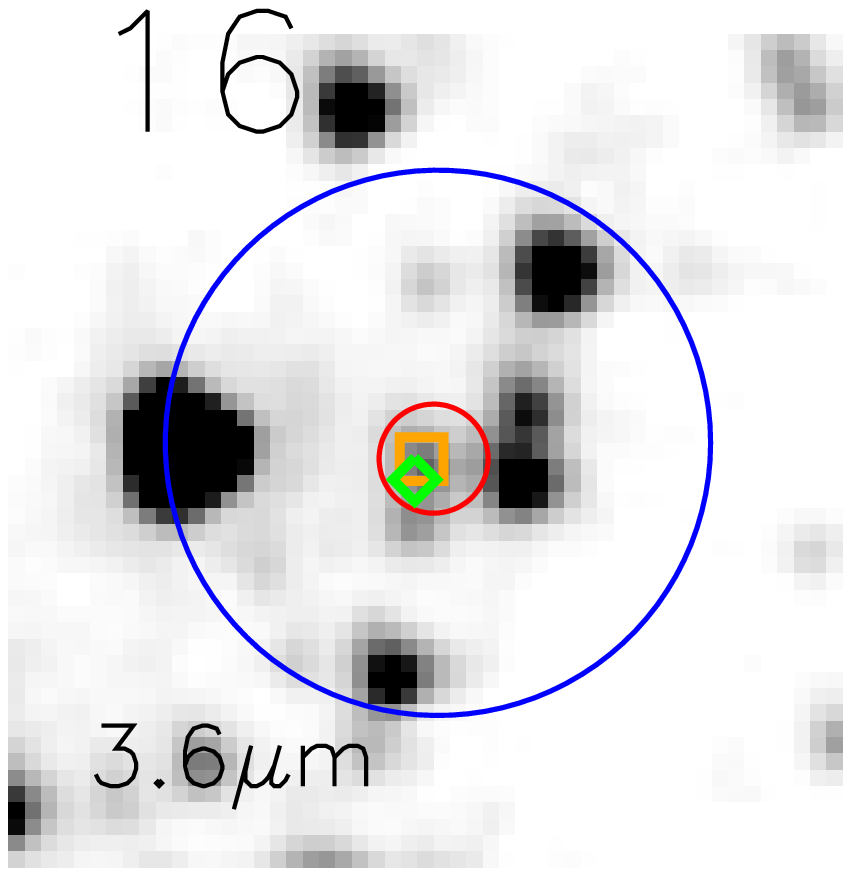}
  }
  \parbox[height=20mm]{20mm}{
    \centering
    \includegraphics[width=20mm]{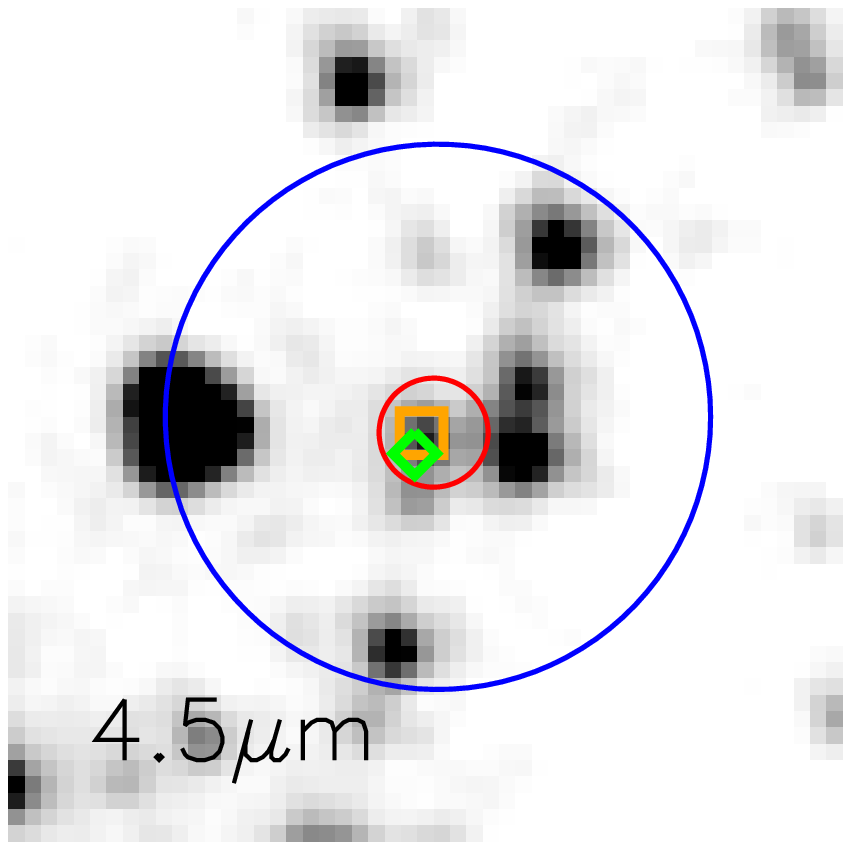}
  }
   \parbox[height=20mm]{20mm}{
    \centering
    \includegraphics[width=20mm]{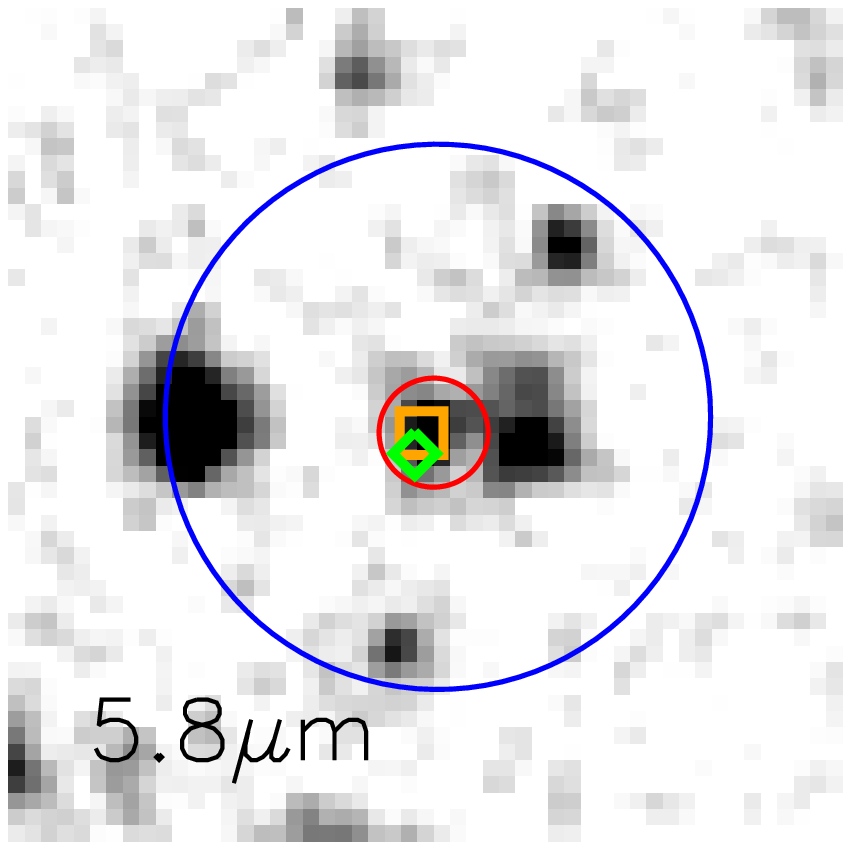}
   }
\parbox[height=20mm]{20mm}{
   \centering
    \includegraphics[width=20mm]{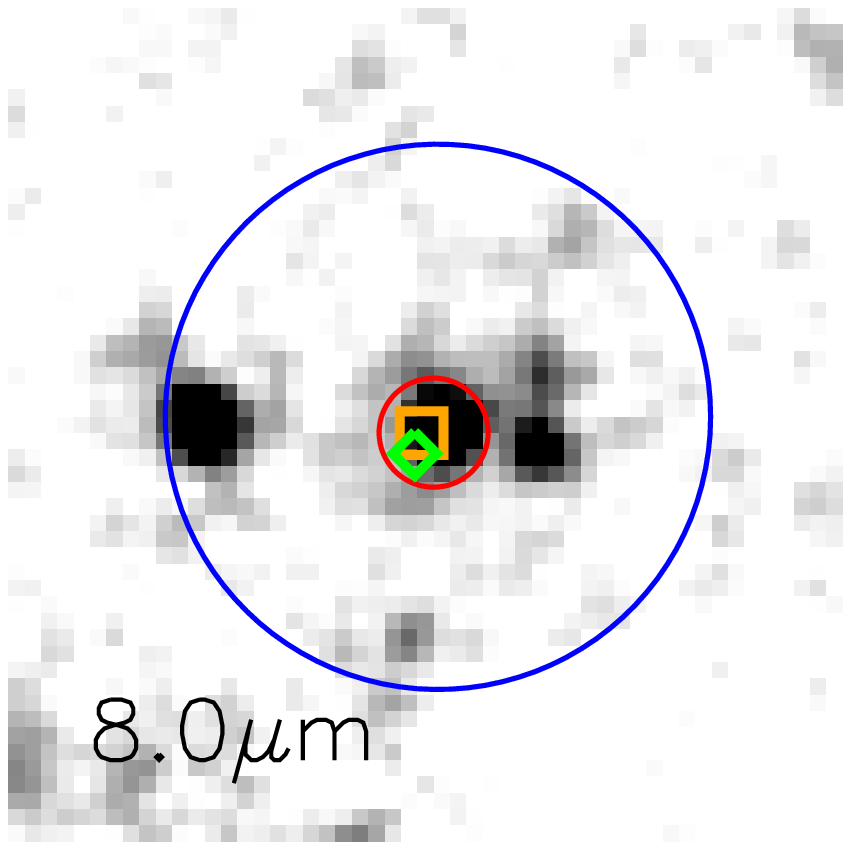}
   }
\parbox[height=20mm]{20mm}{
    \centering
    \includegraphics[width=20mm]{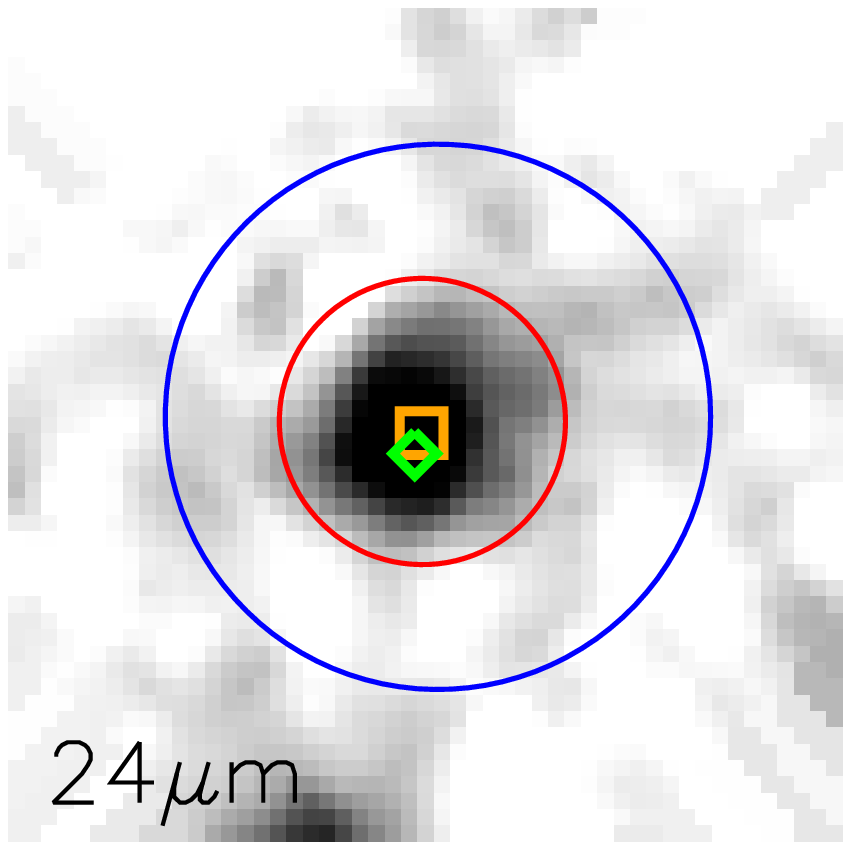}
   }
   
   \parbox[height=20mm]{20mm}{
    \centering\hspace{20mm}
   } 
   \parbox[height=20mm]{20mm}{
    \centering
    \includegraphics[width=20mm]{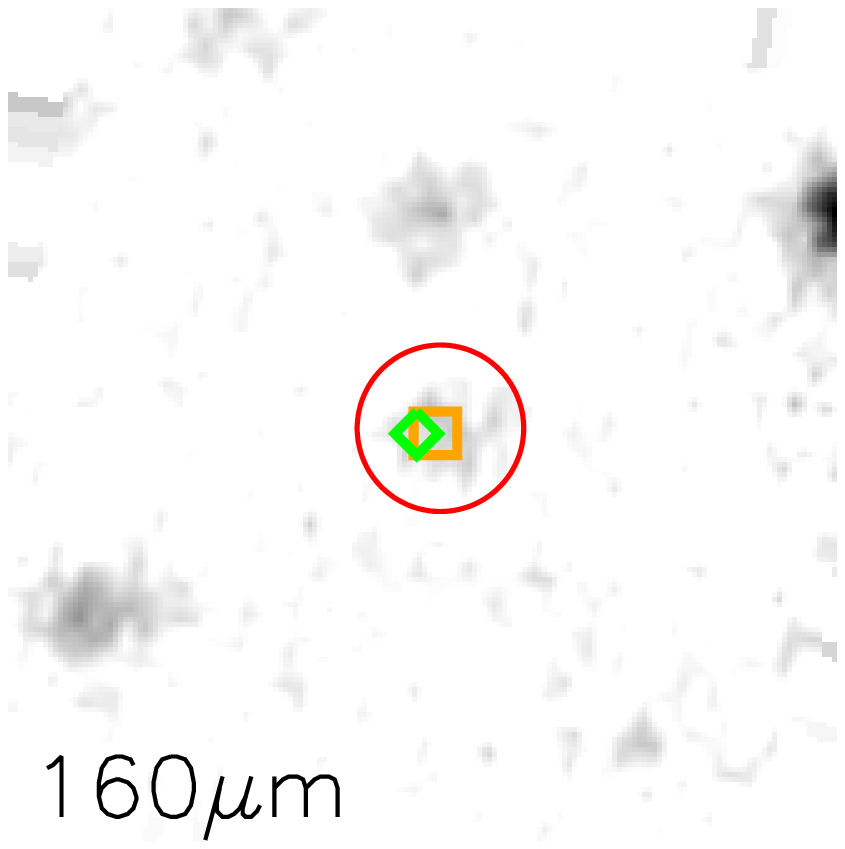}
   } 
   \parbox[height=20mm]{20mm}{
    \centering
    \includegraphics[width=20mm]{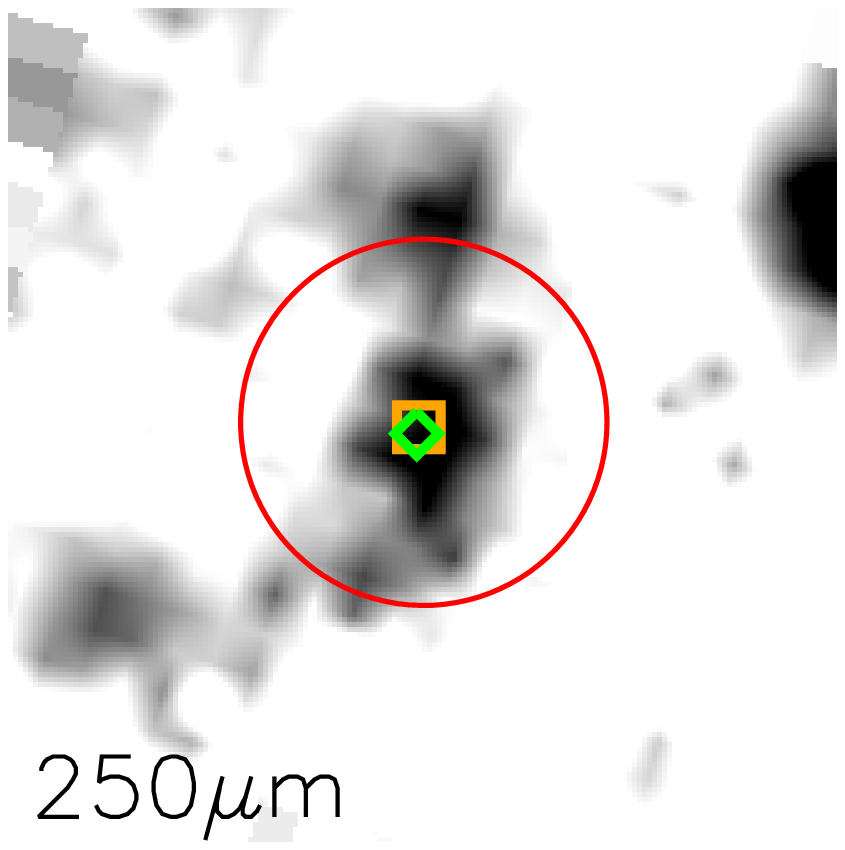}
   } 
   \parbox[height=20mm]{20mm}{
    \centering
    \includegraphics[width=20mm]{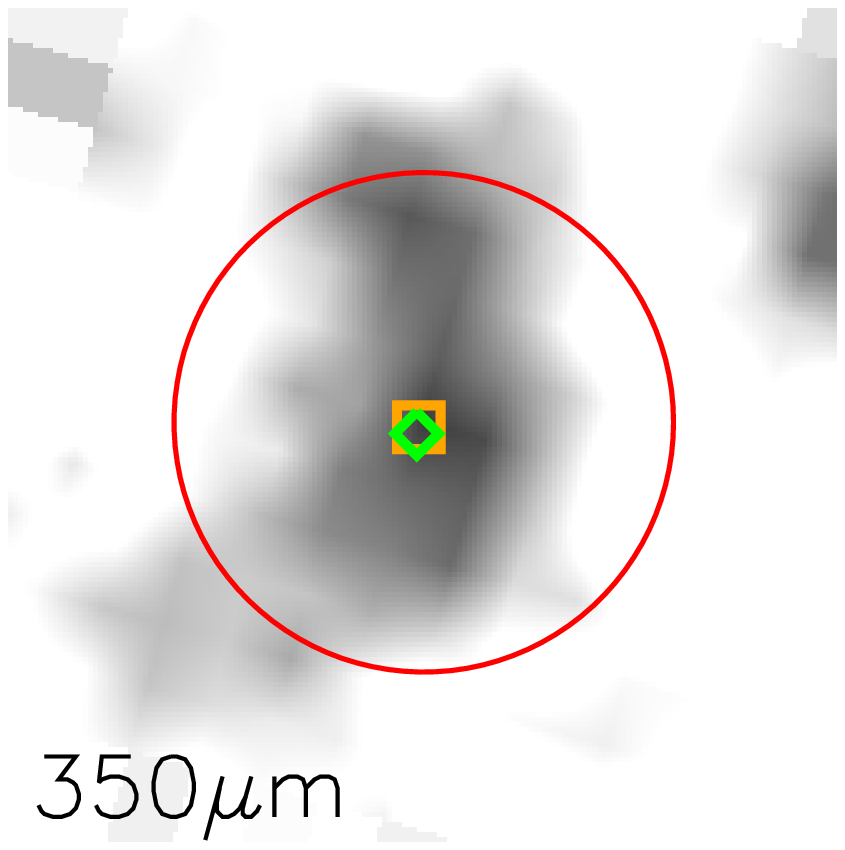}
   } 
   \parbox[height=20mm]{20mm}{
    \centering
    \includegraphics[width=20mm]{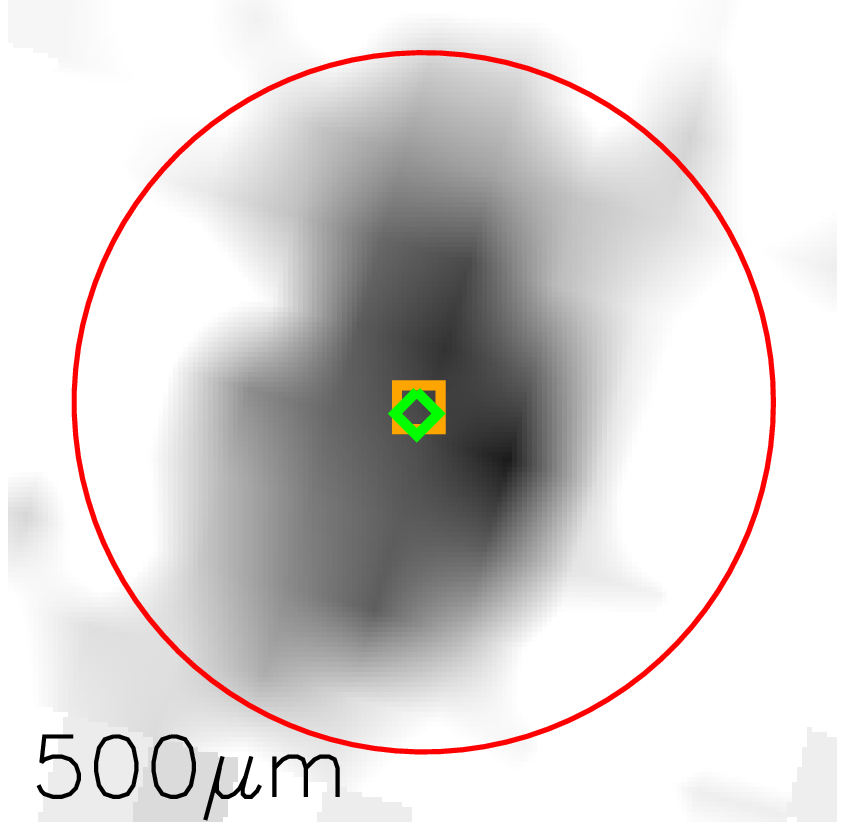}
   } 
\end{figure*}
\begin{figure*}
\parbox{70mm}{
    \centering
    \includegraphics[width=70mm]{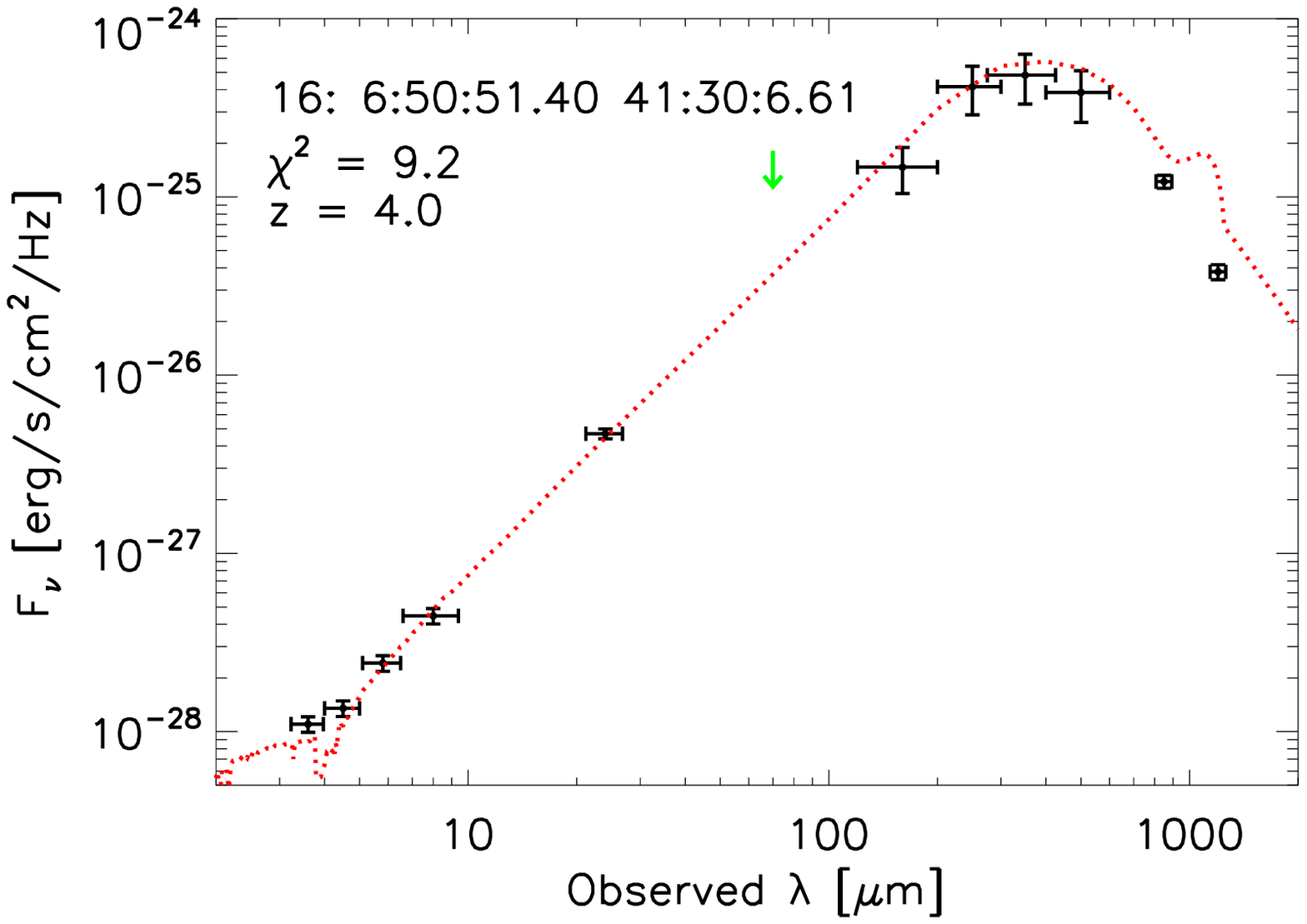}
  }  \parbox{70mm}{
    \centering
    \includegraphics[width=70mm]{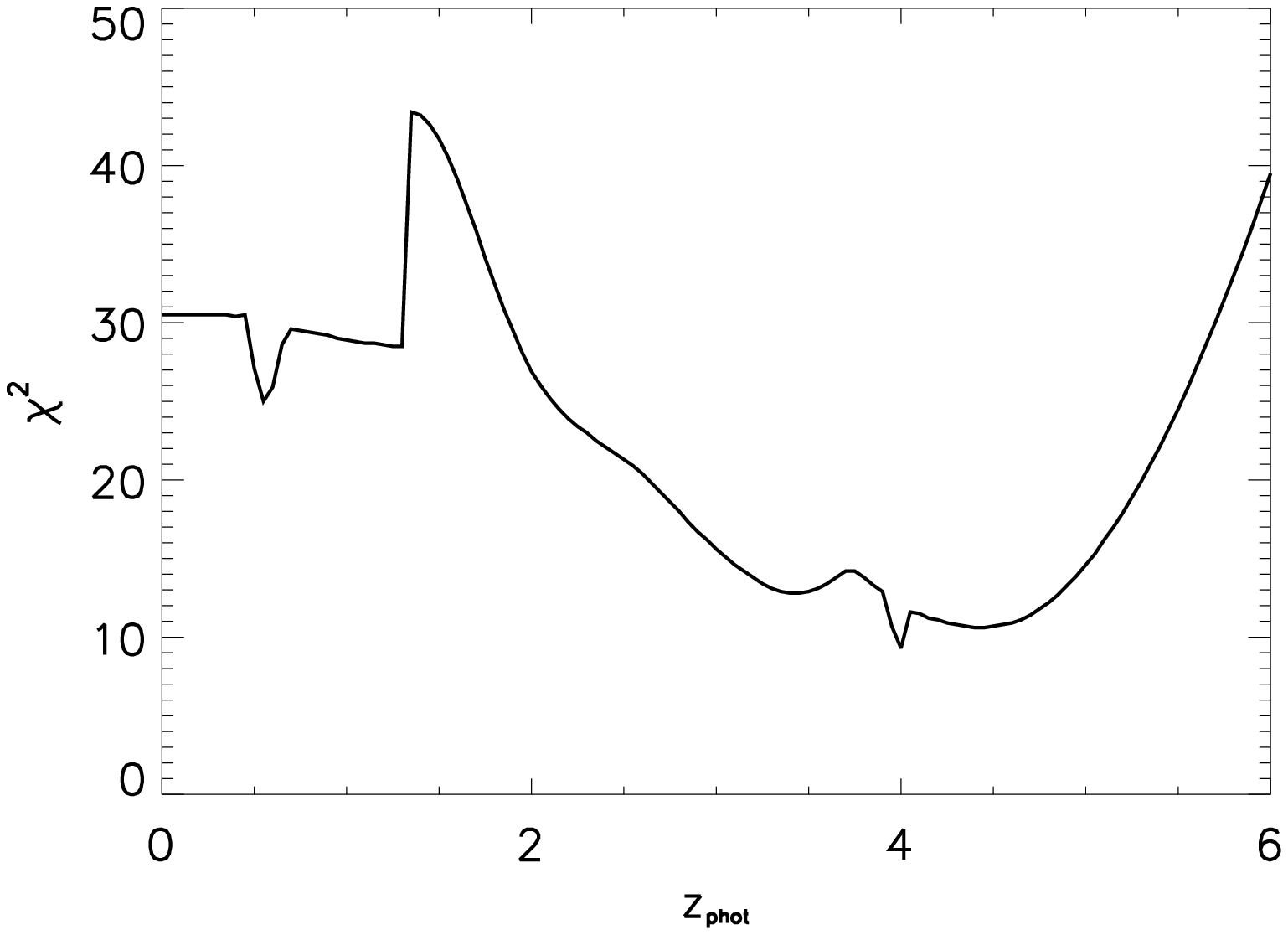}
  }
  \caption{This source is nicely fit by an AGN dominated template, similar to the SED of 4C+41.17. The $\chi^{2}$ distribution shows a clear and prominent dip at the redshift of the radio galaxy. This source is therefore our most likely candidate to be associated with the radio galaxy.} 
   \end{figure*}
 \begin{figure*}
 \parbox[height=20mm]{20mm}{
    \centering
    \includegraphics[width=20mm]{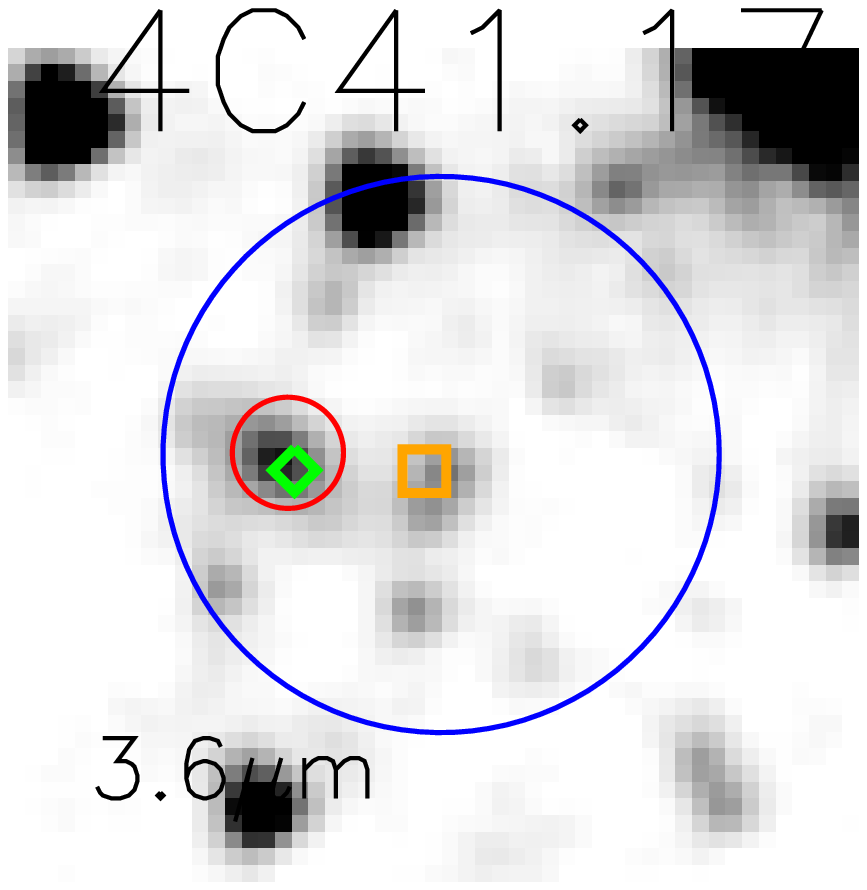}
  }
  \parbox[height=20mm]{20mm}{
    \centering
    \includegraphics[width=20mm]{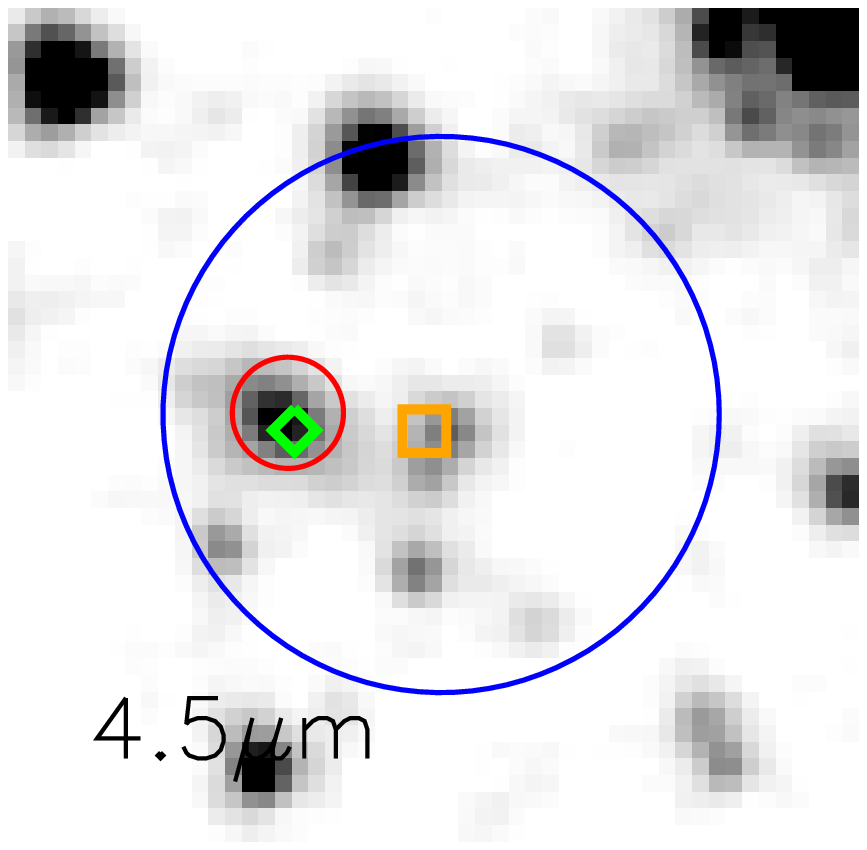}
   }
   \parbox[height=20mm]{20mm}{
    \centering
    \includegraphics[width=20mm]{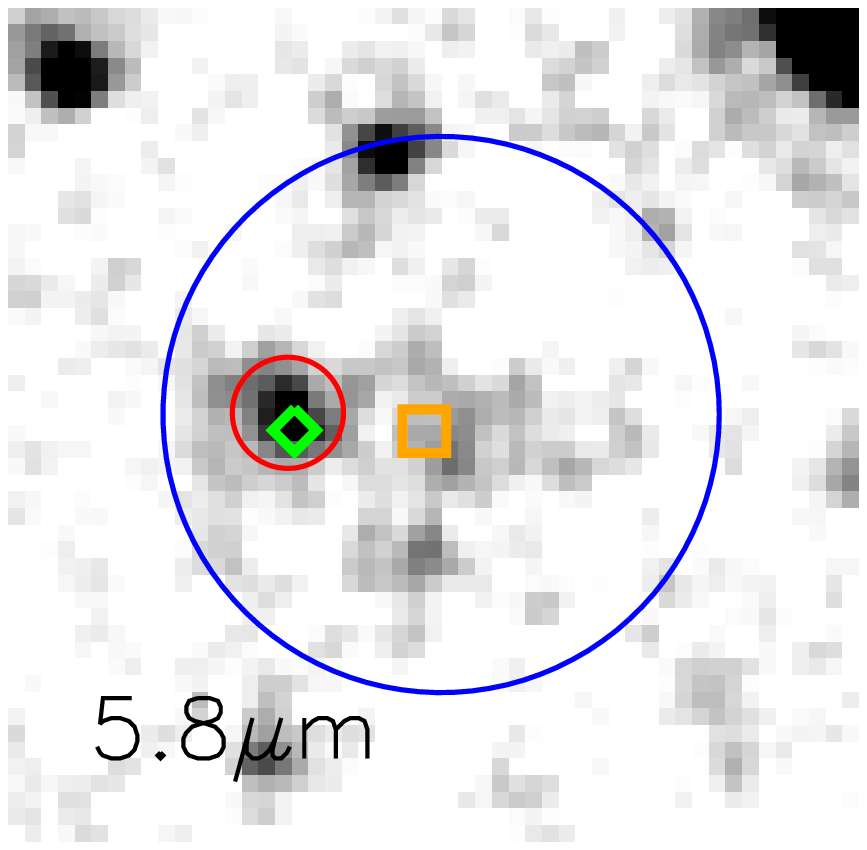}
   }
\parbox[height=20mm]{20mm}{
    \centering
    \includegraphics[width=20mm]{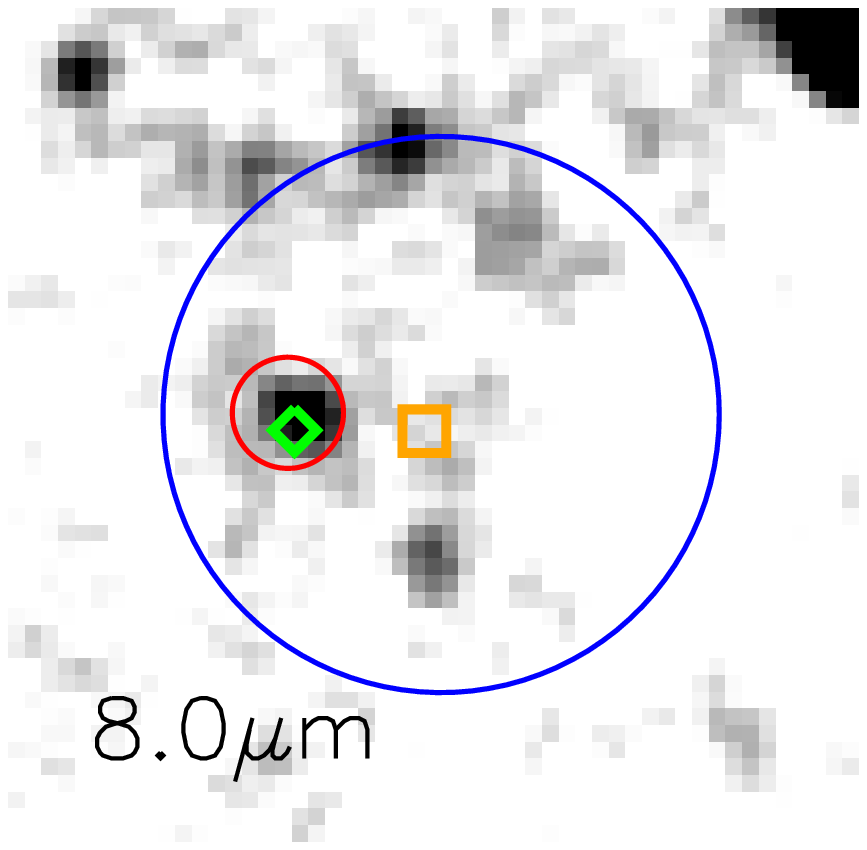}
   }
\parbox[height=20mm]{20mm}{
    \centering
    \includegraphics[width=20mm]{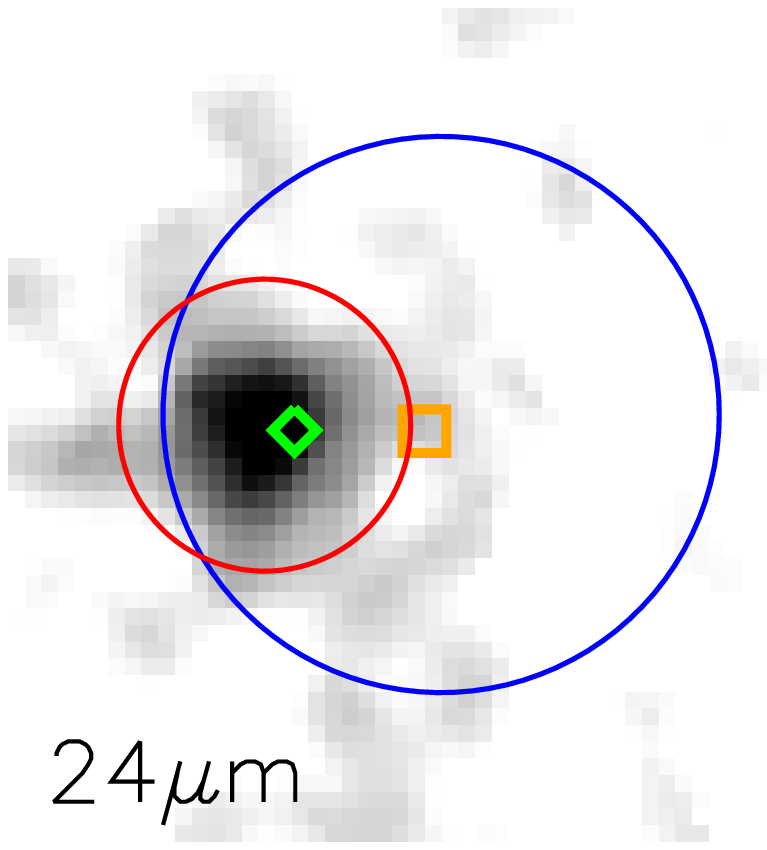}
   } 
   
   \parbox[height=20mm]{20mm}{
    \centering\hspace{20mm}
   }   
     \parbox[height=20mm]{20mm}{
    \centering
    \includegraphics[width=20mm]{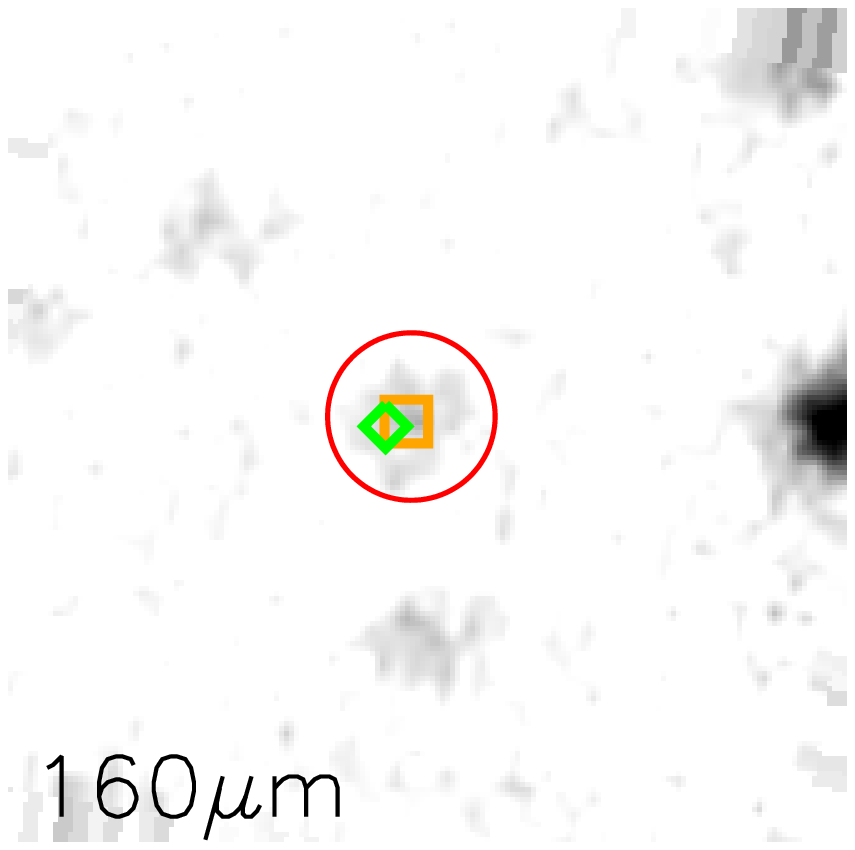}
   }   
     \parbox[height=20mm]{20mm}{
    \centering
    \includegraphics[width=20mm]{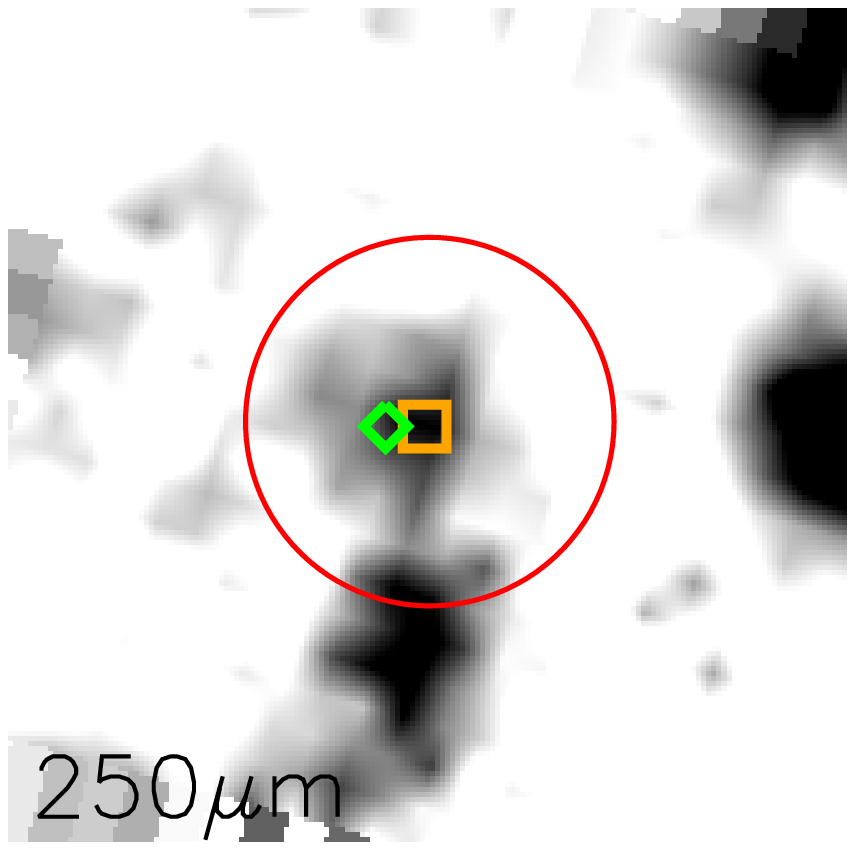}
   }   
     \parbox[height=20mm]{20mm}{
    \centering
    \includegraphics[width=20mm]{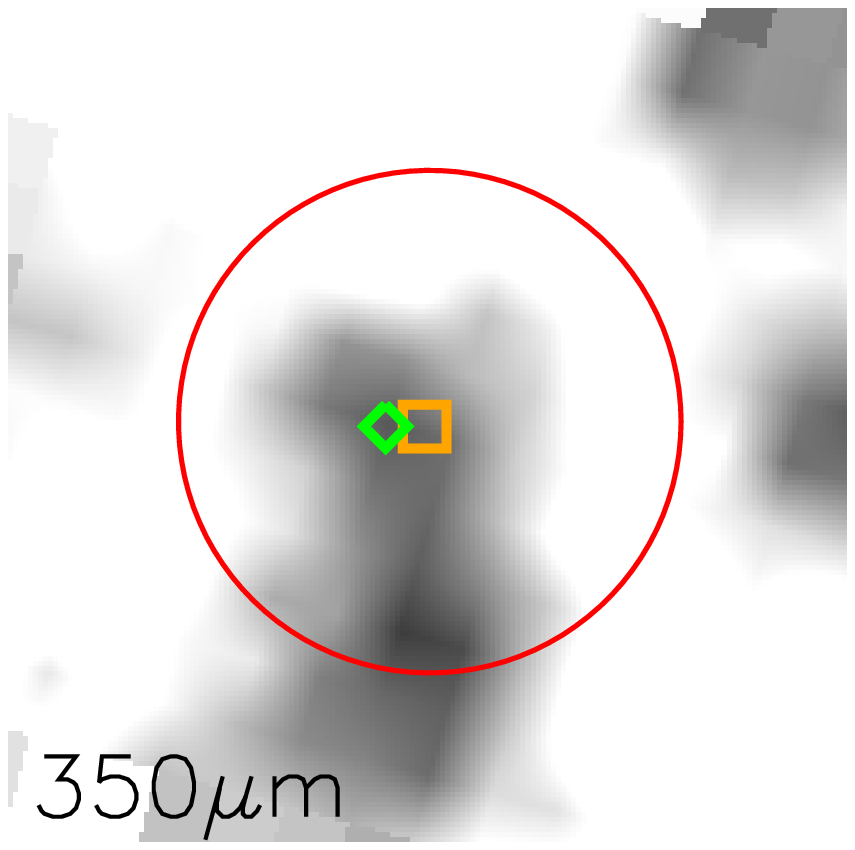}
   }   
     \parbox[height=20mm]{20mm}{
    \centering
    \includegraphics[width=20mm]{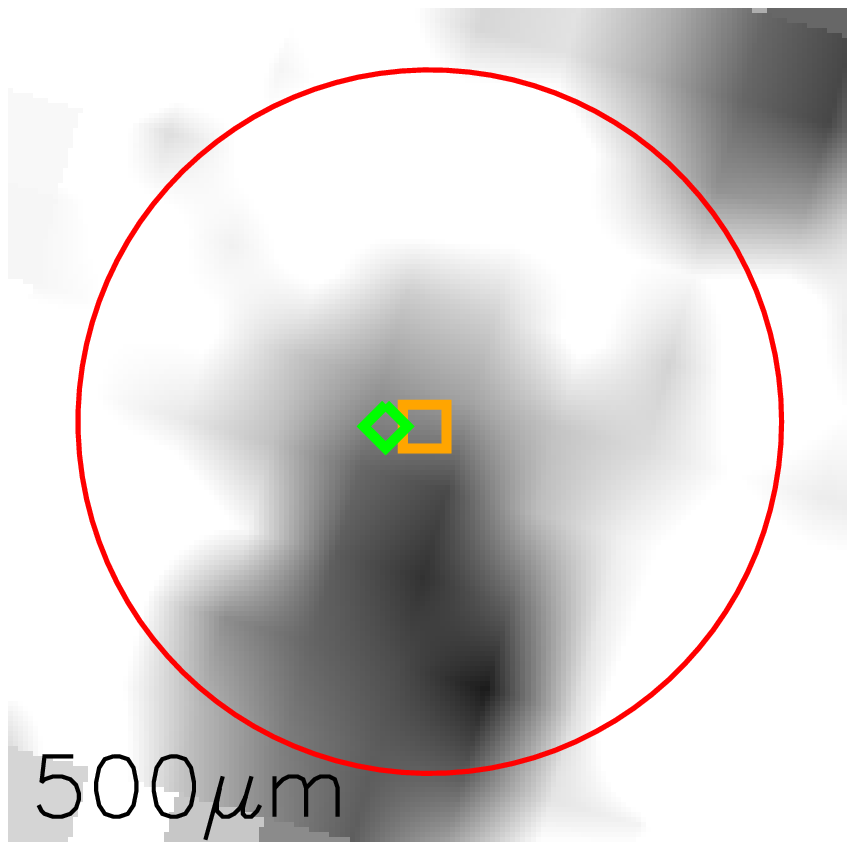}
   }   
   \end{figure*}
   \begin{figure*}
        \parbox{70mm}{
    \centering
    \includegraphics[width=70mm]{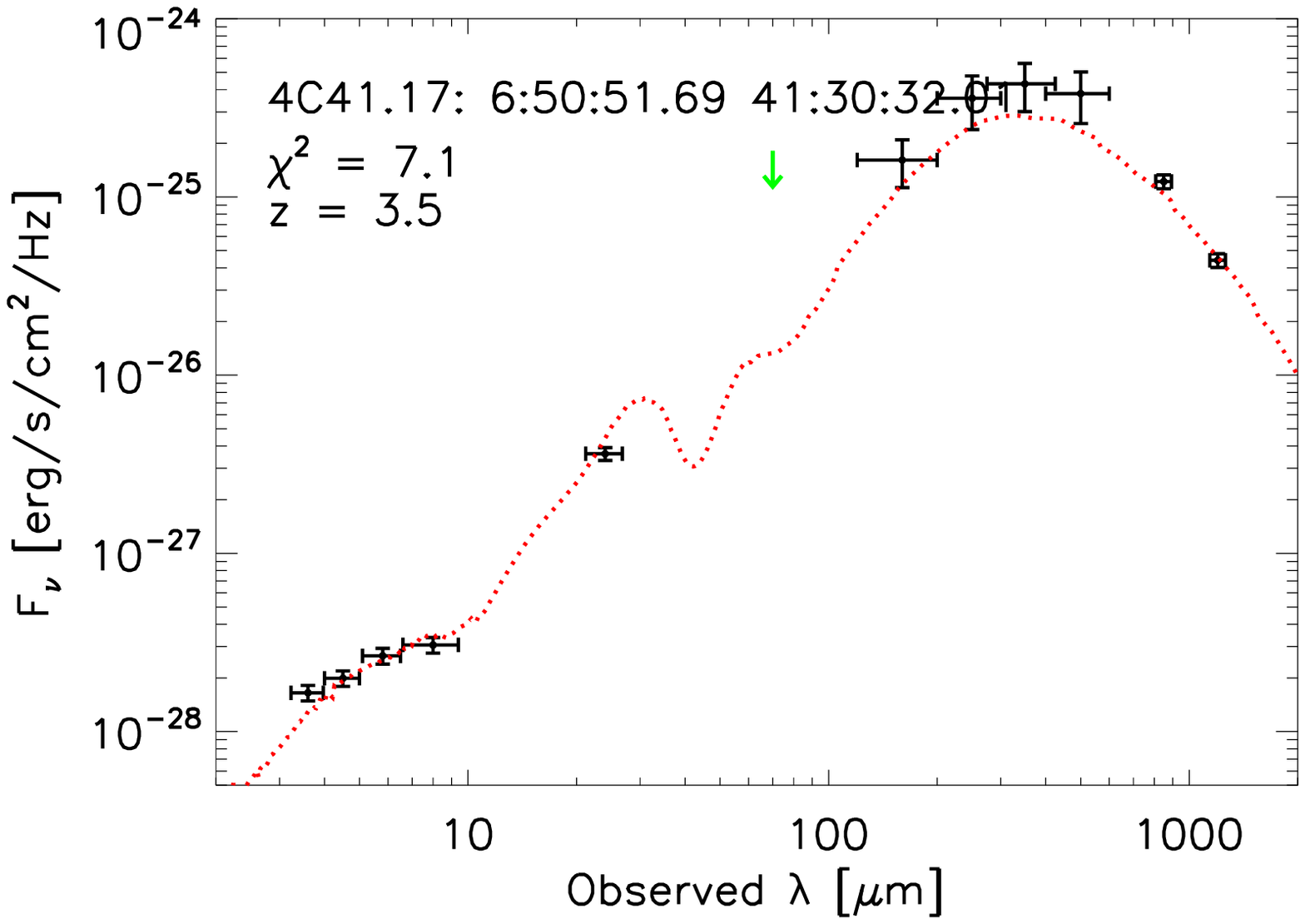}
  } 
    \parbox{70mm}{
    \centering
    \includegraphics[width=70mm]{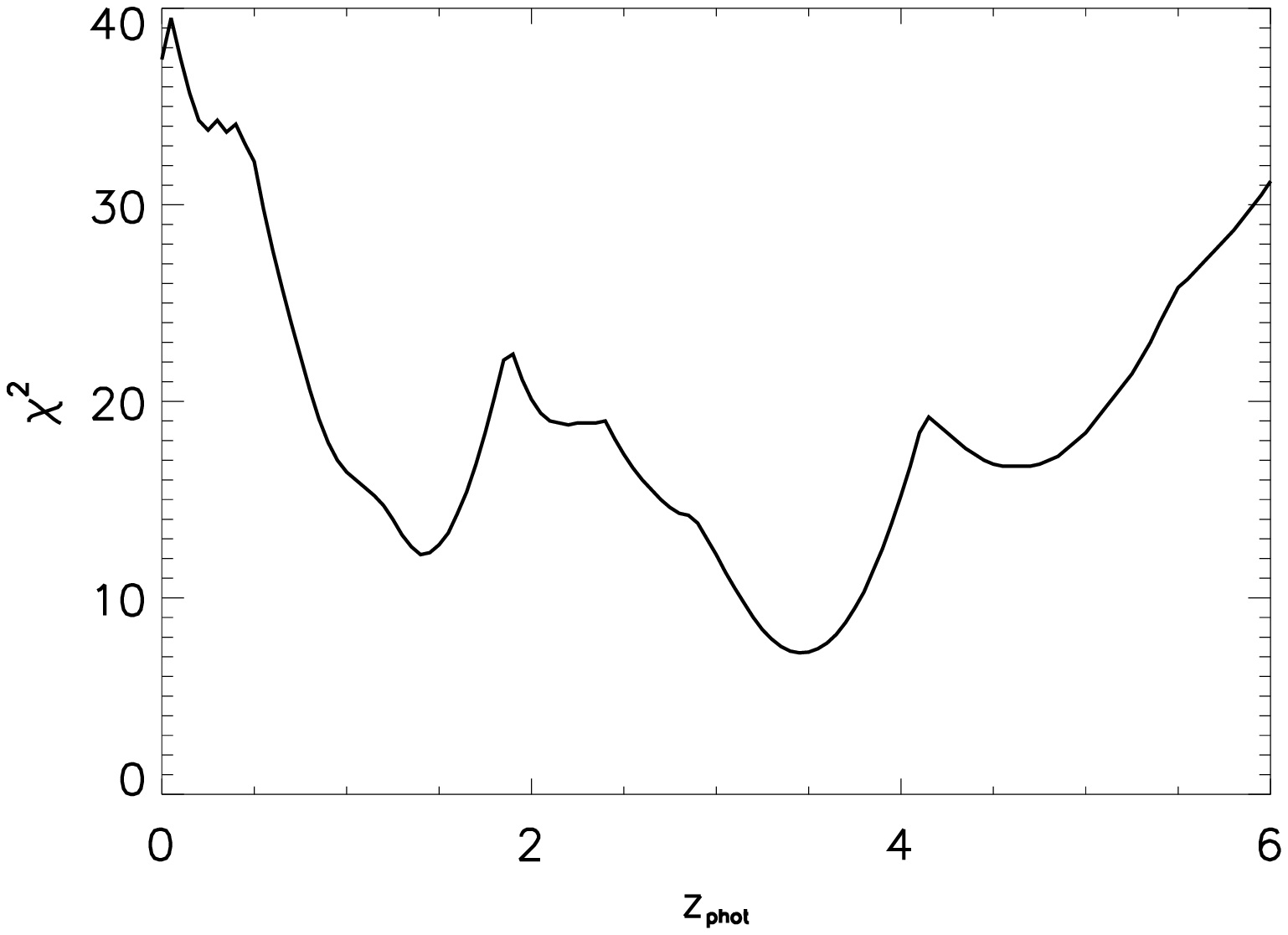}
  }
  \caption{The redshift of the radio galaxy is well constrained by the photometric redshift fitting using a composite AGN+starburst template (I19254). A single significant dip appears at a redshift of $\sim$ 3.5 in the $\chi^{2}$ distribution which is consistent with the spectroscopic redshift of 3.792. Note that this source is not fit by its own template as the stellar to dust peak ratio in those templates is not as optimal as in template 2.} 
 \end{figure*}  
 \clearpage
\begin{figure*}
\parbox[height=20mm]{20mm}{
    \centering
    \includegraphics[width=20mm]{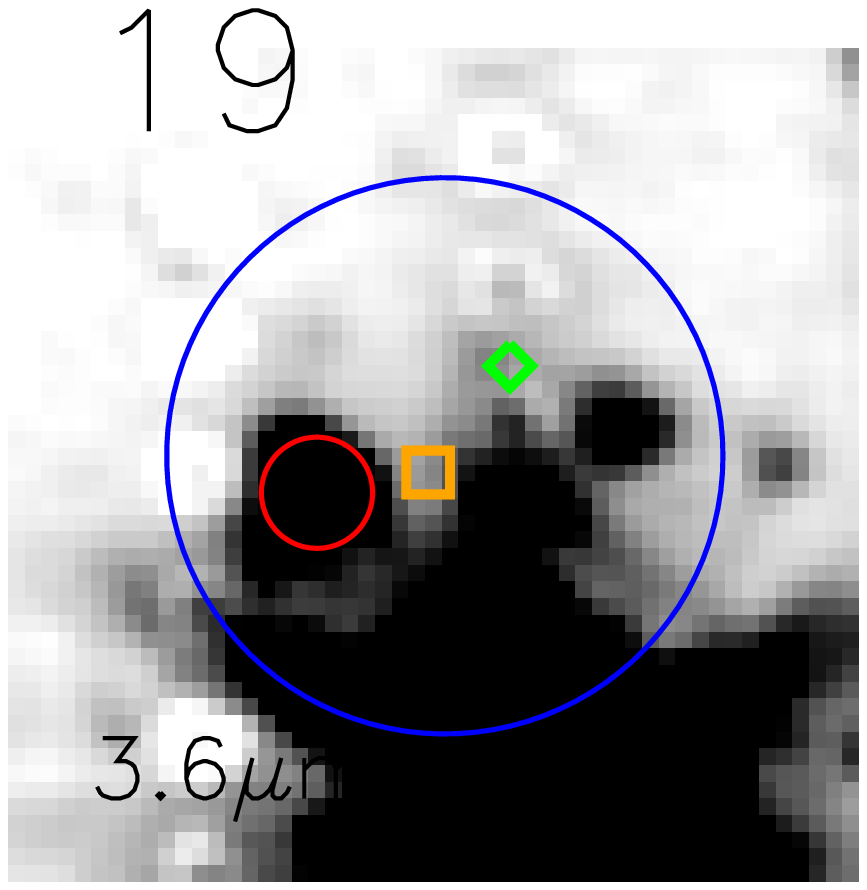}
  }
  \parbox[height=20mm]{20mm}{
    \centering
    \includegraphics[width=20mm]{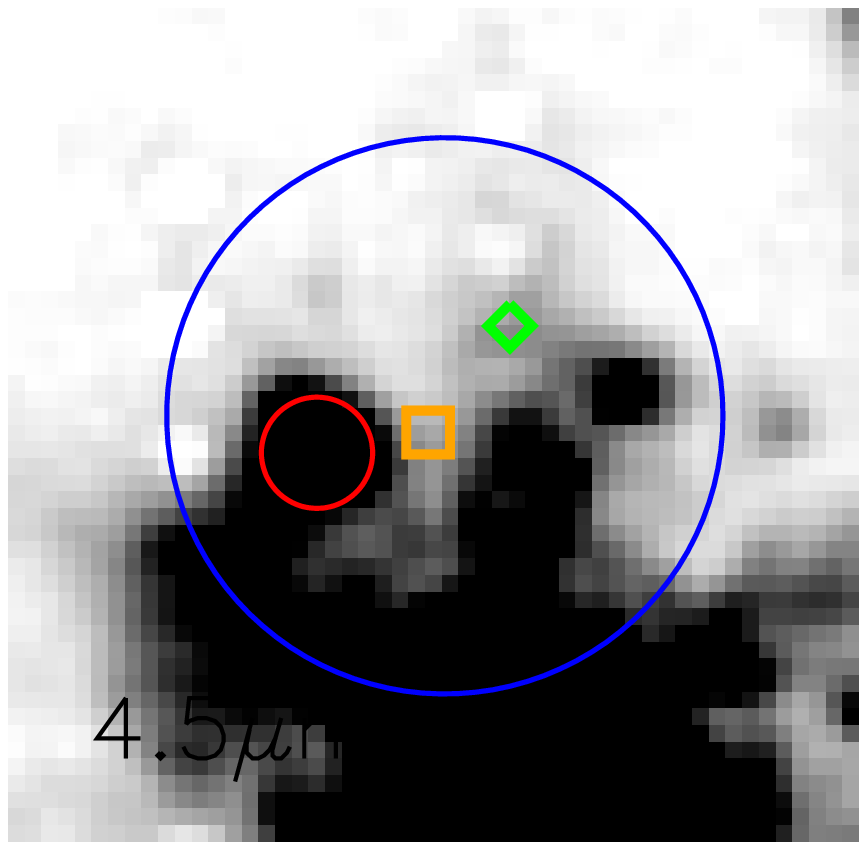}
  }
   \parbox[height=20mm]{20mm}{
    \centering
    \includegraphics[width=20mm]{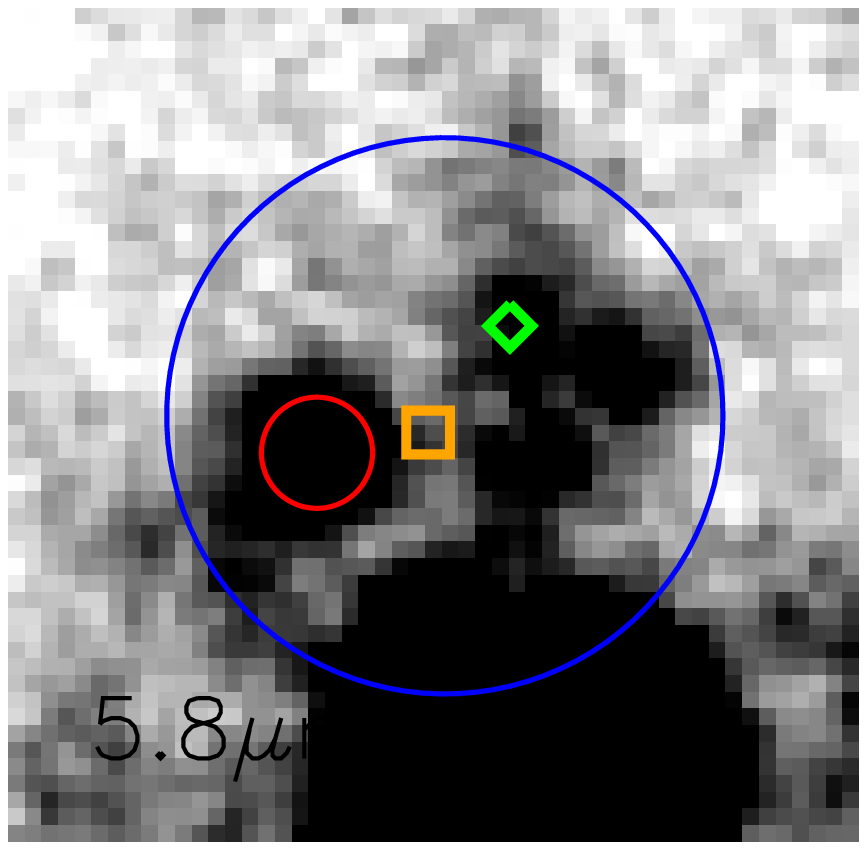}
   }
\parbox[height=20mm]{20mm}{
   \centering
    \includegraphics[width=20mm]{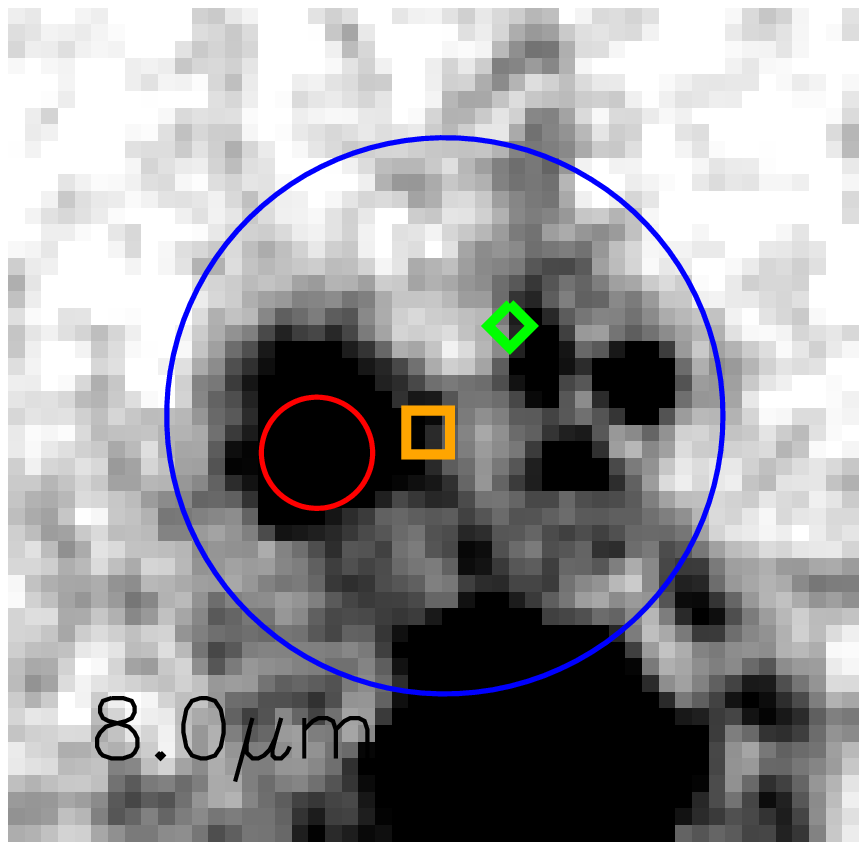}
   }
\parbox[height=20mm]{20mm}{
    \centering
    \includegraphics[width=20mm]{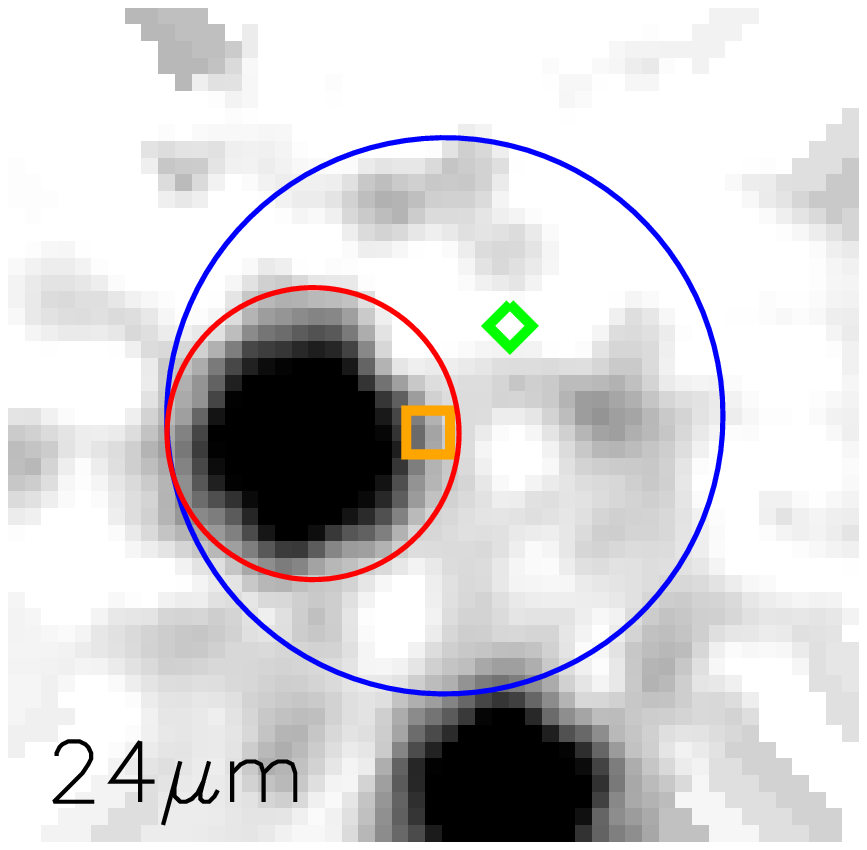}
   }
   
   \parbox[height=20mm]{20mm}{
    \centering
    \includegraphics[width=20mm]{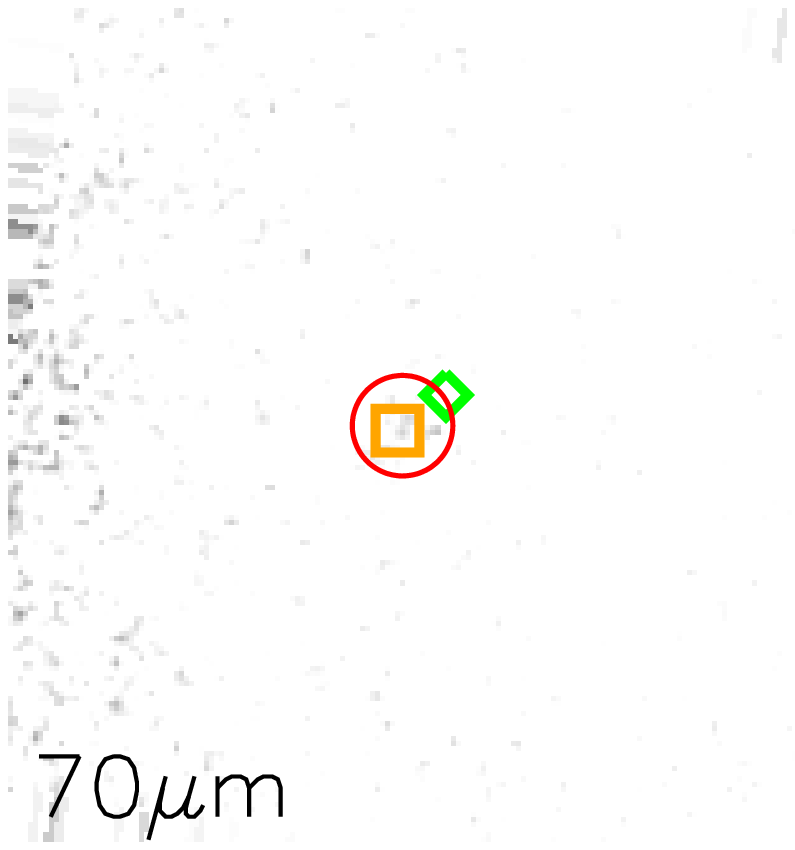}
   } 
 \parbox[height=20mm]{20mm}{
    \centering
    \includegraphics[width=20mm]{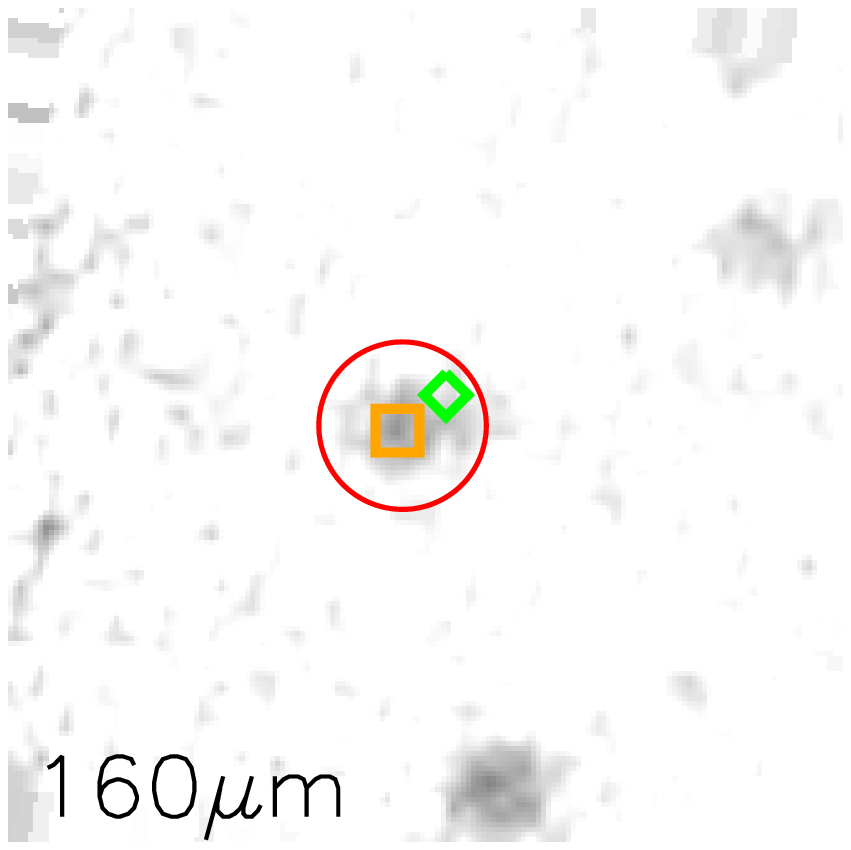}
   } 
 \parbox[height=20mm]{20mm}{
    \centering
    \includegraphics[width=20mm]{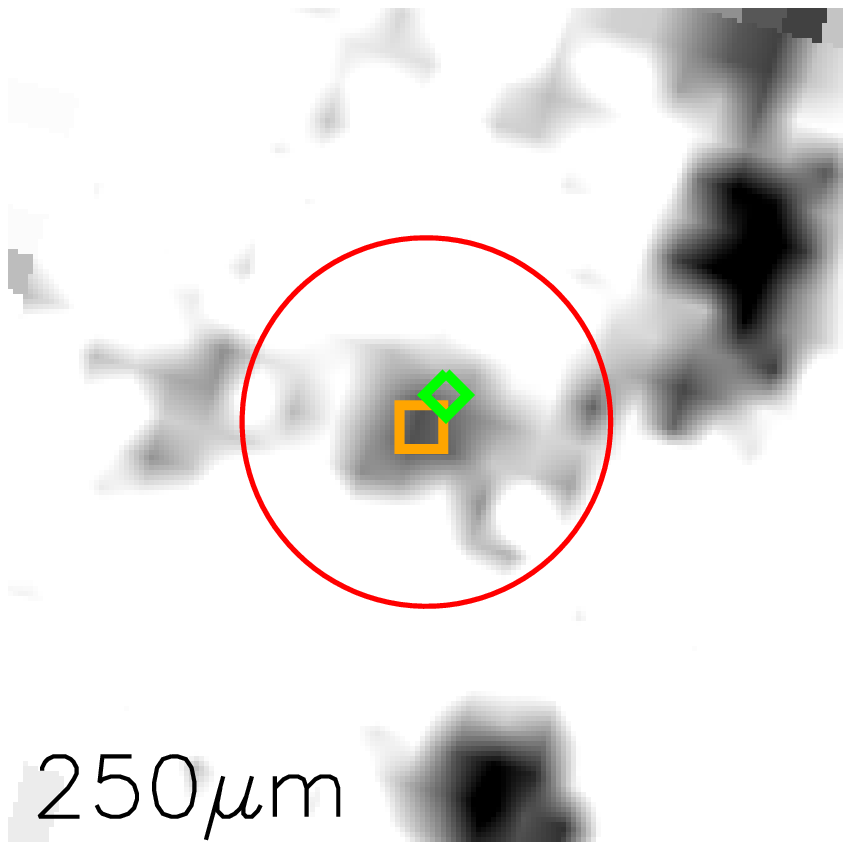}
   } 
 \parbox[height=20mm]{20mm}{
    \centering\hspace{20mm}
   } 
 \parbox[height=20mm]{20mm}{
    \centering\hspace{20mm}
   } 
\end{figure*}
 \begin{figure*}
  \parbox{70mm}{
    \centering
    \includegraphics[width=70mm]{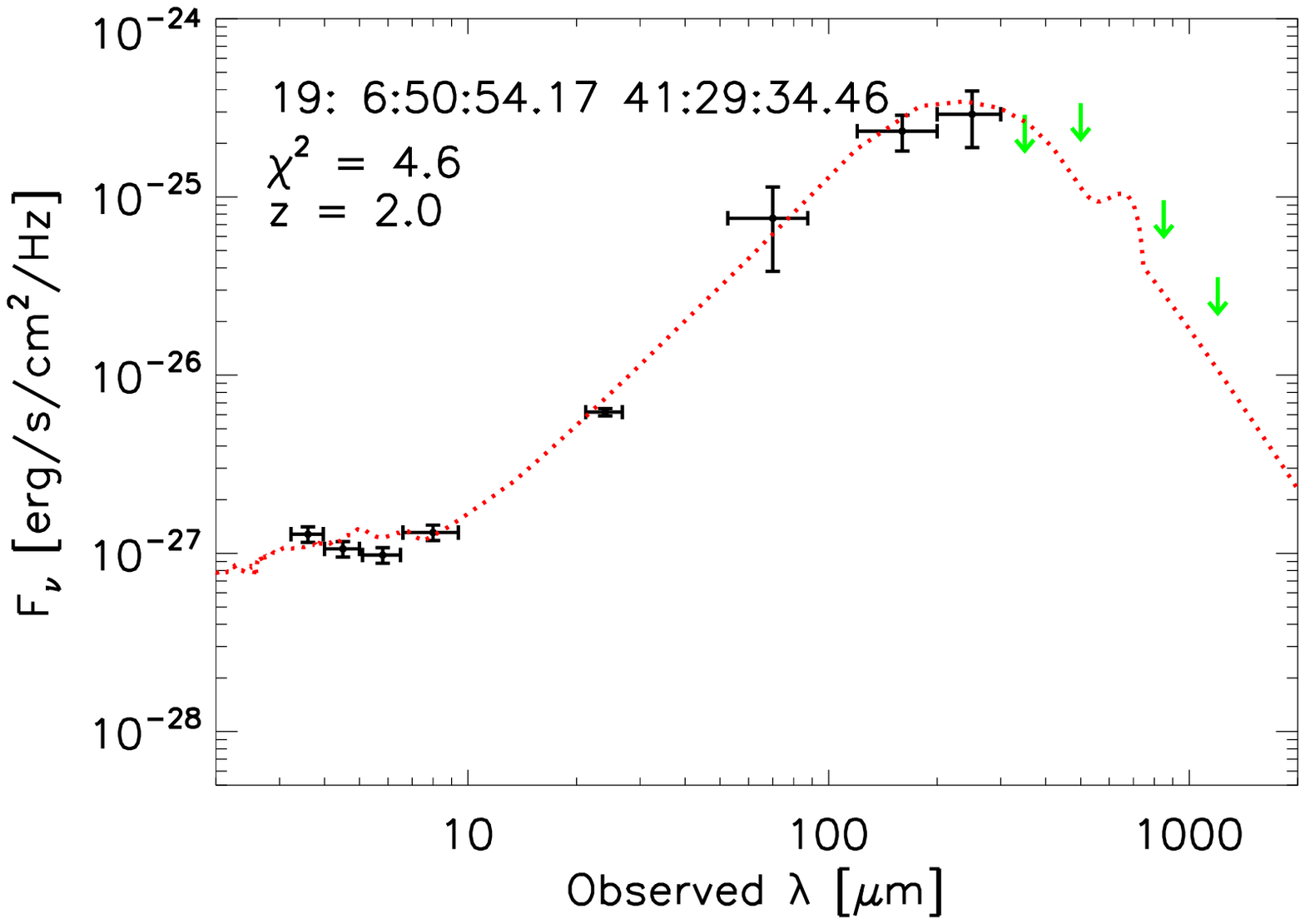}
  }
 \parbox{70mm}{
    \centering
    \includegraphics[width=70mm]{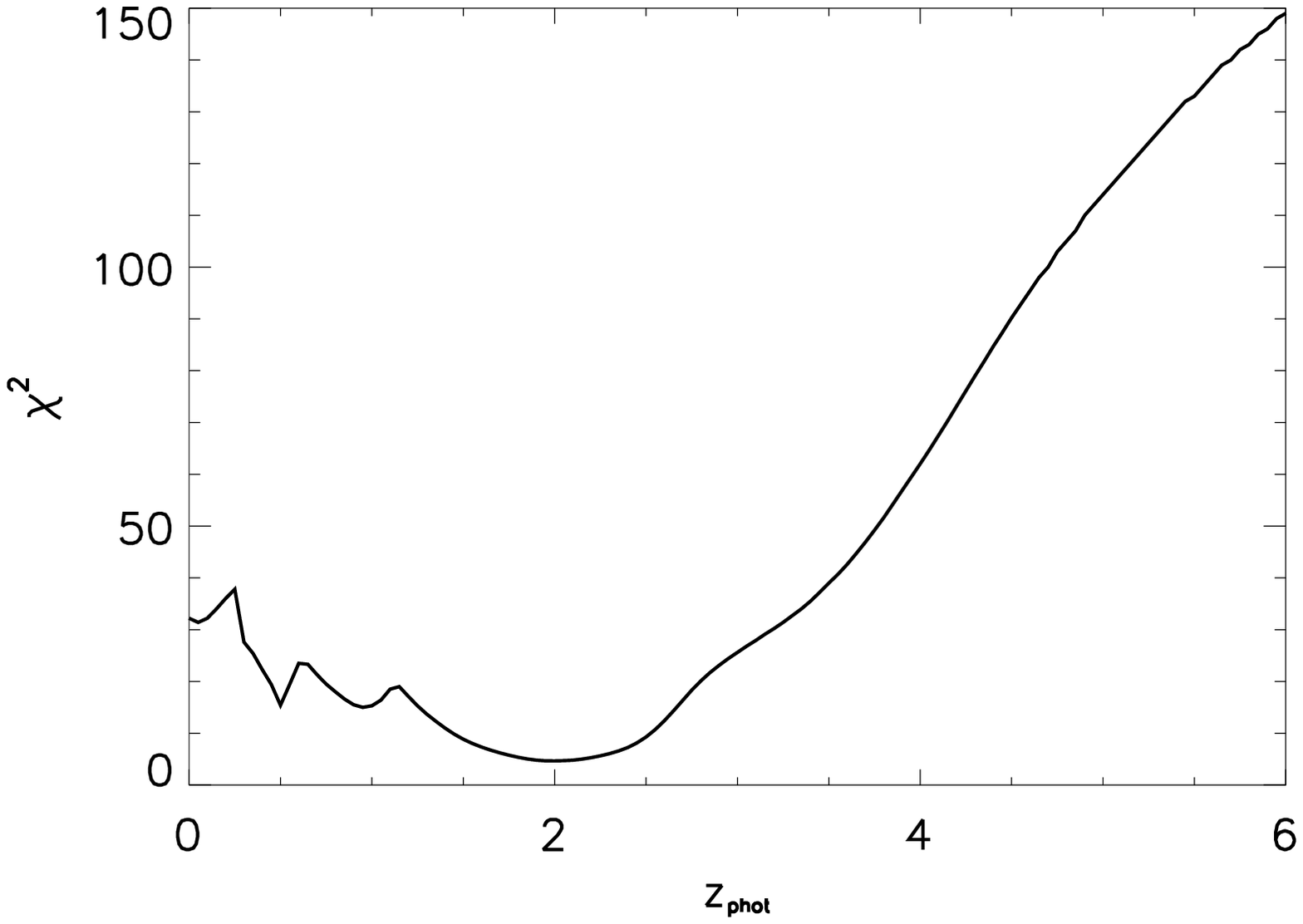}
  }
  \caption{The best $\chi^{2}$ for this source is $z \sim 2.0$, but secondary peaks are more consistent with its spectroscopic redshift ($z_{\rm{spec}}$ = 0.507). Longer wavelength data (e.g. at 850 $\mu$m) are needed to constrain the redshift more accurately. The dust peak at $\sim$ 200 $\mu$m and the upper limits at longer wavelengths strongly constrain the source to be at $z_{\rm{phot}}$ $<$ 3.8.}
\end{figure*}
\begin{figure*}
 \parbox[height=20mm]{20mm}{
    \centering
    \includegraphics[width=20mm]{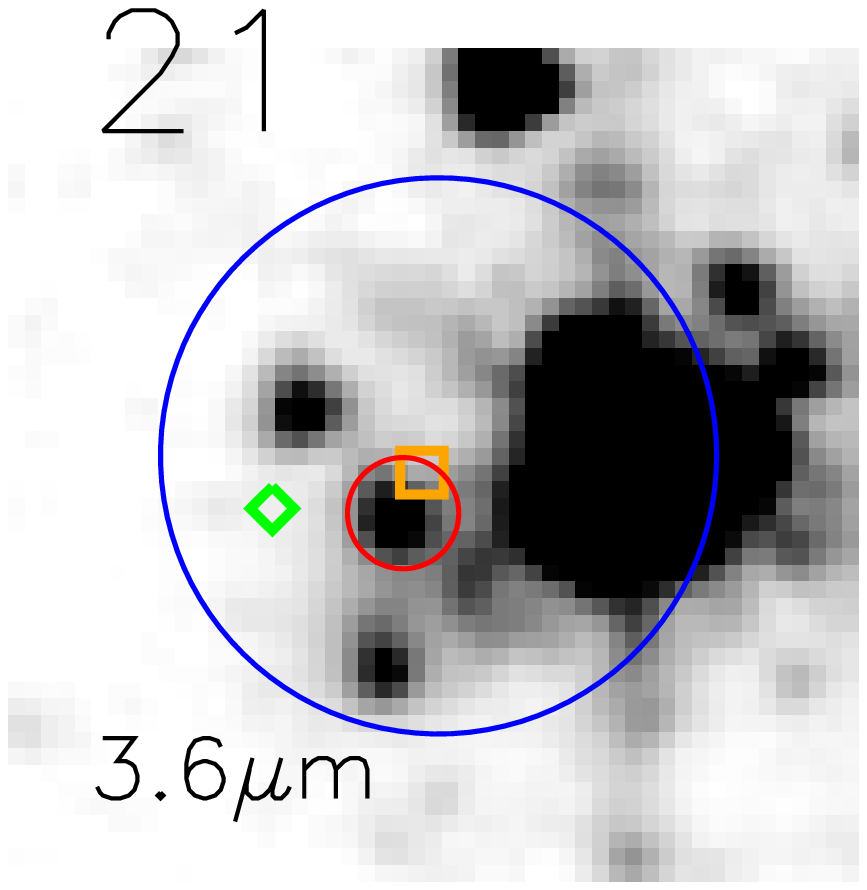}
  }
  \parbox[height=20mm]{20mm}{
    \centering
    \includegraphics[width=20mm]{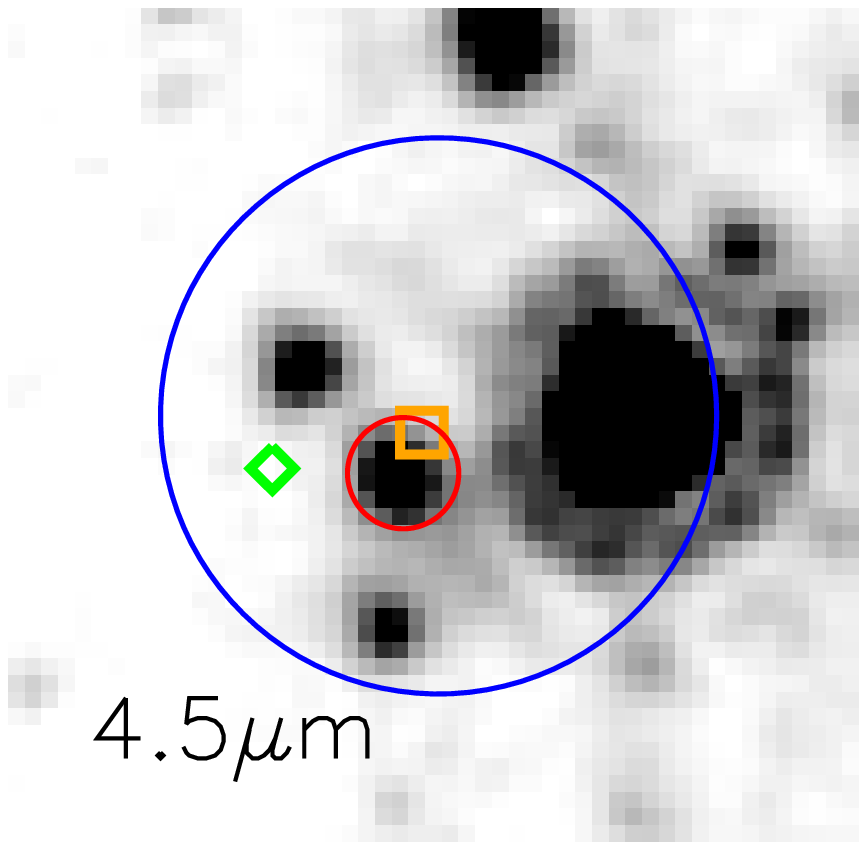}
  }
   \parbox[height=20mm]{20mm}{
    \centering
    \includegraphics[width=20mm]{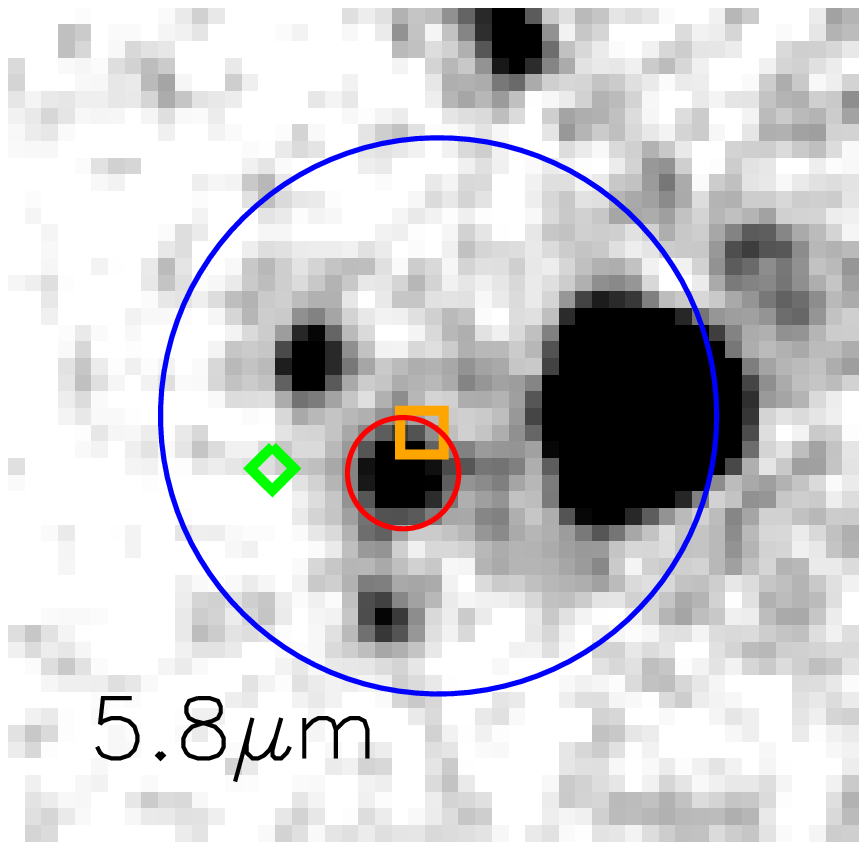}
   }
\parbox[height=20mm]{20mm}{
   \centering
    \includegraphics[width=20mm]{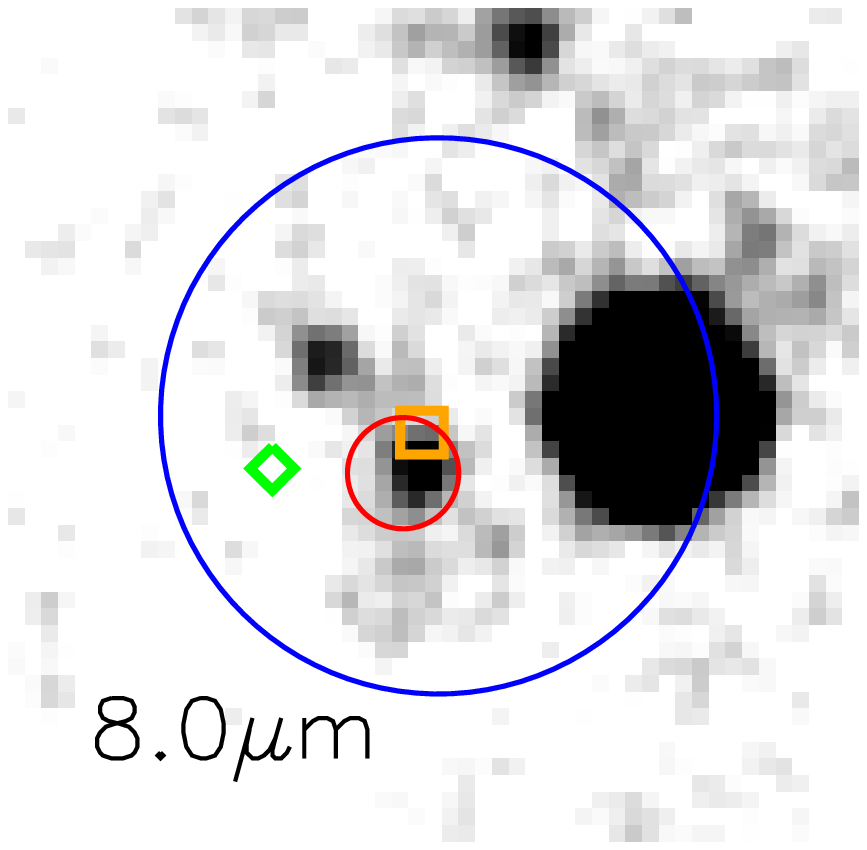}
   }
\parbox[height=20mm]{20mm}{
    \centering\hspace{20mm}
   }
   
   \parbox[height=20mm]{20mm}{
    \centering\hspace{20mm}
   } 
   \parbox[height=20mm]{20mm}{
    \centering\hspace{20mm}
   } 
   \parbox[height=20mm]{20mm}{
    \centering
    \includegraphics[width=20mm]{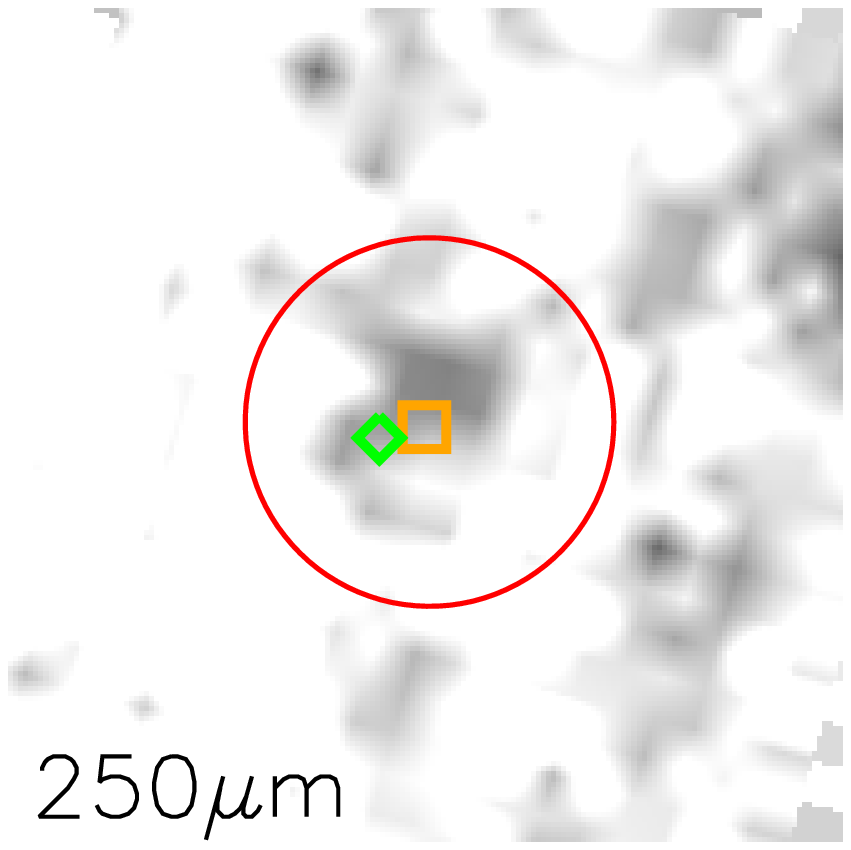}
   } 
   \parbox[height=20mm]{20mm}{
    \centering
    \includegraphics[width=20mm]{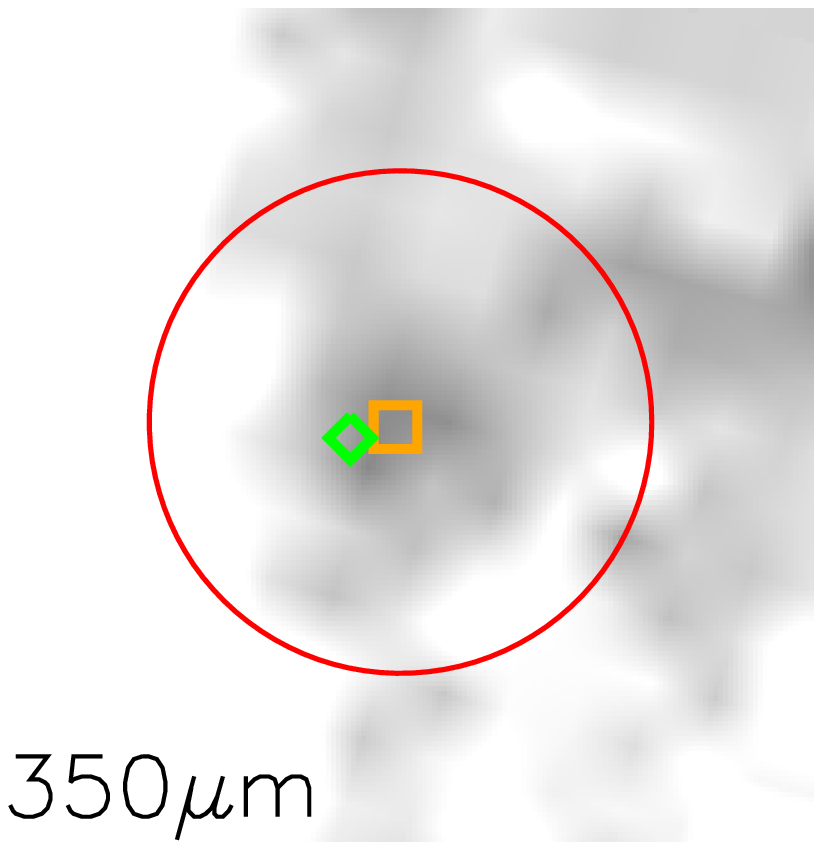}
   } 
   \parbox[height=20mm]{20mm}{
    \centering
    \includegraphics[width=20mm]{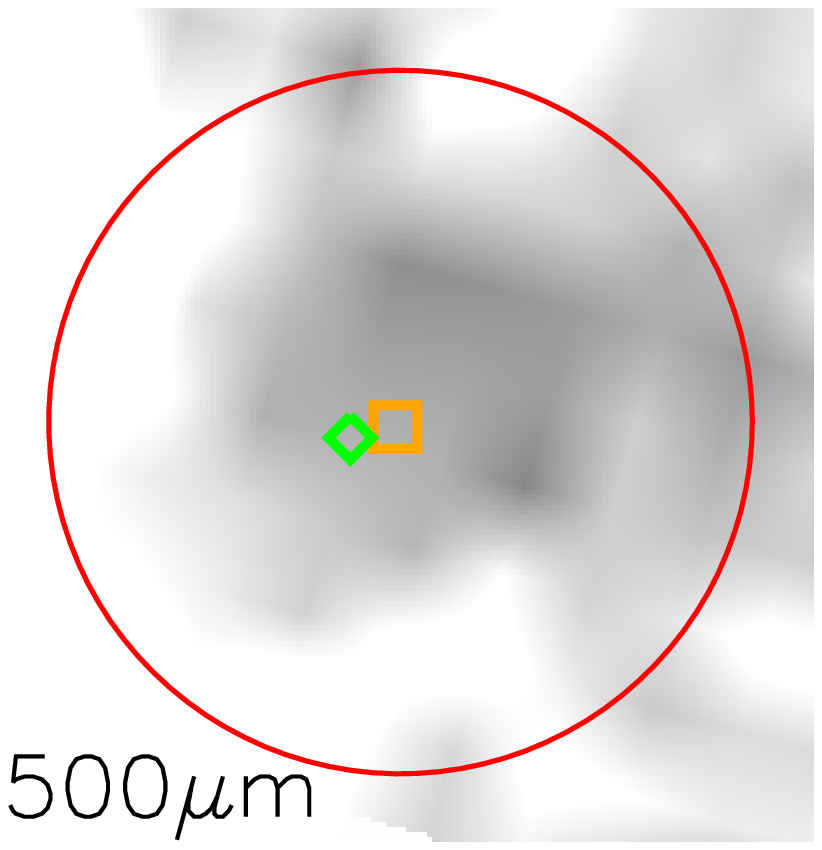}
   } 
\end{figure*}
\begin{figure*}
\parbox{70mm}{
    \centering
    \includegraphics[width=70mm]{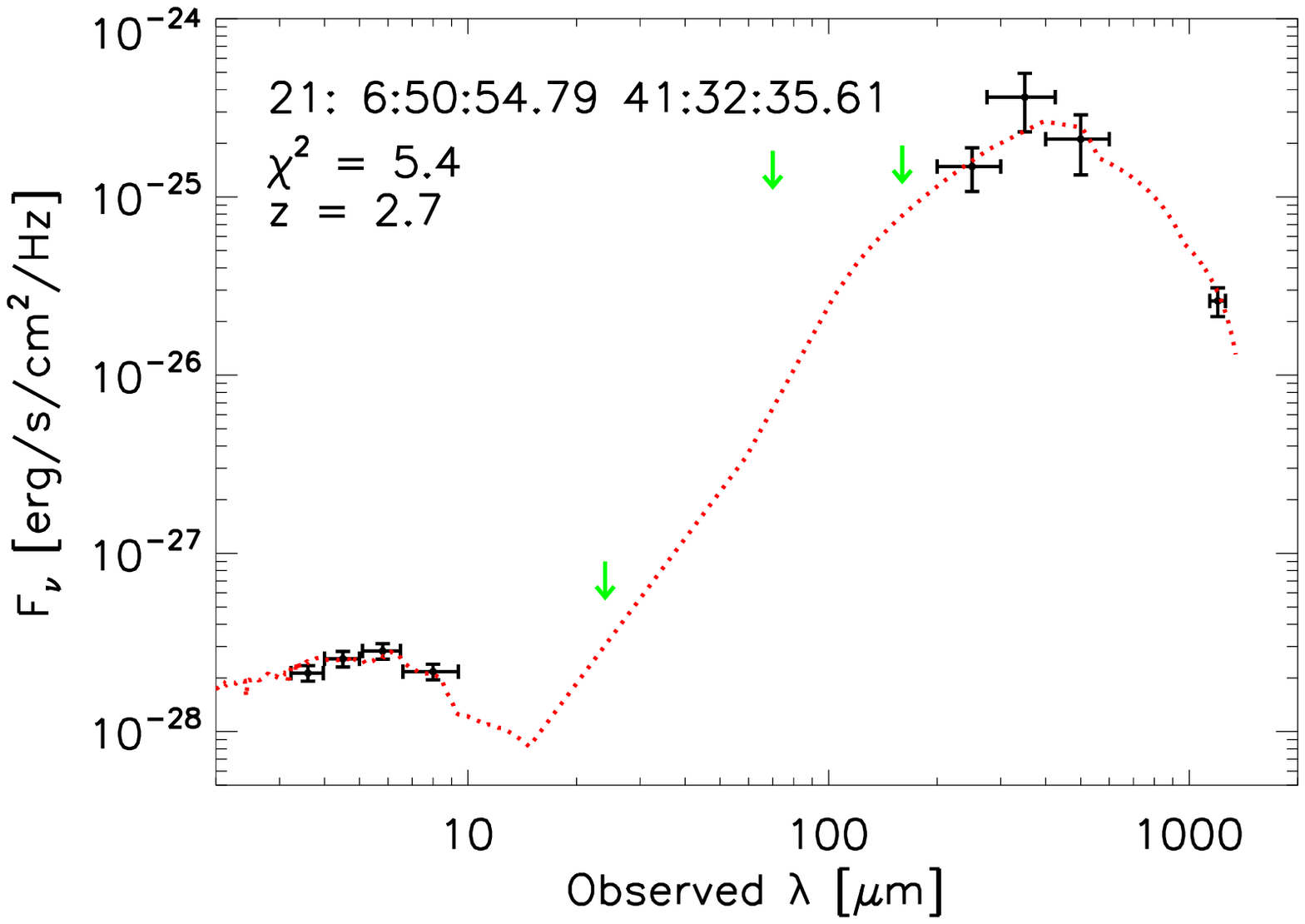} 
  }
\parbox{70mm}{
    \centering
    \includegraphics[width=70mm]{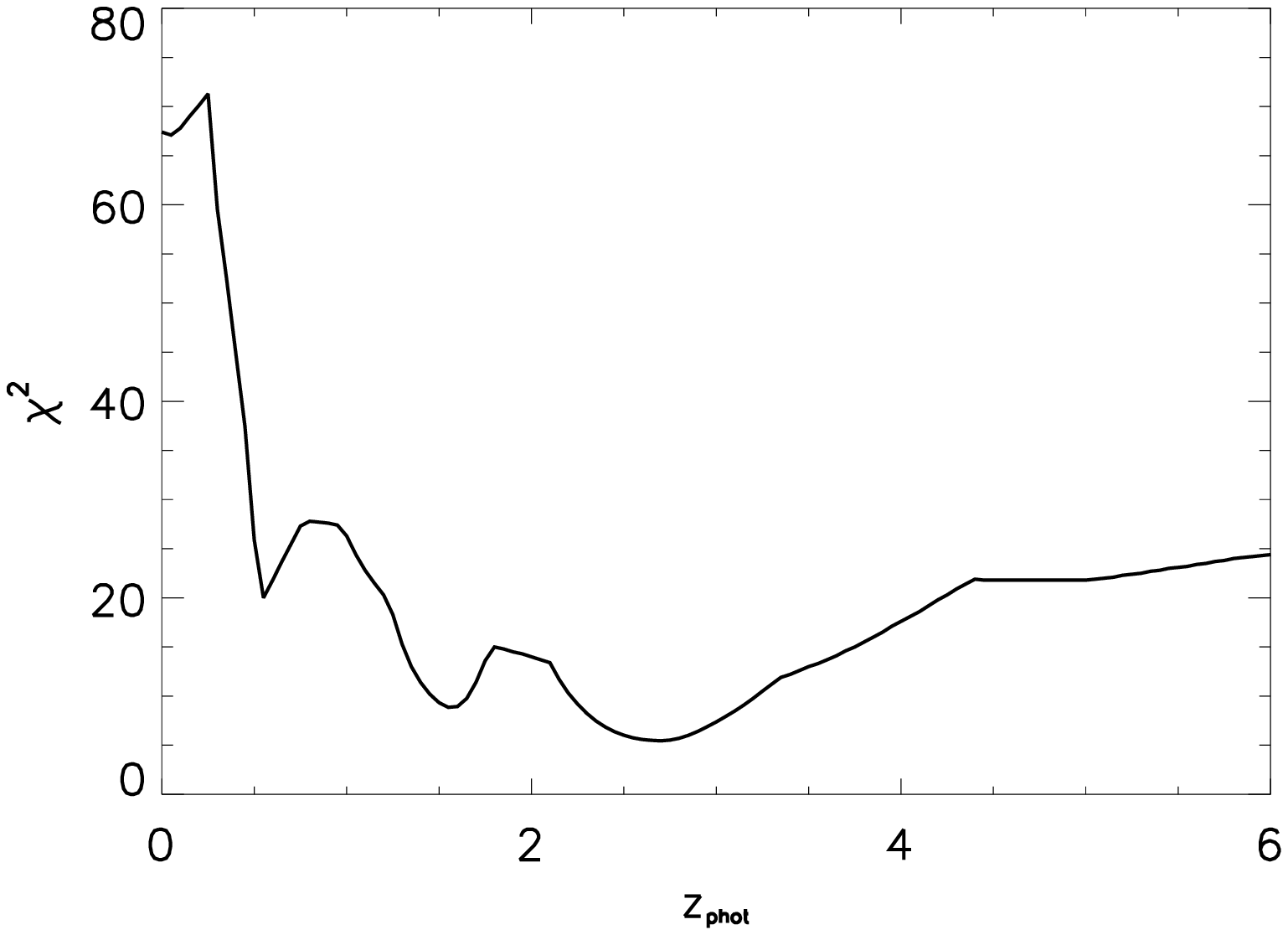} 
  }
  \caption{The stellar bump is clearly observed peaking between the IRAC2 and IRAC3 band. {\tt Hyperz} nicely fits this peak and puts this source at $z_{\rm{phot}}$ $\sim$ 2.7.}
\end{figure*}
\begin{figure*}
\parbox[height=20mm]{20mm}{
    \centering
    \includegraphics[width=20mm]{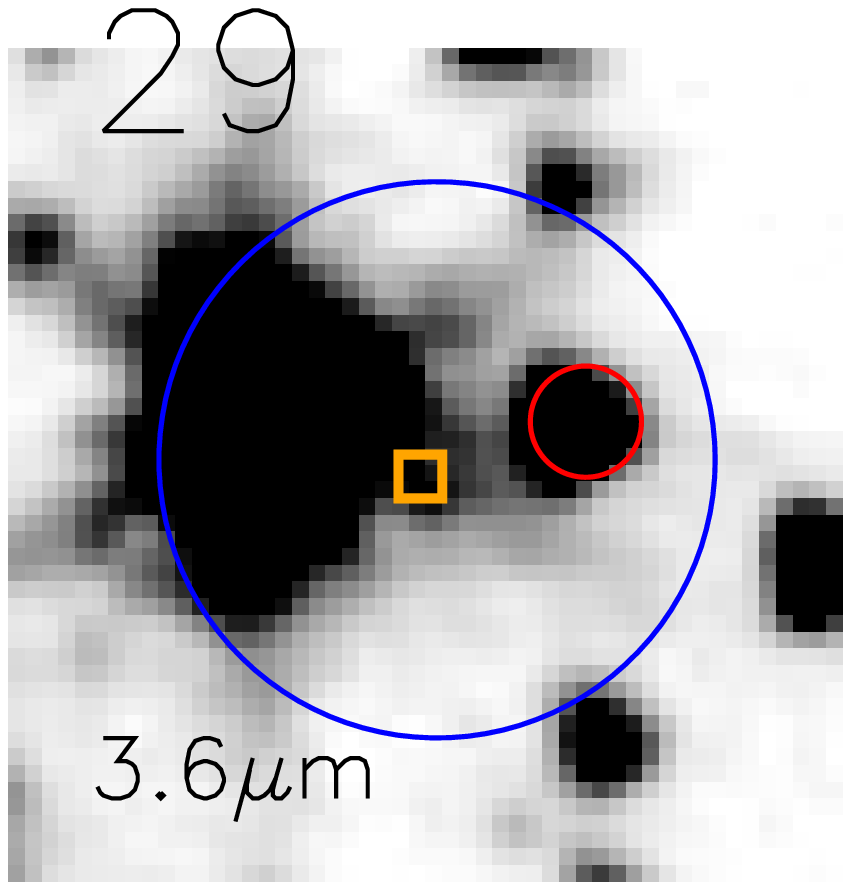}
  }
  \parbox[height=20mm]{20mm}{
    \centering
    \includegraphics[width=20mm]{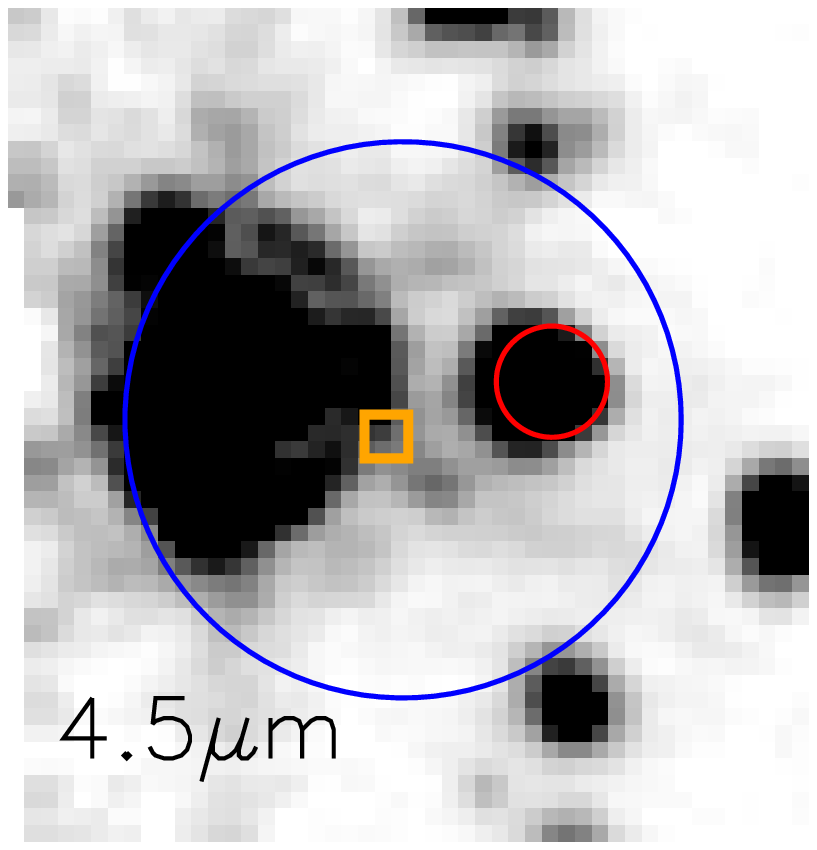}
  }
   \parbox[height=20mm]{20mm}{
    \centering
    \includegraphics[width=20mm]{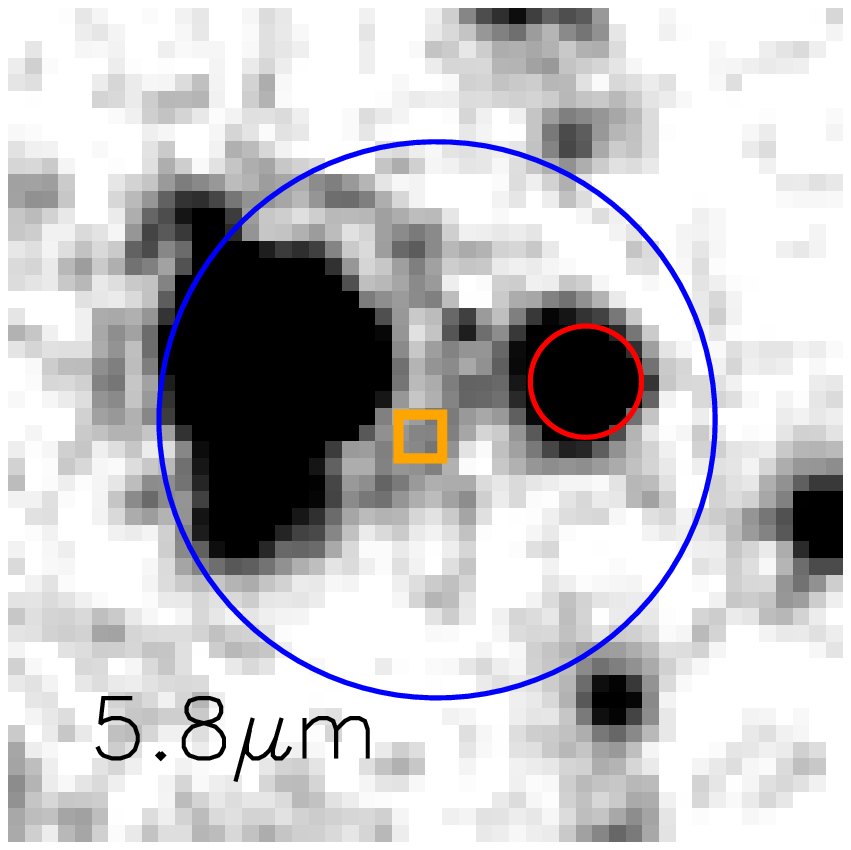}
   }
\parbox[height=20mm]{20mm}{
   \centering
    \includegraphics[width=20mm]{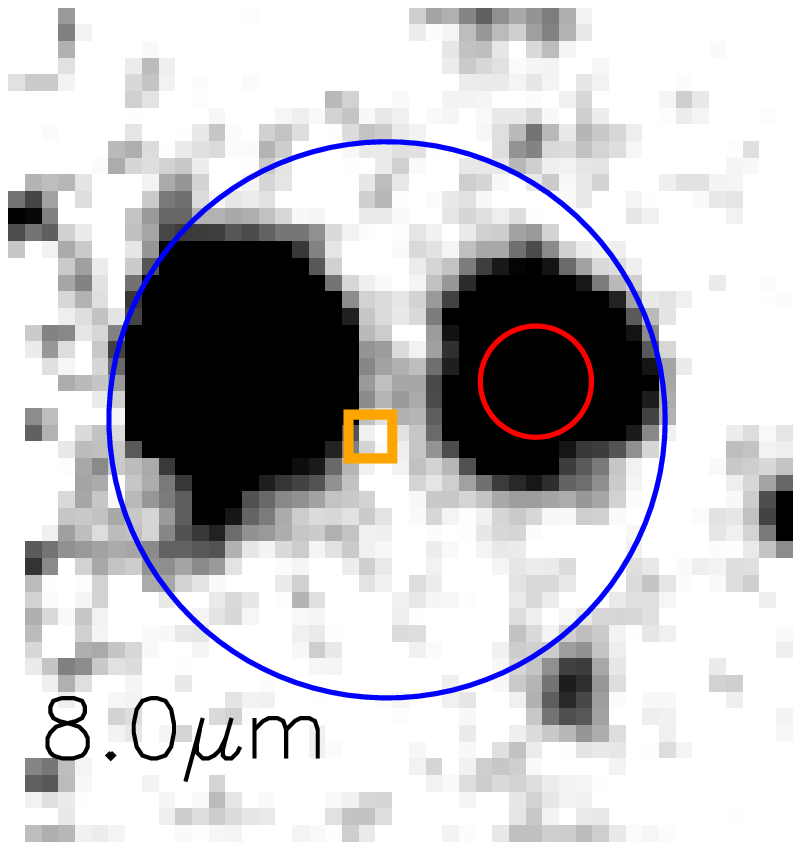}
   }
\parbox[height=20mm]{20mm}{
    \centering
    \includegraphics[width=20mm]{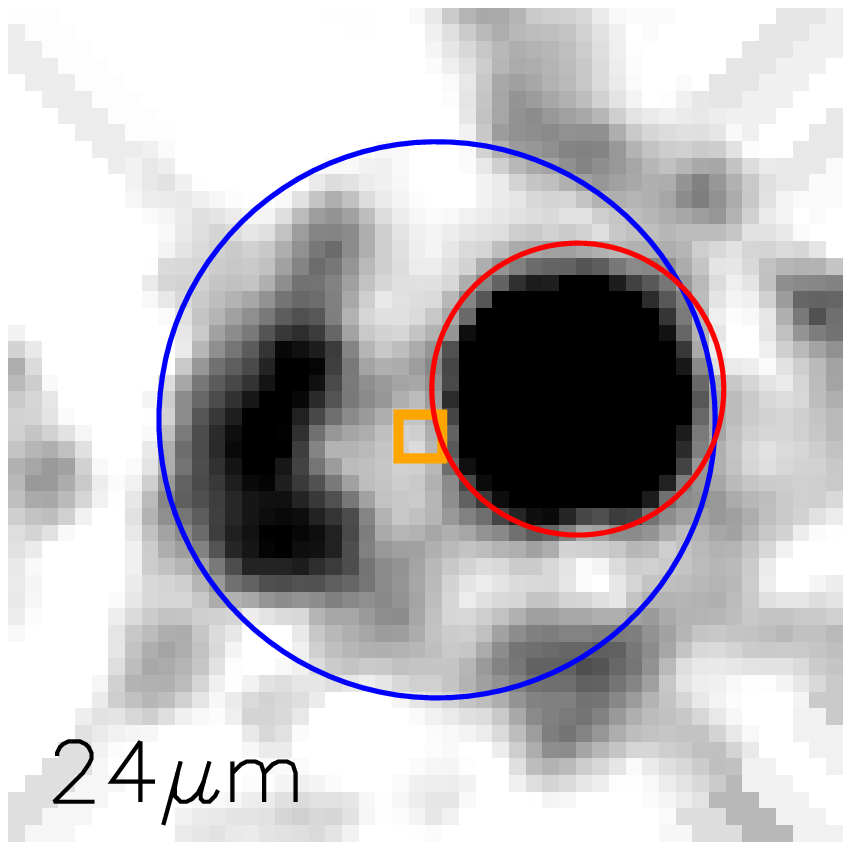}
   }
   
   \parbox[height=20mm]{20mm}{
    \centering\hspace{20mm}
   } 
      \parbox[height=20mm]{20mm}{
    \centering\hspace{20mm}
   } 
      \parbox[height=20mm]{20mm}{
    \centering
    \includegraphics[width=20mm]{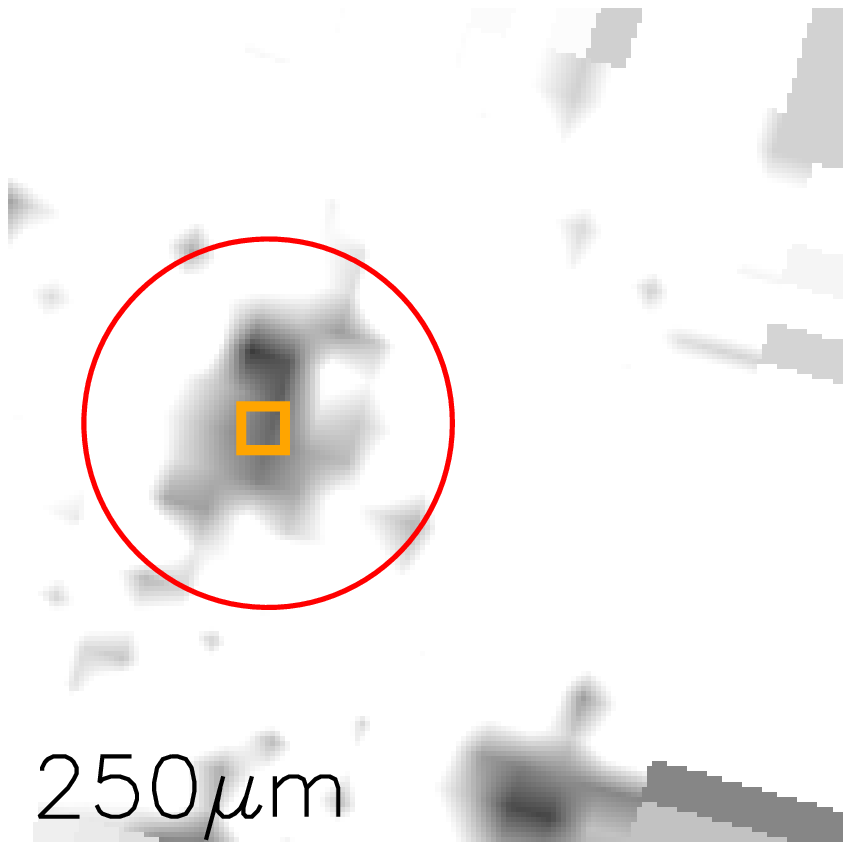}
   } 
      \parbox[height=20mm]{20mm}{
    \centering
    \includegraphics[height=20mm]{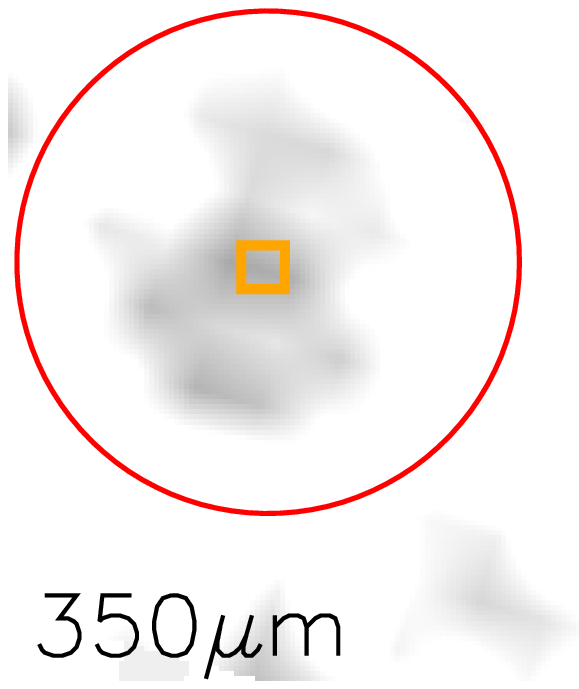}
   } 
      \parbox[height=20mm]{20mm}{
    \centering\hspace{20mm}
   } 
\end{figure*}
\begin{figure*}
  \parbox{70mm}{
    \centering
    \includegraphics[width=70mm]{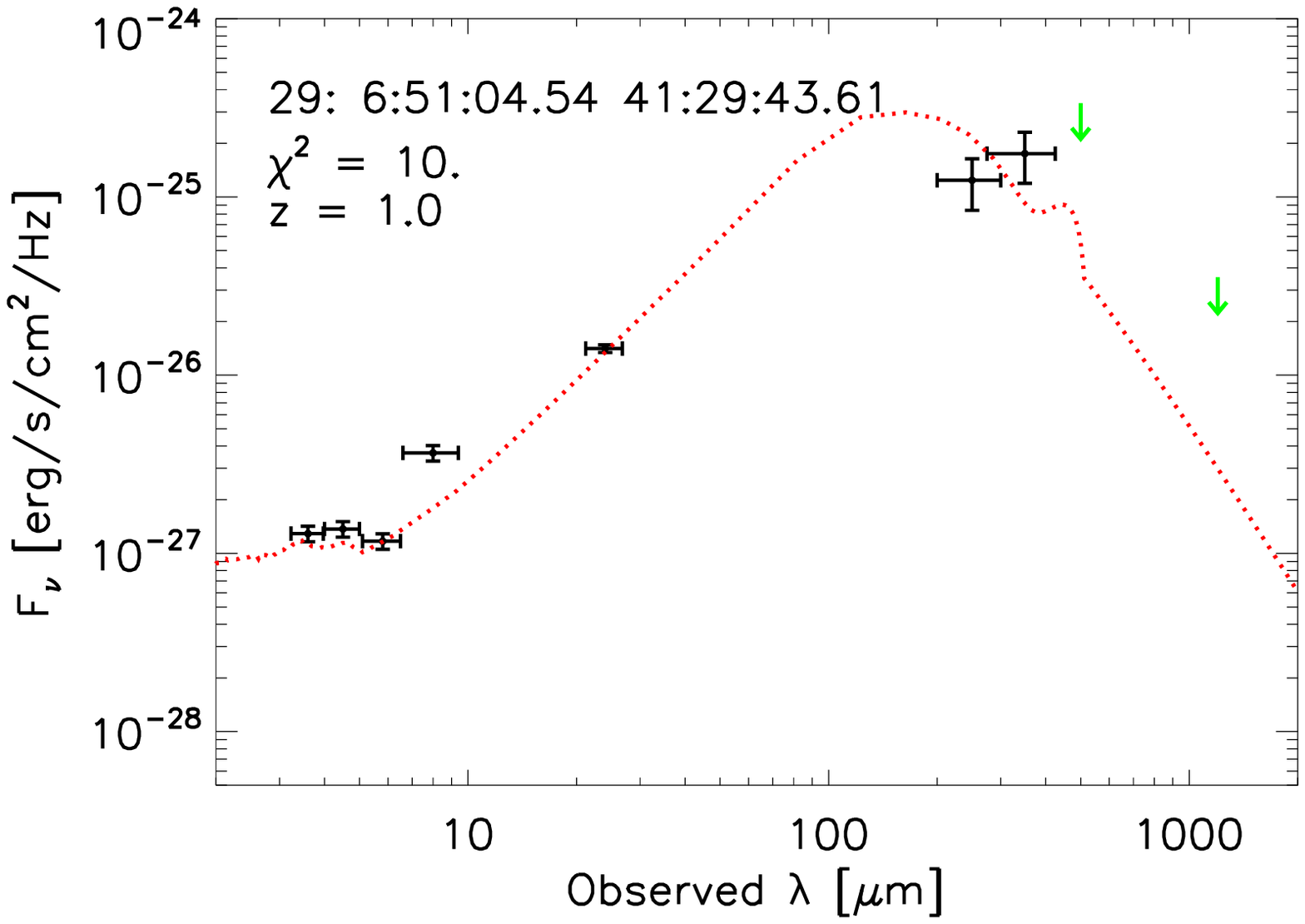}
  }
  \parbox{70mm}{
    \centering
    \includegraphics[width=70mm]{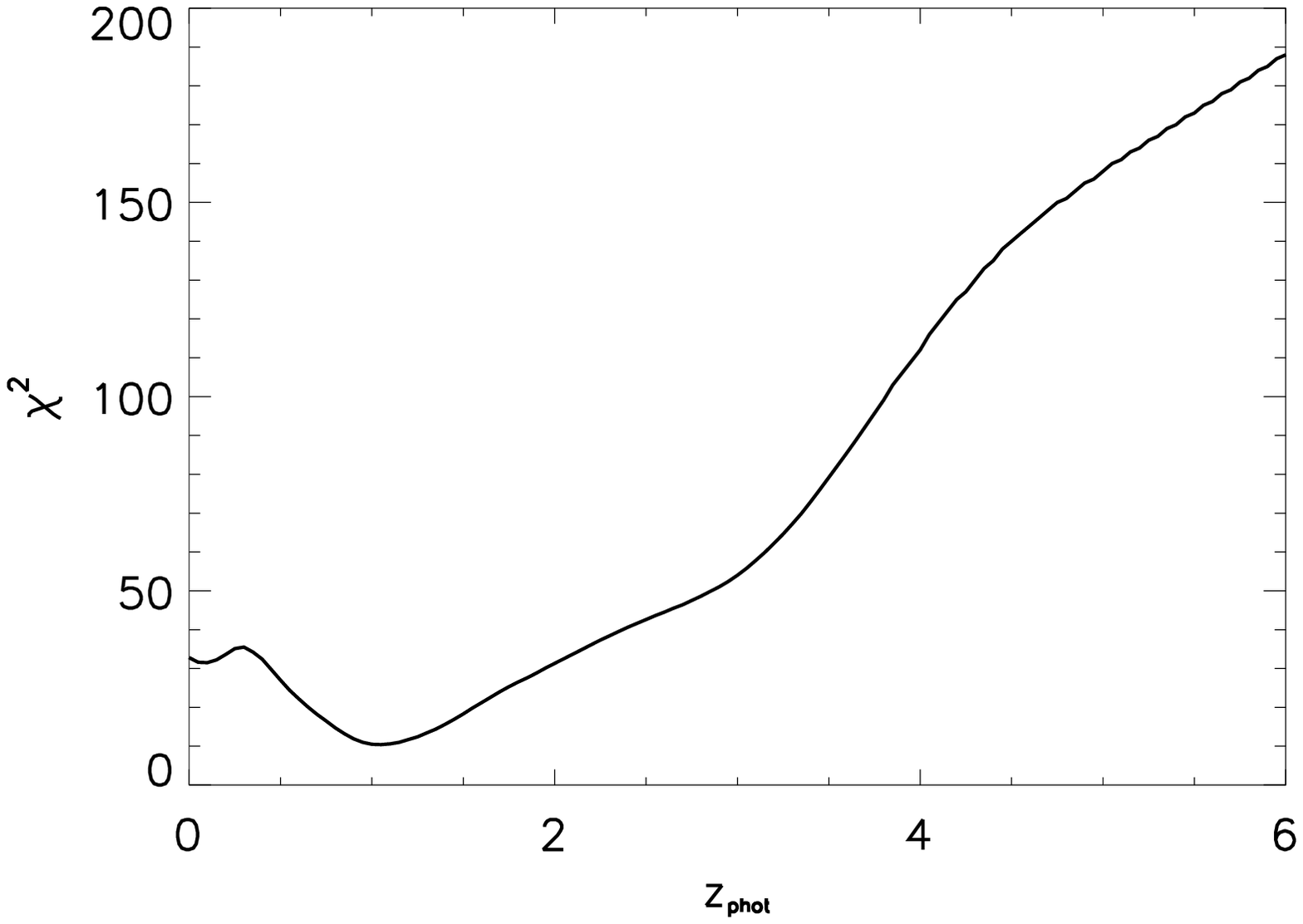}
  }
\caption{A weak stellar bump peaking between the IRAC1 and IRAC2 bands is observed for this source. The overall SED is very similar to source 7 suggesting a low redshift. This is also found by the photometric redshift fitting procedure. The far-IR observations are not well fit but may be due to confusion with another source very bright at 24 $\mu$m and very close ($\sim$ 6 arcsec) to the center of detections. The IRAC photometry, however, is very constraining and the $\chi^{2}$ ditstribution also confirms a clear low redshift for this source.}
\end{figure*}

\bibliographystyle{mn2e}
\bibliography{4c4117}
\label{lastpage}

\end{document}